\documentclass[11pt]{article}
\usepackage{amsmath}
\usepackage{amssymb}
\usepackage{graphics}

\newcommand{\tr}{\operatorname{tr}}
\newcommand{\opname}[1]{\operatorname{#1}}
\newcommand{\Slash}[1]{{#1}\negthickspace{\negthickspace{\slash}}}
\newcommand{\sigmacl}{\sigma_{\text{cl}}}

\newcommand{\Veff}{V_{\text{eff}}}
\newcommand{\hatVeff}{\hat{V}_{\text{eff}}}
\newcommand{\barDEight}{\overline{\text{D8}}}
\newcommand{\cm}{\checkmark}
\newcommand{\yKK}{y_{\text{KK}}}
\newcommand{\uKK}{u_{\text{KK}}}
\newcommand{\UKK}{U_{\text{KK}}}
\newcommand{\BNS}{B_{\text{NS}}}
\newcommand{\tBNS}{\tilde{B}_{\text{NS}}}

\begin{document}

\title{A Comparative Study of NJL and Sakai-Sugimoto Models}
\author{D.~Yamada
        \bigskip
        \\
       {\it Racah Institute of Physics},
       {\it The Hebrew University of Jerusalem},
        \smallskip
       \\
         {\it Givat Ram, Jerusalem, 91904 Israel}
          \bigskip
       \\
         {\tt daisuke@phys.huji.ac.il}}
\date{}
%\date{Starting: August 5th 2009}
%If there's no argument, the date is skipped.
%If date command is ommited, the system date is automatically inserted.
\maketitle

%%%%%%%%%%%%%%%%%%%%%%%%%%%%%%%%%%%%%%%%%%%%%%%%%%%%%%%%%%%%%%%%%%%%%
\begin{abstract}
%%%%%%%%%%%%%%%%%%%%%%%%%%%%%%%%%%%%%%%%%%%%%%%%%%%%%%%%%%%%%%%%%%%%%
The Nambu--Jona-Lasinio and Sakai--Sugimoto models are juxtaposed,
focusing on the models' dynamically generated masses and the phase diagrams.
The models are studied in the parameter space of temperature,
constant electromagnetic fields and the quark chemical potential.
When the chemical potential is zero or small, the models show good
agreement.
However, the responses of the models under the large chemical potential
are very different.
In the presence of the background magnetic field, 
while the large chemical potential triggers
the de Haas-van Alphen oscillation
and the boundstate destabilization effect of the magnetic field
in the NJL model, the Sakai-Sugimoto
model completely fails to capture these phenomena, indicating the
lack of the fermi sphere in the latter model.
Also the orders of the chiral phase transitions are generally different.
\end{abstract}

\pagebreak

%%%%%%%%%%%%%%%%%%%%%%%%%%%%%%%%%%%%%%%%%%%%%%%%%%%%%%%%%%%%%%%%%%%%%
\tableofcontents
%%%%%%%%%%%%%%%%%%%%%%%%%%%%%%%%%%%%%%%%%%%%%%%%%%%%%%%%%%%%%%%%%%%%%

\pagebreak

%%%%%%%%%%%%%%%%%%%%%%%%%%%%%%%%%%%%%%%%%%%%%%%%%%%%%%%%%%%%%%%%%%%%%
\section{Introduction}\label{sec:intro}
%%%%%%%%%%%%%%%%%%%%%%%%%%%%%%%%%%%%%%%%%%%%%%%%%%%%%%%%%%%%%%%%%%%%%
It has been a while since QCD, the color dynamics, was generally accepted
as the theory of strong interaction.
Our firm confidence in it mainly comes from the remarkable successes at
high energy.
In the regime of low energy, however, the degrees of freedom are apparently
different from the QCD fundamental particles and we still do not quite
understand the mechanism that governs this change in the degrees of
freedom.
The essence of the problem is the strongly coupled nature of QCD at low
energy and the lack of powerful and efficient tools to tackle
such theories.
Currently, the best tool available is the lattice simulation of QCD
and it has been providing affirmative results for QCD even in the low
energy regime.
Though the lattice simulation is powerful, it has its own weaknesses.
For example, it suffers from the notorious ``sign problem'' when
chemical potentials are included in the theory.
Therefore, different approaches to the low energy QCD are both essential
and important.
One such approach is the construction of low energy effective theories,
guided by the symmetries of QCD.
The famous models along this line are the Nambu--Jona-Lasinio (NJL)
model \cite{Nambu:1961tp} and the chiral perturbation theory, and
they are both known to be phenomenologically viable.
The prominent feature of the NJL model, in particular, is the dynamical
breaking of the chiral symmetry.
Because of this feature and the sheer simplicity, this model has been
popular, despite the model's deficiencies, such as the lack of the color gauge
fields and confinement, also the model is nonrenormalizable and the
cutoff must be explicitly introduced as a part of the definition of
the theory.

On the other hand, a very different method to handle strongly coupled
field theories emerged as the AdS/CFT correspondence or holography
\cite{Maldacena:1997re,Gubser:1998bc,Witten:1998qj}.
It is the duality between string theory on the AdS space and 
a conformal field theory.
In the low energy limit, the duality boils down to the equivalence
between a gravitational system and a strongly coupled conformal field theory,
and therefore, to obtain the information about the latter, one can instead
study the gravitational theory, which can be easier to analyze.
The dual field theory in the original correspondence is not like QCD,
and the search for the holographic dual of QCD has been naturally
conducted intensively.
One of the most attractive models out of the effort is
the Sakai-Sugimoto (SS) model \cite{Sakai:2004cn}.
What makes this model attractive is the intuitive geometric implementation
of the dynamical chiral symmetry breaking.
Furthermore, it is remarkable that one can obtain the spectrum of the theory, 
such as mesons, in relative ease.
Therefore, even though the model does not exactly correspond to QCD
in many parameters, such as the number of
the color degrees of freedom, the spectrum of the SS model has been
naively compared with the QCD experimental and lattice data.
Surprisingly, the data agree fairly well, and it is currently
popular to suppose a sort of ``universality class'' among QCD-like strongly
coupled theories with various differences in parameters.

Even with those successes, it is difficult to directly compare
the SS model and QCD in more details, such as the response of the theories
under external electromagnetic field and chemical potentials.
This, of course, is because the strongly coupled QCD cannot be solved
analytically and the corresponding experimental setup is hard to achieve.
This makes it hard to tell to what extent the holographic model captures
the features of QCD.
In other words, it is unclear how universal the universality is.
Therefore, in this paper, we turn to the NJL model, instead of QCD, to
contrast with the SS model.
The main motivation is the dynamical chiral symmetry breaking of the models
and this allows us to investigate the chiral phase diagrams of the theories
in various external parameters and compare.
Additionally, one can compute the dynamically generated masses in both theories
and collate the response of the masses under different external conditions.
We focus on those two properties of the models in the comparison.
In order to dig out the similarities and dissimilarities, we scrutinize
the responses of the models in the parameter space of temperature,
background constant electromagnetic fields and the quark chemical potential.
Each model has an extra parameter, the four-fermion coupling for NJL
and the asymptotic inter-brane distance for SS, and we examine the models
with respect to those parameters as well.

We consider both theories for $N_f=1$, this is because of its simplicity,
and more importantly because
the constant background fields introduced
in the SS model in this case exactly correspond to the electromagnetic
fields (we will mention this point more carefully in Section~\ref{sec:SS}).
However, we remark that the generalization to $N_f>1$ is straightforward.
We take the large $N_c$ limit of the NJL model and the cutoff of this
nonrenormalizable theory is left
as a parameter and we measure all the quantities in terms of this scale.
Similarly, the quantities in the SS model are all measured in the units
of the curvature scale, $R$.
As for the NJL coupling, we investigate its relation to the inter-brane
distance of the SS model.
(The parameters of the NJL model are usually adjusted to be consistent 
with the experimental data.
Most commonly, the NJL coupling and the cutoff
are tuned to reproduce the scalar density and the pion decay
constant from the experiment.
See {\it e.g.}, References~\cite{Hatsuda:1994pi,Klevansky:1992qe}.)

The general approaches to the models are the followings.
The effective potential of the NJL model is derived by utilizing
the worldline formalism \cite{Strassler:1992zr,Schubert:2001he}.
This is known as ``the string inspired method'' and the reader might
feel more comfortable than other methods.
Finite temperature is introduced by the imaginary time formalism.
We pay attention to the subtlety with the gauge invariance in the presence
of the 
topologically non-trivial time circle, and show that the careful treatment
results in the automatic introduction of the chemical potential.
For the SS model, the background fields and the chemical potential
are introduced from the constant NSNS-$B$ field, similar to
Reference~\cite{Erdmenger:2007bn}.
We touch on the issue of the topological constraints on the constant fields
and argue that the violation of the constraint directly leads to the
instability of the system.

Since both models are studied extensively, it is worthwhile to summarize
what are known and what are the novel aspects of this work.
First for the NJL part.
Among many others, the work by Shovkovy in Reference~\cite{Shovkovy:1998xw}
comes closest to our goal.
(See Reference~\cite{Haack:1998uy} for the similar work.)
The worldline formalism is adopted in the reference for
the derivation of the one-loop effective potential at finite temperature
in the context of QED.
The one-loop computation is exact in the large $N_c$ limit and
the resulting effective potential is identical to the NJL model.
However, as pointed out by Gies \cite{Gies:1998vt}, Shovkovy glosses over
the subtlety of the gauge invariance at finite temperature.
As mentioned above, we treat this issue carefully and
reproduce Gies' result in the worldline formalism.
In addition to reproducing the result of Reference~\cite{Gies:1998vt},
we apply the Poisson (re)summation formula to obtain the effective potential
in more physically transparent and numerically useful form and 
explain how to treat large values of the chemical potential.
Then we proceed to actually solve the theory numerically as the NJL model
(and not as QED).

The notable references for the worldline formalism at zero temperature
are \cite{Strassler:1992zr,Reuter:1996zm} and the review paper
by Schubert \cite{Schubert:2001he}.
We mention that the more popular approach than the worldline formalism
is the propertime formalism
and it originates from Schwinger's paper \cite{Schwinger:1951nm}.
The notable and relevant works based on the propertime formalism include
References~\cite{Harrington:1974xh,Elmfors:1993bm,Elmfors:1994fw,Gusynin:1995nb,Gies:1998vt}.

As for the numerical evaluation of the NJL model, the work by
Inagaki {\it et al.} in Reference~\cite{Inagaki:2003yi}
explores the model in the parameter space
of temperature, constant background magnetic field and the chemical
potential.
Though their derivation of the effective potential is based on
the propertime formalism, the results significantly overlap with ours.
However, since we are not comparing the model with the experimental data
but with the SS model, the parameter space explored in our work
includes the coupling constant.
Moreover, we provide considerably deeper investigation 
and emphasize the importance
of the fermi sphere, which causes the major differences between the NJL
and SS models.

Another relevant work is Reference~\cite{Klevansky:1989vi} of
Klevansky and Lemmer, which
deals specifically with
the pure electric field background at zero temperature, and
the result overlaps with some of ours.
They find the second order chiral phase transition with respect
to the external electric field and we reproduce this result.
However, in addition to the
second order phase transition, we uncover the first order
transition at a higher value of the NJL coupling and
we believe that this first order chiral phase transition is reported
for the first time.
Furthermore, we examine the real and imaginary parts of the effective
potential and discuss the validity of the conclusion drawn from
the real part.
(The imaginary part is not discussed in Reference~\cite{Klevansky:1989vi}.)
We then turn on the temperature, in the background of the electric field,
and handle this case by the weak background field expansion.
Though we are not able to go too far in this direction, we still obtain
some information about the dynamical mass and the phase structure.
The attempt here is largely new, though the weak field expansion
has been discussed briefly in Reference~\cite{Elmfors:1994fw}.

In above, we have listed papers of the direct relevance to our aim, but
there are, of course, many more papers on the NJL model.
For those papers, we refer to the review papers
\cite{Hatsuda:1994pi,Klevansky:1992qe}.

Now for the SS part.
Aharony {\it et al.} have introduced finite temperature
in the SS model in the paper \cite{Aharony:2006da} 
and we adopt their setup.
The most important result of the paper for us is the existence of
the chiral symmetry breaking in the deconfined phase.
We will see that the NJL model has direct counterparts to the chiral
behaviors of the SS model in the deconfined phase.
The constituent mass of the holographic theories was first discussed
by Casero {\it et al.} in Reference~\cite{Casero:2005se}
for slightly different model from the SS model but exactly the same idea
applies to SS and this is pointed out and examined in
References~\cite{Aharony:2006da,Peeters:2006iu}.

The references listed thus far deal with the single external parameter,
temperature.
The introduction of the chemical potential in the SS model was first
discussed by Horigome and Tanii in Reference~\cite{Horigome:2006xu}.
Even though the way we introduce the chemical potential is slightly
different from theirs, we essentially follow and reproduce their work
as a special case.

The SS model under the influence of the constant background electromagnetic
field has been extensively studied by the group in
Haifa~\cite{Bergman:2008sg,Bergman:2008qv,Lifschytz:2009si}
and by others \cite{Johnson:2008vna,Kim:2008zn,Johnson:2009ev}.
Reference~\cite{Bergman:2008sg} obtains the $E$-$T$ and $B$-$T$ chiral phase
diagrams and Reference~\cite{Johnson:2008vna} works out
the $B$-$T$ phase diagram and the response of the dynamical quark
mass ($m^*$) with respect to the $B$ field.
In addition to reproducing these results, we also produce $\mu$-$T$,
$B$-$\mu$ chiral phase diagrams and $T$-$m^*$, $E$-$m^*$ graphs.
In Other words, we work out all the properties that have counterpart
in the NJL model.

The juxtaposition of the NJL and SS models is the aim of this paper
and this comparative study is the main novelty of this work.

Finally, the table of contents
is a good guidance to the structure of this article.

%%%%%%%%%%%%%%%%%%%%%%%%%%%%%%%%%%%%%%%%%%%%%%%%%%%%%%%%%%%%%%%%%%%%%
\section{NJL Model}\label{sec:NJL}
%%%%%%%%%%%%%%%%%%%%%%%%%%%%%%%%%%%%%%%%%%%%%%%%%%%%%%%%%%%%%%%%%%%%%
We start with the NJL model.
Good review papers on this model are 
References~\cite{Hatsuda:1994pi,Klevansky:1992qe}, but the following
discussions will be self-contained for our purposes.
As mentioned in the introduction, we consider the NJL model with
one flavor in large $N_c$ limit and
some parts of the discussion are specialized for this case.
We also note that throughout this work we use the Euclidean sign convention,
including the $\gamma$ matrices.

The NJL Lagrangian is
\begin{align}\label{eq:1stL}
  \mathcal{L}= \bar q i \Slash{\partial} q
    + \frac{g^2}{2N_c\Lambda^2} 
      \big[ (\bar q q)^2 + (\bar q i\gamma_5 q)^2 \big]
  \;,
\end{align}
where we introduced the dimensionless coupling $g$ and
an arbitrary momentum scale $\Lambda$,
which we will take to be the cutoff scale of this 
nonrenormalizable theory.%
\footnote{
  Commonly, the coupling of the four-quark interaction is defined
  to be $G=g^2\Lambda^{-2}/2$.
  We find it more convenient to introduce the dimensionless coupling
  $g$ with the cutoff scale.
  We will adopt the worldline formalism, so $\Lambda$ will be
  the ``propertime cutoff'' scale, as opposed to the three- or four-
  momentum cutoff, and this choice is a part of the definition of
  the theory.
}
We have abbreviated the summation over the color indices and
explicitly, for instance, we have
\begin{align}
  (\bar q q)^2 := \bigg( \sum_{a=1}^{N_c} \bar q^a q^q \bigg)^2
  \;.
\end{align}
We emphasize that $g$ is the (effective) four-fermion
coupling and it does not correspond to the usual ``strong coupling''
of AdS/CFT.
Comparing to the standard NJL Lagrangian,
our Lagrangian lacks the flavor structure, because we are
concentrating on the one-flavor case.

This Lagrangian has the $U(1)_L\times U(1)_R$ symmetry,
that is, it is invariant under the transformations
\begin{align}
  q_{L,R} \to e^{i\theta_{L,R}/2} q_{L,R}
  \;,\quad
  \bar q_{L,R} \to e^{-i\theta_{L,R}/2} \bar q_{L,R}
  \;,
\end{align}
where $\theta_{L,R}$ are arbitrary real numbers and we have defined
\begin{align}
  q_{L,R} := \frac{1}{2}(1 \mp \gamma_5) q
  \;,\quad
  \bar q_{L,R} := \bar q \frac{1}{2}(1 \pm \gamma_5)
  \;,
\end{align}
with the convention $\gamma_5^2=1$.
With those definitions, we have
\begin{align}
  \bar q q = \bar q_R q_L + \bar q_L q_R 
  \;,\quad
  \bar q i \gamma_5 q = -i(\bar q_R q_L - \bar q_L q_R)
  \;,
\end{align}
and above relations make it clear that
the invariance is achieved only when we have {\it both} quartic
interactions of Equation~(\ref{eq:1stL});
each quartic interaction is not chirally invariant.

For the special case where we have $\theta_V:=\theta_L \equiv \theta_R$,
the symmetry is the diagonal vector part $U(1)_V$ and the corresponding
transformation is
\begin{align}
  q \to e^{i\theta_V/2} q
  \;,\quad
  \bar q \to \bar q e^{-i\theta_V/2}
  \;,
\end{align}
which leads to the quark number conservation.
On the other hand, when $\theta_A := -\theta_L\equiv\theta_R$,
we have the $U(1)_A$ transformation
\begin{align}\label{eq:AxialTransf}
  q \to e^{-i\theta_A\gamma_5/2} q
  \;,\quad
  \bar q \to \bar q e^{-i\theta_A\gamma_5/2}
  \;.
\end{align}
This invariance, if exists, gives the axial current conservation.

We now want to introduce the background abelian gauge field and
its interaction with the quarks, with respect to the $U(1)_V$ symmetry.
This means that we include the kinetic term of the gauge field in
Equation~(\ref{eq:1stL}), though it is not dynamical,
and replace the ordinary derivative with the covariant derivative
\begin{align}
  D_\mu = \partial_\mu + iA_\mu
  \;,
\end{align}
where we absorbed the coupling constant in the gauge field.
We then have the Lagrangian
\begin{align}
   \mathcal{L} = \bar q i \Slash{D} q
     + \frac{1}{4} (F_{\mu\nu})^2
     +\frac{g^2}{2N_c\Lambda^2}
        \big[ (\bar q q)^2 + (\bar q i\gamma_5 q)^2 \big]
  \;.
\end{align}
Note that the sign of the $F^2$ term is for the Euclidean signature
which we are adopting.
The explicit expression of $F$ is given in Appendix~\ref{sec:F}
and we investigate the cases with {\it constant} electric and magnetic
fields.

It will be convenient to introduce the auxiliary fields $\sigma$
and $\pi$ so that the Lagrangian appears as
\begin{align}\label{eq:auxL}
  \mathcal{L} =& \bar q (i\Slash{D}+g\sigma+g\pi i\gamma_5) q
    + \frac{1}{4} (F_{\mu\nu})^2
    - \frac{1}{2}N_c\Lambda^2 (\sigma^2+\pi^2)
  \;.
\end{align}
Upon path-integration with respect to the fields $\sigma$ and $\pi$,
we recover the original Lagrangian and we have the relations
\begin{align}
  \sigma = \frac{g}{N_c\Lambda^2}(\bar q q)
  \quad\text{and}\quad
  \pi = \frac{g}{N_c\Lambda^2}(\bar q i \gamma_5 q)
  \;.
\end{align}

%%%%%%%%%%%%%%%%%%%%%%%%%%%%%%%%%%%%%%%%%%%%%%%%%%%%%%%%%%%%%%%%%%%%%
\subsection{Axial Anomaly}\label{subsec:anomaly}
The key feature of the NJL model is the dynamical breakdown of
the chiral symmetry, and
for our simplified one-flavor model, the corresponding symmetry
should be $U(1)_A$.
However, since we have included the gauge fields, this symmetry does not
exist in general.%
\footnote{
  The large $N_c$ does not cure this problem, because the gauge fields in
  focus are not the color gauge fields.
}
In fact, the divergence of the
axial current is proportional to $F\tilde F$, where
$\tilde F$ is the dual field strength tensor as defined explicitly
in Appendix~\ref{sec:F}.
Therefore, our model is meaningful only when
\begin{align}
  F\tilde F \propto \vec{E}\cdot\vec{B} =0
  \;,
\end{align}
where $\vec E$ and $\vec B$ are the background
electric and magnetic fields, respectively.

When we carry out the actual numerical evaluation, we concentrate on
this case.
However, up until then, we discuss the model in generality, because
it is easy to generalize to the case with $N_f>1$ and for the sake
of the comparison with the SS model.

To see what would be the order parameter of the $U(1)_A$ breaking,
we note that the bilinears $\bar q q$ and $\bar q i \gamma_5 q$
transform under (\ref{eq:AxialTransf}) as
\begin{align}
  \bar q q \to \cos\theta_A \bar q q - \sin\theta_A \bar q i \gamma_5 q
  \quad\text{and}\quad
  \bar q i \gamma_5 q \to
  \sin\theta_A \bar q q + \cos\theta_A \bar q i \gamma_5 q
  \;.
\end{align}
We thus see that both $\sigma$ and $\pi$ are good order parameters, but
what we will do is to use the symmetry to rotate away $\pi$ and consider
(the expectation value of) $\sigma$ as the order parameter of the dynamical
$U(1)_A$ breaking.

%%%%%%%%%%%%%%%%%%%%%%%%%%%%%%%%%%%%%%%%%%%%%%%%%%%%%%%%%%%%%%%%%%%%%
\subsection{Effective Action}\label{subsec:effAction}
Given a theory, all the physical information can be conveniently extracted
from its effective action and this is especially powerful in determining
the groundstate of the theory.
We derive the effective action for our NJL model
up to the one-loop order, which is exact in the large $N_c$ limit.
There are several ways to construct the effective action, but
we closely follow the classic paper by Coleman and 
Weinberg~\cite{Coleman:1973jx} (see also Reference~\cite{Harrington:1974xh}).
Provided that we avoid the anomalous case,
the effective action, of course, is invariant under the full 
$U(1)_L\times U(1)_R$ symmetry.
That is, it depends only on $(\sigma^2+\pi^2)$, therefore,
as mentioned at the end of the previous subsection,
we set $\pi=0$ and derive the effective action with respect to
the classical value of $\sigma$.

We first quickly review how the effective action can determine the groundstate
of the theory.
Define the connected (Euclidean)
Green's functional $W[J]$
by the relation
\begin{align}
  Z = e^{-W[J]} 
    = \int\mathcal{D}q\mathcal{D}\bar q\mathcal{D}\sigma
    \exp [ - \int dx^4 (\mathcal{L} + J\sigma) ]
  \;,
\end{align}
where $\mathcal{L}$ is given in Equation~(\ref{eq:auxL}), without
the $\pi$ field, and
the function $J(x)$ is the source for the field $\sigma(x)$.
Since we are interested in a constant background electromagnetic field,
we have omitted the path integral over $A_\mu$.
We then have
\begin{align}
  \frac{\delta W[J]}{\delta J(x)} = - \frac{\delta\ln Z}{\delta J(x)}
  = \langle\Omega|\sigma(x)|\Omega\rangle_{J}
  \;,
\end{align}
where $|\Omega\rangle_J$ is the groundstate in the presence of
the source $J$.
This implies that we have the classical field $\sigmacl(x)$
\begin{align}
  \sigmacl(x) := \frac{\delta W[J]}{\delta J(x)}
  \;.
\end{align}
We can think of this $\sigmacl$ as the conjugate of $J$.
Hence we change the variable of the connected Green's function by
the functional Legendre transformation
\begin{align}\label{eq:GammaCL}
  \Gamma[\sigmacl] =& W[J] - \int dy^4 \sigmacl(y)J(y)
  \nonumber\\
  =& - \ln \int\mathcal{D}q\mathcal{D}\bar q\mathcal{D}\sigma
    \exp [ - \int dx^4 \big\{ \mathcal{L} + J(\sigma-\sigmacl) \big\} ]
  \;.
\end{align}
This functional, $\Gamma$, satisfies
\begin{align}\label{eq:selfConsis}
  \frac{\delta\Gamma[\sigmacl]}{\delta\sigmacl(x)}
  =& \frac{\delta W[J]}{\delta \sigmacl(x)} 
  - \int dy^4 \frac{\delta J(y)}{\delta\sigmacl(x)}\sigmacl(y) - J(x)
  \nonumber\\
  =& \int dy^4 \frac{\delta J(y)}{\delta\sigmacl(x)}
              \frac{\delta W[J]}{\delta J(y)}
  - \int dy^4 \frac{\delta J(y)}{\delta\sigmacl(x)}\sigmacl(y) - J(x)
  \nonumber\\
  =& -J(x)
  \;.
\end{align}
Therefore, as we turn off the source $J$, we obtain
\begin{align}\label{eq:ExtG}
  \frac{\delta\Gamma[\sigmacl]}{\delta\sigmacl(x)} = 0
  \;,
\end{align}
and this implies that the groundstate of the theory can be obtained by
extremizing the functional $\Gamma$  with respect to the 
classical field $\sigmacl$.
This can, in turn, be thought to determine the classical field $\sigmacl$
in the groundstate.

Now our task is to compute the effective action (\ref{eq:GammaCL})
and we do this to the one-loop order.
Note that the path integral over the quark fields are quadratic and
can be carried out exactly, yielding the factor
\begin{align}\label{eq:g5Trick}
  \ln \det (i\Slash{D}+g\sigma)
  = \frac{1}{2}\ln\det[(\Slash{D})^2+g^2\sigma^2]
  \;,
\end{align}
where we used the usual ``$\gamma_5$-trick''.
Now, since we have
$\gamma_\mu\gamma_\nu = \delta_{\mu\nu} + [\gamma_\mu,\gamma_\nu]/2$,
we get
\begin{align}
  (\Slash{D})^2 =& D^2 + \frac{1}{2}[\gamma_\mu,\gamma_\nu]D_\mu D_\nu
                = D^2 + \frac{1}{2}\gamma_\mu\gamma_\nu[D_\mu,D_\nu]
  \nonumber\\
                =& D^2 + \frac{i}{2}\gamma_\mu\gamma_\nu F_{\mu\nu}
  \;,
\end{align}
where we used the fact that $[D_\mu,D_\nu] = iF_{\mu\nu}$.
Using this relation to Equation~(\ref{eq:g5Trick}),
% \begin{align}
%   \ln\det (i\Slash{D}+g\sigma)
%   = \frac{1}{2}\ln\det[D^2 + \frac{i}{2} \gamma_\mu\gamma_\nu F_{\mu\nu}
%                     +g^2\sigma^2]
%   \;.
% \end{align}
we end up with the expression for the effective action
\begin{align}\label{eq:effact}
  \Gamma[\sigmacl(x)] =& - \ln\int\mathcal{D}\sigma
    \exp\bigg[-\int dx^4 \bigg\{
           \frac{1}{4} (F_{\mu\nu})^2
          - \frac{1}{2}N_c\Lambda^2 \sigma^2
  \nonumber\\
        &+\frac{1}{2}\tr\ln[D^2 + \frac{i}{2} \gamma_\mu\gamma_\nu F_{\mu\nu}
                    +g^2\sigma^2]
        + J(\sigma-\sigmacl) \bigg\} \bigg]
  \;,
\end{align}
where $\tr$ is over the Clifford algebra representation space,
the color space and the spacetime momentum.
Notice that the extra minus sign from the fermion loop makes the sign
in front of the ``trace-log'' plus.

Now as $N_c$ becomes large, 
the path integral above is more dominated by the saddle point value at 
\begin{align}\label{eq:clDominance}
  \sigma=\sigmacl
  \;.
\end{align}
This statement becomes exact at $N_c=\infty$, on which case we are focusing.
In general, we need to perform the derivative expansion at this stage
with respect to $\sigmacl(x)$, as is done in the paper by Coleman and
Weinberg~\cite{Coleman:1973jx}.
However, since we are interested in the
dynamical quark mass, we only need the leading order
in the expansion, {\it i.e.}, the constant $\sigmacl$.
In other words, we only look for the translationally invariant groundstate.
The effective action $\Gamma$ now is a function with
respect to the $c$-number $\sigmacl$ and
to this end, the spacetime integral in the effective action becomes
trivial.
Thus, it is common to define the effective potential {\it via}
the relation
\begin{align}\label{eq:Gamma}
  \Gamma[\sigmacl] = -\int d^4x \Veff(\sigmacl)
  \;,
\end{align}
yielding
\begin{align}\label{eq:Veff}
  \Veff(\sigmacl) = -\frac{1}{4} (F_{\mu\nu})^2 
        + \frac{1}{2}N_c\Lambda^2\sigmacl^2
        -\frac{1}{2}\tr\ln[D^2 + \frac{i}{2} \gamma_\mu\gamma_\nu F_{\mu\nu}
        +g^2\sigmacl^2 ]
  \;.
\end{align}
Then the functional relation
(\ref{eq:ExtG}) reduces to a simple differential equation
\begin{align}\label{eq:tadCancel}
  \frac{d\Veff}{d\sigmacl} = 0
  \;.
\end{align}

\begin{figure}[ht]
{
\centerline{\scalebox{1.5}{\includegraphics{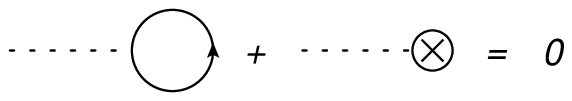}}}
\caption{\footnotesize
  The tadpole cancellation condition.
  The dotted and solid lines represent the $s$ and the quark fields,
  respectively.
  The second term on the left-hand side comes from the ``vertex''
  $N_c\Lambda^2\sigmacl s(x)$.
}\label{fig:tadCancel}
}
\end{figure}

Another way to obtain the effective potential is to utilize
the background field method, by shifting the field as
$\sigma(x)\to s(x)+\sigmacl$ and integrating out the quark field.
Then it is clear that Equation~(\ref{eq:tadCancel}) is precisely
the tadpole cancellation condition as shown in Figure~\ref{fig:tadCancel}.
In what follows, we refer this condition as the ``tadpole condition''.

%%%%%%%%%%%%%%%%%%%%%%%%%%%%%%%%%%%%%%%%%%%%%%%%%%%%%%%%%%%%%%%%%%%%%
\subsection{Worldline Formalism}\label{subsec:worldline}
We would like to evaluate the last term in the effective potential
(\ref{eq:Veff}) and for the convenience, we redefine this term as
\begin{align}\label{eq:defVhat}
  \hat V(m) :=
  \frac{1}{2}\tr\ln[D^2 + \frac{i}{2} \gamma_\mu\gamma_\nu F_{\mu\nu} +m^2 ]
  \;,
\end{align}
where we recall that the trace is over the color, Dirac and loop momentum
spaces, and
we have defined the mass parameter
\begin{align}\label{eq:defm}
  m:=g\sigmacl
  \;.
\end{align}
To evaluate the effective potential, 
we utilize the worldline formalism that goes along
the ideas of string theory~\cite{Strassler:1992zr}.

To begin with, we express the function $\hat V$ in Schwinger's
propertime~\cite{Schwinger:1951nm}.
By using the relation%
\footnote{
  This is the analytically continued version of the propertime and is
  the ``imaginary propertime''.
  Later, we will come back to this point concerning the real and imaginary
  propertime.
  We note that this is essentially different from the analytic continuation
  of the spacetime coordinate.
}
\begin{align}\label{eq:imPTime}
  \ln X = - \int_0^\infty \frac{e^{-Xs}}{s} ds
  \;,
\end{align}
we can write
\begin{align}
  \hat V(m) =&
  -\frac{1}{2}\int_0^\infty\frac{ds}{s}\tr
        \exp[-s \big( D^2 + \frac{i}{2} \gamma_\mu\gamma_\nu F_{\mu\nu}
        +m^2 \big) ]
  \nonumber\\
  =& -\frac{1}{2}N_c\int_0^\infty\frac{ds}{s} e^{-m^2s}
      \tr \langle x|
      \exp[-s \big( D^2 
               + \frac{i}{2} \gamma_\mu\gamma_\nu F_{\mu\nu} \big) ]
      |x \rangle
  \;,
\end{align}
where the operator
$\tr$ now is only over the representation space of Clifford algebra
and the factor $N_c$ came out of the trace over the color space.

Now in the worldline formalism, we want to regard $x_\mu$ as fields
in the one-dimensional space of 
circular worldline, parametrized by, say, $\tau$,
and express the effective potential in terms of the path integral over
the fields $x_\mu(\tau)$.
The proper length of the circle's circumference
is the parameter $s$, which is
called ``modulus'' in the string literature, and the factor $1/s$ in $\hat V$
removes the redundancy from the translation and
reversal of the worldline coordinate.
In order to deal with the last trace over the Clifford representation space,
we need one more ingredient.
What we need to do is very similar to the treatment of worldsheet fermions
in superstring theory.
We introduce constant Grassmann fields $\psi_\mu$ that satisfy
\begin{align}
  \{ \psi_\mu,\psi_\nu \} = \delta_{\mu\nu}
  \;,
\end{align}
and introduce a basis of the Clifford representation space
$|\alpha\rangle$ so that
\begin{align}\label{eq:ClifRep}
  \psi_\mu |\alpha\rangle = 
  \frac{1}{\sqrt{2}} (\gamma_\mu)_{\alpha\beta} |\beta\rangle
  \;.
\end{align}
We then can write
\begin{align}\label{eq:pathInt}
  \hat V(m) =&
  -\frac{1}{2}N_c\int_0^\infty\frac{ds}{s} e^{-m^2s}
      \langle \alpha, x|
      \exp[-s \big( D^2 
               + i \psi_\mu\psi_\nu F_{\mu\nu}  \big) ]
      |\alpha, x \rangle
  \nonumber\\
  =&  -\frac{1}{2}N_c\int_0^\infty\frac{ds}{s} e^{-m^2s}
     \mathcal{N} \int_P\mathcal{D}x(\tau)\int_A\mathcal{D}\psi(\tau)
      \exp\big[-\int_0^sd\tau \big(
        \mathcal{L}_x + \mathcal{L}_\psi  \big) \big]
  \;,
\end{align}
with
\begin{align}\label{eq:EucL}
  \mathcal{L}_x := \frac{1}{2e'}\dot x^2 + iA \cdot \dot x
  \quad\text{and}\quad
  \mathcal{L}_\psi := \frac{1}{2}\psi\cdot\dot\psi
                   - i\frac{e'}{2} \psi \cdot F \cdot \psi
  \;.
\end{align}
Here the dots on the fields denote the derivative with respect to $\tau$ and
$e'$ is the einbein of the worldline circle, which can be an arbitrary
fixed number because we have decided to encode the moduli in the parameter $s$
(see, for example, Chapter 5 of Polchinski~\cite{Polchinski:1998rq}).
Hence from now on, we set
$e' \equiv 2$,
which is the common choice in the literature.
As usual, the factor $\mathcal{N}$ arises from the integration over
the conjugate momenta to obtain the path integral in the Lagrangian form,
and it satisfies the relation
\begin{align}\label{eq:Nrelation}
  \mathcal{N}\int_P\mathcal{D}x \exp\big[
                -\int_0^sd\tau\frac{1}{4}\dot x^2 \big]
  = (4\pi s)^{-d/2}
  \;,
\end{align}
where $d$ is the spacetime dimension, {\it i.e.}, $4$ in our case.
Finally, the subscripts $P$ and $A$ on the path integral symbols
imply the boundary conditions of the fields
\begin{align}
  x(0) = x(s)
  \quad\text{and}\quad
  \psi(0) = -\psi(s)
  \;,
\end{align}
for all the components.

Before we proceed further, we discuss the treatment of finite temperature
in the context of the worldline formalism.

%%%%%%%%%%%%%%%%%%%%%%%%%%%%%%%%%%%%%%%%%%%%%%%%%%%%%%%%%%%%%%%%%%%%%
\subsection{Finite Temperature and Chemical Potential}
\label{subsec:finiteTemp}
\begin{figure}[ht]
{
\centerline{\scalebox{0.5}{\includegraphics{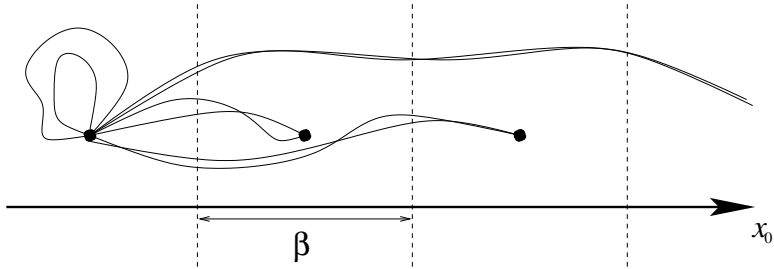}}}
\caption{\footnotesize
  Conceptual picture of the path integral at finite temperature.
  The left-most loop paths corresponds to $n=0$, {\it i.e.}, the
  $T=0$ contribution.
}\label{fig:paths}
}
\end{figure}

As we have adopted the Euclidean signature, we use the imaginary
time formalism to handle finite temperature.
This means that the time direction
in the spacetime is compactified with the circumference $\beta$,
and a worldline can wind around this direction.
Thus, the path integral should be carried out in each winding sector,
and then they must be summed with appropriate statistical factor
\cite{Schulman:1981vu} (see also Refs.~\cite{McKeon:1992if,McKeon:1993sh}).
This can be written as
\begin{align}\label{eq:FTPriscrip}
  P^{\beta}_{xx} = \sum_{n=-\infty}^{\infty} (-1)^n P^{\infty}_{x(x_0+n\beta,x_i)}
  \;,
\end{align}
where $P_{xy}$ represents the path integral over the paths from a spacetime
point $x$ to $y$ and
$P^{\infty}$ means the path integral in $\mathbb{R}^4$, {\it i.e.} without
the topological condition.
We emphasize that the alternating sign $(-1)^n$ comes from the {\it spacetime}
fermions $q$, and they should not be confused with the worldline fields.
See Figure~\ref{fig:paths} for a conceptual picture of the path
integral.

Let us look at the path integral of $x(\tau)$ 
in Equation~(\ref{eq:pathInt}).
We have the Lagrangian $\mathcal{L}_x$ as shown in Equation~(\ref{eq:EucL})
and this does not appear to be quadratic in the field $x_\mu(\tau)$.
At this point, it is very common that a specific gauge, called
the Schwinger-Fock gauge:
\begin{align}\label{eq:SFgauge}
  A_\mu = -\frac{1}{2}F_{\mu\nu}x_\nu
  \;,
\end{align}
is adopted to bring the Lagrangian into the quadratic form.
However, one immediately realizes the problem with this gauge at finite
temperature by 
re-examining Schwinger's original paper \cite{Schwinger:1951nm}.
There, the holonomy factor,
\begin{align}
  \Phi(x,y)=\exp\big[i\int_x^ydx_\mu A_\mu(x)\big]
  \;,
\end{align}
appears in the two-point Green's function, while in the gauge
(\ref{eq:SFgauge}) this factor is absent.
This factor has no relevance if one is interested in the effective potential
where $x=y$, and {\it if the spacetime is topologically trivial}.
In the imaginary time formalism of the theory at finite temperature,
the latter condition is false.
In the geometrical language, the transition function defined as a map
from the $S^1$ time circle to the $U(1)$ gauge group
is non-trivial and classified by $\pi_1[U(1)]=\mathbb{Z}$.
This in turn implies that the gauge transformation is severely restricted
to a certain class and one must be very careful with the gauge fixing.
In other words, the consequence of the time circle is
(geometrically) exactly the same as the charge quantization of
the Dirac monopole and the Aharonov-Bohm effect 
(see for instance Reference~\cite{Nakahara:1990th}).
As in those cases, the parallel transport of the wavefunction
around the time circle, {\it i.e.}, the Polyakov loop, obtains
a non-trivial physical significance.
As we will see below, this automatically introduces the chemical potential
into the system.

Gies~\cite{Gies:1998vt} appears to be the first one to pointed out
this subtlety and he generalized Schwinger's propertime formalism
to finite temperature with the careful treatment of the holonomy factor.
We are going to show how this works in the worldline formalism.

To take advantage of the Schwinger-Fock gauge, we split
the Lagrangian into the Schwinger-Fock gauge part and the rest
\begin{align}
  \mathcal{L}_x = \frac{1}{4}\dot x^2 + i \dot x \cdot
                        (A + \frac{1}{2}F \cdot x - \frac{1}{2}F \cdot x )
        = \bigg(\frac{1}{4}\dot x^2 - \frac{i}{2} \dot x \cdot F \cdot x\bigg)
            + i \dot x \cdot (A + \frac{1}{2}F \cdot x)
  \;.
\end{align}
Now the important point is that
\begin{align}
  d(A+\frac{1}{2}F \cdot x) = 0
  \;,
\end{align}
and this implies that {\it within a homotopy class}, the integral
\begin{align}\label{eq:closed}
  \int_0^sd\tau \dot x \cdot (A+\frac{1}{2}F \cdot x) 
        = \int_{x(0)}^{x(s)} dx \cdot (A+\frac{1}{2}F \cdot x)
\end{align}
is independent of the path.
Therefore, this factors out of the path-integral.
Furthermore, without loss of generality, we may evaluate the $x$ integral
of Equation~(\ref{eq:closed})
along the straight line in $x_0$-direction.
In this case, the integral over $F\cdot x$ vanishes due to the anti-symmetry
of $F$ and we are left with the integral of $A_0$.
In what follows,
we are going to consider the case with constant $A_0$ and we define
a real number
\begin{align}
  \mu := iA_0
  \;.
\end{align}
Then we can write
\begin{align}\label{eq:preVeff}
  \hat V(m) =& -\frac{1}{2}N_c\int_0^\infty\frac{ds}{s} e^{-m^2s}
    \sum_{n=-\infty}^{\infty}(-1)^n e^{-n\beta\mu}
  \nonumber\\
     &\times\mathcal{N}
     \int_{x(0)}^{x(s)}\mathcal{D}x(\tau)\int_A\mathcal{D}\psi(\tau)
      \exp\big\{-\int_0^sd\tau \big(
        \mathcal{L}_{xSF} + \mathcal{L}_\psi  \big) \big\}
  \;,
\end{align}
with
\begin{align}
  \mathcal{L}_{xSF} := \frac{1}{4}\dot x^2-\frac{i}{2} \dot x \cdot F \cdot x
  \quad\text{and}\quad
  \mathcal{L}_\psi := \frac{1}{2}\psi\cdot\dot\psi
                   - i \psi \cdot F \cdot \psi
  \;.
\end{align}
The subscript ``SF'' of the bosonic Lagrangian signifies the form
in the Schwinger-Fock gauge.
The integer $n$ is the winding number in the $x_0$ direction and
it is also implicitly in $x(s)=x(0)+n\beta\hat x_0$,
where $\hat x_0$ is the unit vector in the time direction.

%%%%%%%%%%%%%%%%%%%%%%%%%%%%%%%%%%%%%%%%%%%%%%%%%%%%%%%%%%%%%%%%%%%%%
\subsection{The Effective Potential}\label{subsec:Veff}
In this subsection, we derive the expression for the effective potential
that is ready for the numerical evaluation.

In Equation~(\ref{eq:preVeff}),
we are interested in the constant $F$, {\it i.e.}, it is
independent of the paths $x(\tau)$.
Thus we can separately calculate the path integrals
for the worldline bosonic and fermionic parts,
\begin{align}
    I_B :=& \mathcal{N}\int_{x(0)}^{x(s)}\mathcal{D}x(\tau)
      \exp\big(-\int_0^sd\tau \mathcal{L}_{xSF} \big)
  \nonumber\\
    I_F :=& \int_A\mathcal{D}\psi(\tau)
      \exp\big(-\int_0^sd\tau \mathcal{L}_\psi  \big)
  \;,
\end{align}
where the factor $\mathcal{N}$ naturally comes with the bosonic part
because of the relation (\ref{eq:Nrelation}).
The evaluation of them is rather lengthy and is
given in Appendix~\ref{sec:IBIF}.
The results are
\begin{align}\label{eq:IFfinal}
  I_B =&  (4\pi s)^{-2} 
          \exp\big[ -(n\beta/2)^2 \{F\cot(Fs)\}_{00} \big]
          \bigg({\det}_L\bigg[\frac{\sin(Fs)}{Fs}\bigg]\bigg)^{-1/2}
  \nonumber\\
  I_F =& 4({\det}_L[\cos(Fs)])^{1/2}
  \;,
\end{align}
where the functions of the matrix ${F_\mu}^\nu$ are defined through the formal
power series and ${\det}_L$ implies the determinant with respect to
the matrices with the components in the Lorentz indices.
We notice that the Lorentz invariance is explicitly broken in $I_B$,
due to the finite temperature.

At this point, $\hat V$ has the form
\begin{align}\label{eq:preResum}
  \hat V(m) =&
  -\frac{N_c}{8\pi^2}\int_0^\infty ds s^{-3} e^{-m^2s}
     ({\det}_L[\cos(Fs)])^{1/2}
     \bigg({\det}_L\bigg[\frac{\sin(Fs)}{Fs}\bigg]\bigg)^{-1/2}
  \nonumber\\
     &\times\sum_{n=-\infty}^{\infty} (-1)^n e^{-n\beta\mu}
     \exp\big[ -(n\beta/2)^2 \{F\cot(Fs)\}_{00} \big]
  \;.
\end{align}
In this form, the physical interpretation of the parameter $\mu$ is still
not clear.
Moreover, it turns out that this form is not suitable for the numerical
evaluation in a certain range of parameters.
Thus, we use the Poisson (re)summation formula
\begin{align}
  \sum_{n=-\infty}^\infty \exp(-\pi a n^2 + 2\pi i b n)
  = a^{-1/2} \sum_{n=-\infty}^\infty \exp\big[-\frac{\pi}{a}(n-b)^2 \big]
  \;,
\end{align}
to convert the above expression to
\begin{align}\label{eq:postResum}
  \hat V(m)/N_c =&
  -\frac{T}{4\pi^{3/2}}\int_0^\infty ds s^{-5/2} \{(Fs)\cot(Fs)\}_{00}^{-1/2}
     \big\{{\det}_L\big[(Fs)\cot(Fs)\big]\big\}^{1/2}
  \nonumber\\
     &\times\sum_{l\in\mathbb{Z}_{1/2}} 
     \exp\big[ - \{(Fs)\cot(Fs)\}_{00}^{-1} (2\pi T l - i\mu)^2s - m^2s \big]
  \;,
\end{align}
where $T:=1/\beta$ and $l$ runs over the half integers
$\mathbb{Z}_{1/2}:=\{n+1/2:n\in\mathbb{Z}\}$.
In this form, we see that the parameter $\mu$ is the chemical potential,
as advertised before.
In the standard method, the chemical potential is included from the beginning
as a part of the grand partition function
(in the end, this amounts to the shift $p_0\to p_0-i\mu$).
Contrast to this, the automatic inclusion of the chemical potential
we see here is rather remarkable.

We now have to evaluate the terms involving the matrix $F$.
The determinant can be calculated relatively easily by using Schwinger's
method \cite{Schwinger:1951nm}, but $\{(Fs)\cot(Fs)\}_{00}$ is not
straightforward and for this, we adopt the method of
Reference~\cite{Batalin:1971au} (see also \cite{Gusynin:1998bt}).
They are computed in Appendix~\ref{sec:F} and the results are
\begin{align}
  \big\{{\det}_L\big[(Fs)\cot(Fs)\big]\big\}^{1/2}
       = s^2 \left|  \mathcal{G}
       \coth\bigg(s\sqrt{\mathcal{F}+\sqrt{\mathcal{F}^2+\mathcal{G}^2}}\bigg)
       \cot\bigg(s\sqrt{-\mathcal{F}+\sqrt{\mathcal{F}^2+\mathcal{G}^2}}\bigg)
       \right|
  \;,
\end{align}
and
\begin{align}\label{eq:tautau}
  &\{(Fs)\cot(Fs)\}_{00} =\frac{1}{2\sqrt{\mathcal{F}^2+\mathcal{G}^2}}
    \bigg[ 
  \nonumber\\
        &\bigg\{-\frac{1}{2}(\vec B^2+\vec E^2) 
                 + \sqrt{\mathcal{F}^2+\mathcal{G}^2} 
           \bigg\}
        \sqrt{\mathcal{F}+\sqrt{\mathcal{F}^2+\mathcal{G}^2}}
        \coth\bigg(s\sqrt{\mathcal{F}+\sqrt{\mathcal{F}^2+\mathcal{G}^2}}\bigg)
  \nonumber\\
        &+
        \bigg\{\frac{1}{2}(\vec B^2+\vec E^2) 
                 + \sqrt{\mathcal{F}^2+\mathcal{G}^2} 
           \bigg\}
        \sqrt{-\mathcal{F}+\sqrt{\mathcal{F}^2+\mathcal{G}^2}}
        \cot\bigg(s\sqrt{-\mathcal{F}+\sqrt{\mathcal{F}^2+\mathcal{G}^2}}\bigg)
   \bigg]
  \;,
\end{align}
where the Lorentz invariants are defined as
\begin{align}
  \mathcal{F} := -\frac{1}{4}F_{\mu\nu}F_{\mu\nu}
               = \frac{1}{2} (\vec B^2 - \vec E^2)
  \quad\text{and}\quad
  \mathcal{G} := -\frac{i}{4} F_{\mu\nu}\tilde F_{\mu\nu}
               = \vec E \cdot \vec B
  \;.
\end{align}

Those expressions are very clumsy, so we consider a few special cases.
As discussed in Section~\ref{subsec:anomaly}, our $N_f=1$ model
does not make sense unless $\mathcal{G}=0$ due to the $U(1)_A$ anomaly.
The case with $\mathcal{G}\neq 0$ is of interest when $N_f>1$, but
in what follows, we concentrate on the non-anomalous case.

At zero temperature, the Lorentz invariance should be recovered
and we expect the expression for the effective potential becomes
considerably simple.
In fact, we can go back to Equation~(\ref{eq:preResum}) and simply
set $n=0$.%
\footnote{
  At zero temperature, the dependence on the chemical potential
  drops out, as the time direction is noncompact,
  and introduction of the chemical potential at zero temperature
  requires the standard method mentioned previously.
  We will generally consider the zero temperature case without chemical
  potential.
}
We then have
\begin{align}\label{eq:zeroTF}
  \hat V(m) = -\frac{N_c}{8\pi^2}\int_0^\infty ds s^{-3} e^{-m^2s}
              \big\{s\sqrt{2\mathcal{F}}\coth(s\sqrt{2\mathcal{F}})\big\}
  \;.
\end{align}
From this, we see that for the cases with $|\vec B|>|\vec E|$ and
$|\vec B|<|\vec E|$, the system behaves as if there is only
$B$-field and $E$-field, respectively.
Also as is well-known, the pure $E$-field case can be obtained from
the pure $B$-field case just by sending $B\to iE$, and {\it vice versa}.

When we turn on the temperature, all those features are lost.
Examining Equation~(\ref{eq:tautau}), one sees that the response
of the system for the cases $|\vec B|\lessgtr |\vec E|$ is not as
simple as the zero temperature case.
Moreover, the pure $B$ and $E$ background systems are not related
through $B\to iE$.
These are all due to the lack of the Lorentz invariance at finite temperature.
Because of this complexity, we find it best to handle the finite
temperature system with only $B$ and $E$ fields separately.

%%%%%%%%%%%%%%%%%%%%%%%%%%%%%%%%%%%%%%%%%%%%%%%%%%%%%%%%%%%%%%%%%%%%%
\subsubsection{Pure $B$ Background}\label{subsec:pureB}
We have for this case
\begin{align}
  \big\{{\det}_L\big[(Fs)\cot(Fs)\big]\big\}^{1/2}
  \to\; s|\vec B|\coth(s|\vec B|)
  \quad\text{and}\quad
  \{(Fs)\cot(Fs)\}_{00}  \to\; 1
  \;,
\end{align}
which lead to the expression of the effective potential
\begin{align}
  \Veff(m) =& \frac{1}{2}\vec B^2 + \frac{1}{2}N_c\Lambda^2g^{-2}m^2
    + \frac{N_cT}{4\pi^{3/2}}\int_0^\infty ds s^{-5/2}
      \{ s|\vec B|\coth(s|\vec B|) \}
  \nonumber\\
     &\times\sum_{l\in\mathbb{Z}_{1/2}}
     \exp\big[ -s \big\{ (2\pi Tl-i\mu)^2 + m^2 \big\} \big]
  \;.
\end{align}

First, notice that the first term on the right-hand side is $N_c$-suppressed,
therefore, we drop this term from now on.
Next, observe that the $s$ integral is ill-defined at $s=0$ and
we regulate this by introducing the propertime cutoff $s=1/\Lambda^2$
for the lower bound of the integral.%
\footnote{
  Reference~\cite{Klevansky:1989vi} assumes that the cutoff $\Lambda$ can depend
  on the parameter $m$.
  This assumption requires another equation, in addition to the
  tadpole condition (\ref{eq:tadCancel}).
  For this, they derive a consistency condition, which corresponds to
  Equation~(\ref{eq:selfConsis}) in our path-integral treatment
  of Section~\ref{subsec:effAction}.
  However, the large $N_c$ makes the classical path
  totally dominant in the path integration as discussed around
  Equation~(\ref{eq:clDominance}). 
  This results in the fact that the effective action is
  independent of the source $J$ and the relation
  (\ref{eq:selfConsis}) gives nothing other than the tadpole condition
  (\ref{eq:tadCancel}), still lacking an equation that determines 
  the function $\Lambda(m)$.
  It seems that the consistency condition of Reference~\cite{Klevansky:1989vi}
  comes out of the lack of the proper change of variable {\it via}
  the Legendre transformation, in order to obtain the energy of the system.
}
(Up to this point, the scale $\Lambda$ has been an arbitrary momentum scale
but now it is taken to be the propertime cutoff scale.)
We would like to measure all the quantities in the units of the cutoff
scale $\Lambda$.
This means that we rescale all the dimensionful parameters with $\Lambda$,
for instance, $s\to s\Lambda^2$ and $B\to B/\Lambda^2$.
Then defining the dimensionless quantity $\hatVeff$, we have
\begin{align}\label{eq:VeffB}
  \hatVeff(m) :=& \frac{\Veff(m)}{N_c\Lambda^4}
  \nonumber\\
   =& \frac{m^2}{2g^2}
    + \frac{|\vec B|T}{4\pi^{3/2}}\int_1^\infty ds s^{-3/2}
     \coth(s|\vec B|)
     \sum_{l\in\mathbb{Z}_{1/2}}
     \exp\big[ -s \big\{ (2\pi Tl-i\mu)^2 + m^2 \big\} \big]
  \;,
\end{align}
where all the parameters are now dimensionless.

Though we have regulated the lower end of the $s$ integral, it is finite
at the upper end, only when
\begin{align}\label{eq:muCond}
  \opname{Re}\big[(2\pi Tl-i\mu)^2 + m^2 \big] =
  (2\pi Tl)^2 - \mu^2 + m^2 > 0
  \;.
\end{align}
This condition crucially depends on the value of the chemical potential.
It is well-known that the chemical potential of free (complex) scalar
field theory may not exceed the mass of the particle, because the chemical
potential effectively acts as a negative mass squared.
One might naively think that the similar situation is occurring in our
theory when the chemical potential is too large.
However, this is false.
We must recall that we have been adopting the analytically
continued ``imaginary propertime'' in Equation~(\ref{eq:imPTime}).
We have just found that this analytic continuation is appropriate only
when the condition (\ref{eq:muCond}) is met.
If we consider Equation~(\ref{eq:imPTime}) as the continuation to
the {\it positive} imaginary axis, there is nothing that prevents
us to continue to the {\it negative} imaginary axis to make the propertime
integral well-defined.
In other words, when the condition (\ref{eq:muCond}) is violated,
we must analytically continue to negative $s$.

This is the idea of treating the case with a large chemical potential,
but there are complications due to the existence of the cutoff,
the cut in the complex $s$-plane from the factor of $s^{-3/2}$
and the poles from the hyperbolic cotangent along the imaginary $s$-axis.
We work out the details in Appendix~\ref{sec:largeMu}.
The result is
\begin{align}\label{eq:fVeffB}
  &\hatVeff =
  \nonumber\\
  &\frac{m^2}{2g^2}
  + \frac{T}{2\pi^{3/2}}\opname{Re} \bigg[
    \int_1^\infty ds (sB)\coth(sB) s^{-5/2}\sum_{l>\bar l}
    \exp\big[ -s \big\{ (2\pi Tl-i\mu)^2+m^2 \big\} \big]
  \nonumber\\
   &+i\int_1^\infty ds (sB)\coth(sB) s^{-5/2}\sum_{l=1/2}^{\bar l}
   \exp\big[ s \big\{ (2\pi Tl-i\mu)^2+m^2 \big\} \big]
  \nonumber\\
   &+i\int_0^\pi d\phi (e^{i\phi}B)\coth(e^{i\phi}B)
   e^{-\frac{3}{2}i\phi}\sum_{l=1/2}^{\bar l}
   \exp\big[ -e^{i\phi} \big\{ (2\pi Tl-i\mu)^2+m^2 \big\} \big]
  \nonumber\\
  &+(2\pi i)\sum_{l=1/2}^{\bar l}\sum_{k=1}^\infty e^{-\frac{3}{4}\pi i}
    \bigg(\frac{B}{k\pi}\bigg)^{3/2}
    \exp\big[ -ik\frac{\pi}{B} \big\{ (2\pi Tl-i\mu)^2+m^2 \big\} \big]
  \bigg]
  \;,
\end{align}
where $B:=|\vec B|$ and we have the condition $B<\pi$,
indicating that $B$ may not exceed the cutoff scale squared $\Lambda^2$ 
too much.
The parameter $\bar l$ is defined to be
\begin{align}
  \bar l := \big[ (2\pi T)^{-1}\theta(\mu-m)\sqrt{\mu^2-m^2} \big]_G
  \;,
\end{align}
where without loss of generality $\mu$ and $m$
are assumed to be greater than or equal to zero,
$\theta(x)$ is the Heaviside theta and the symbol
$[x]_G$ is the half-integer version of the Gauss symbol, {\it i.e.},
it is the largest half integer less than or equal to $x$.
Obtaining Equation~(\ref{eq:fVeffB}) involves contour integrations,
and referring to Figure~\ref{fig:ACont4}, the first line is the contribution
from the positive horizontal axis, the second line is from the negative
horizontal axis, third one is from the small circular contour 
(coming from the existence of the cutoff) and the last
term is the contribution from the poles mentioned above.
One can see that the last term is oscillatory and we will see that this
part gives rise to the well-known de Haas-van Alphen effect
\cite{Elmfors:1993bm}
and it complicates the phase structure.

The effective potential obtained above contains all the necessary
information and is ready for the numerical evaluation.
Regarding $\hatVeff$ as a function of $m^2$, the tadpole condition
(\ref{eq:tadCancel}) becomes
\begin{align}
    0 = \frac{d\hatVeff}{d\sigmacl}
    = \frac{d(m^2)}{d\sigmacl}
      \frac{d\hatVeff}{d(m^2)}
    = 2g^2\sigmacl\frac{d\hatVeff}{d(m^2)}
    \;,
\end{align}
where we used the definition (\ref{eq:defm}).
We denote the solution to the tadpole condition by $m^*$.
We see that $m^*=0$ is always a solution and the solution other than this
satisfy
\begin{align}
  \frac{d\hatVeff}{d(m^2)}\bigg|_{m=m^*} =0
  \;.
\end{align}
When there are more than one solution, energetically preferred one
must be chosen by consulting with $\hatVeff$.
In this way, we can determine the dynamical mass and the phase structure
of the theory.

%%%%%%%%%%%%%%%%%%%%%%%%%%%%%%%%%%%%%%%%%%%%%%%%%%%%%%%%%%%%%%%%%%%%%
\subsubsection{Pure $E$ Background}\label{subsec:pureE}
We have for this case
\begin{align}
  \big\{{\det}_L\big[(Fs)\cot(Fs)\big]\big\}^{1/2}
  \to\;& s |\vec E|\,|\cot(s |\vec E|)|
  \nonumber\\
  \{(Fs)\cot(Fs)\}_{00}  \to\;& s|\vec E|\cot(s |\vec E|)
  \;,
\end{align}
which lead to the expression of the effective potential
\begin{align}\label{eq:VeffE}
  \hatVeff(m) =& \frac{m^2}{2g^2}
    + \frac{T}{4\pi^{3/2}}\int_1^\infty ds s^{-5/2} \sqrt{sE\cot(sE)}
  \nonumber\\
     &\times\sum_{l\in\mathbb{Z}_{1/2}}
     \exp\big[ -s \big\{\big(sE\cot(sE)\big)^{-1}
        (2\pi Tl-i\mu)^2 + m^2 \big\} \big]
  \;,
\end{align}
where as in the previous subsection, $\hatVeff$ and all the parameters
are made dimensionless with respect to the cutoff $\Lambda$
and we redefined $E:=|\vec E|$.
Because of the cotangent in the exponent, we have essential singularities
and this makes the examination of this case notoriously difficult.
In the following, we make two attempts: $T\to0$ limit and weak $E$-field
expansion.
\\
\pagebreak

\noindent {\it Zero Temperature}

\noindent In the zero temperature limit, instead of summing over $l$,
we should {\it integrate} Equation~(\ref{eq:VeffE}) with respect to
the combination $Tl$.
This yields
\begin{align}\label{eq:VeffET0}
  \hatVeff = \frac{m^2}{2g^2}
           + \frac{1}{8\pi^2}\int_1^\infty ds s^{-3}
           \big\{ (sE)\cot(sE) \big\} e^{-m^2s}
  \;.
\end{align}
As noted in the footnote near Equation~(\ref{eq:zeroTF}),
the chemical potential drops out
and we consider the system without it.
We find that only the simple poles of the cotangent are present.
Thus, it is commonly evaluated through the convention
\begin{align}\label{eq:PVreplacement}
  \cot(sE) \to \opname{P.V.}\cot(sE)
          + \frac{i\pi}{E} \sum_{n=1}^\infty \delta\big(s-\pi n/E\big)
  \;,
\end{align}
where P.V. denotes the principal value.

The principal value gives the real part of the effective potential
and this has been evaluated in the literature.
We quote the result from Reference~\cite{Klevansky:1989vi},
\begin{align}\label{eq:ReVeffET0}
  \opname{Re}\big[\hatVeff\big] = \frac{m^2}{2g^2}
  + \frac{1}{8\pi^2} \bigg\{
  \opname{Re}[Q] - a_E + E_3(m^2) - \frac{1}{3}E^2 E_1(m^2) \bigg\}
  \;,
\end{align}
where
\begin{align}
  Q := m^4 \bigg\{ -\frac{1}{4}-\frac{1}{3}\bigg(\frac{E}{m^2}\bigg)^2
     + \frac{1}{2}\bigg( 1-\frac{1}{z}+\frac{1}{6z^2} \bigg)\ln z
     - \frac{1}{z^2}\partial_x \zeta(x,z)|_{x=-1} \bigg\}
  \;,
\end{align}
with $z:=im^2/(2E)$,
\begin{align}
  a_E := \int_0^1 ds s^{-3} e^{-m^2s}\big\{ (sE)\cot(sE) -1
       + (sE)^2/3 \big\}
  \;,
\end{align}
and the functions $E_n(x)$ are the exponential integrals defined as
\begin{align}
  E_n(x) := \int_1^\infty \frac{dt}{t^n}e^{-xt}
  \;.
\end{align}
Using the real part of the effective potential, Klevansky and Lemmer
\cite{Klevansky:1989vi} examines the dynamical mass and the phase structure
of the system under the stress of the constant electric field.
(We will reproduce their results in the next subsection.)

The second term in the replacement (\ref{eq:PVreplacement}) yields the
imaginary part of the effective potential
\begin{align}\label{eq:ImVeffET0}
  \opname{Im}\big[\hatVeff\big] =
  \frac{1}{8\pi}\sum_{n=1}^\infty \bigg(\frac{E}{n\pi}\bigg)^2
  e^{-\frac{n\pi}{E}m^2}
  \;.
\end{align}
The imaginary part is interpreted as the pair creation rate of the quarks
in Reference~\cite{Schwinger:1951nm}.
The pair creation inevitably implies the instability of the groundstate
and in our system, this occurs at any finite value of $E$.
Therefore, the conclusions about the mass and the phase structure
drawn from the real part may not have any physical significance.
However, as long as the imaginary part is small compared to the real part,
we can assume that the instability sets in slowly, compared to the dynamical
mass $m^*$, and we can still obtain the physical insights.
In the numerical evaluation, therefore, we will simultaneously examine
the real and the imaginary parts of the effective potential.
\\

\noindent {\it Weak $E$-Field Expansion}

\noindent We attempt to circumvent the essential singularities
in Equation~(\ref{eq:VeffE}) by expanding in small $E$.
This, of course, is a very dangerous thing to do.
However, we extract some qualitative conclusions by comparing it to the zero
temperature case and also we check if the higher order terms in the expansion
give less contributions to the effective potential.

The expansion itself is straightforward and we have
\begin{align}\label{eq:Eexpansion}
  \hatVeff(m) =& \frac{m^2}{2g^2}
    + \frac{T}{4\pi^{3/2}}\int_1^\infty ds s^{-5/2}
     \sum_{l\in\mathbb{Z}_{1/2}}
     \exp\big[ -s \big\{(2\pi Tl-i\mu)^2 + m^2 \big\} \big]
  \nonumber\\
    -& \frac{E^2T}{24\pi^{3/2}}\int_1^\infty ds s^{-1/2}
      \sum_{l\in\mathbb{Z}_{1/2}}
      \big\{ 1+2s(2\pi Tl-i\mu)^2 \big\}
      \exp\big[ -s \big\{(2\pi Tl-i\mu)^2 + m^2 \big\} \big]
  \nonumber\\
    +& \mathcal{O}(E^4)
  \;.
\end{align}
One can easily go up to higher orders as desired.
We must evaluate the sum and the integral term by term in the expansion
as in Appendix~\ref{sec:largeMu}.
However, assuming that the parameter $|l|$ is large enough to satisfy
the convergence condition (\ref{eq:muCond}),
one can carry out the integral before the sum and obtain a closed form.
As discussed in the appendix, one can then analytically continue the
integral beyond the
validity range of the parameter $l$ to the whole half integers
and this considerably eases the practical evaluation.%
\footnote{
  For the pure $B$ background, we have not been able to obtain the
  closed form of the $s$ integral, so needed to evaluate the very
  complicated expression in Equation~(\ref{eq:fVeffB}).
}
Thus in practice, one should exchange the order of the integral and
the sum in Equation~(\ref{eq:Eexpansion}), without worrying about
the convergence condition (\ref{eq:muCond}).%
\footnote{
  Also it is practically convenient to realize that the negative
  $l$ is exactly the complex conjugate of the positive counterpart.
}

%%%%%%%%%%%%%%%%%%%%%%%%%%%%%%%%%%%%%%%%%%%%%%%%%%%%%%%%%%%%%%%%%%%%%
\subsection{Numerical Evaluation}\label{sec:NJLNum}
We are interested in the dynamical mass $m^*$ and the effective potential
$\Veff$ in the five-dimensional parameter space $(T,\mu,B,E,g)$.
It is, of course, not practical to cover entire space at once, so
we examine slices of the parameter space.
As a warm up and to illustrate the general behavior of the NJL model,
we first investigate the slice of $B=0=E$.
We then turn on those parameters one at a time.

%%%%%%%%%%%%%%%%%%%%%%%%%%%%%%%%%%%%%%%%%%%%%%%%%%%%%%%%%%%%%%%%%%%%%
\subsubsection{General Observation without Electromagnetic Fields}\label{sec:genNum}
We learn in this section that the temperature and the chemical potential
both behave as the ``destabilizer'' of the quark boundstate,
and that when the condition (\ref{eq:muCond}) is violated and the analytic
continuation is required, the effective potential shows more complicated
behavior than when the condition is satisfied.
Those are the generic properties of the model and remain essentially
the same when we turn on the external electromagnetic fields.

The tadpole condition (\ref{eq:tadCancel}) for this case can be
written as
\begin{align}\label{eq:tadCondF0}
  \frac{\pi^{3/2}}{g^2} =
  T\opname{Re} \bigg[
    \sum_{l>0} \int_1^\infty ds s^{-3/2}
    \exp\big[ -s \big\{ (2\pi Tl-i\mu)^2+m^{*2} \big\} \big]
  \bigg]
  \;.
\end{align}
We have indicated that the integral should be done before
the sum.
This is because the integral for this case is simple enough to yield
a closed form, and
as discussed in Appendix~\ref{sec:largeMu} and Section~\ref{subsec:pureE},
we can then analytically continue the integral to the values
of $l$ that violates the condition (\ref{eq:muCond}).

\begin{figure}[ht]
{
\centerline{\scalebox{0.9}{\includegraphics{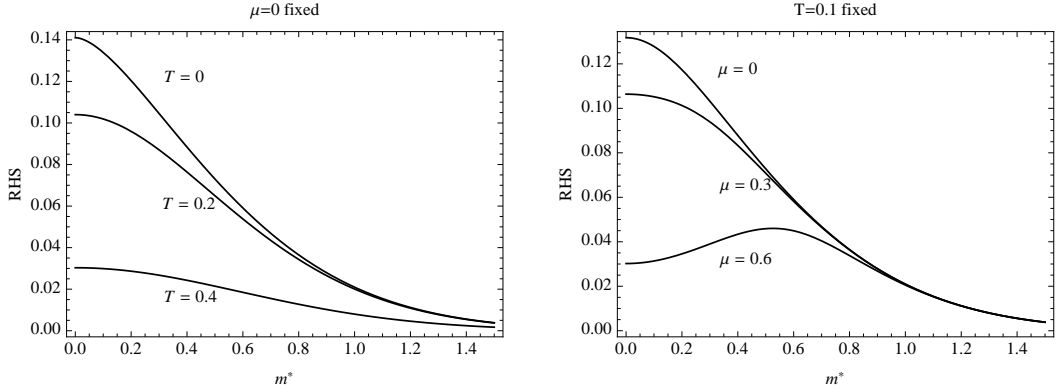}}}
\caption{\footnotesize
  The right-hand side of Equation~(\ref{eq:tadCondF0}), {\it i.e.}, 
  $\pi^{3/2}/g^2$, against $m^*$.
  For $T=0=\mu$ the value of RHS at $m^*=0$ is $1/4\pi^{1/2}\approx 0.141$
  and this corresponds to the famous critical coupling $g_c=2\pi$.
  Note the different behavior appearing for $(\pi T)^2-\mu^2<0$,
  which is the case for $T=0.1$ and $\mu=0.6$.
  In this case, there can be two solutions with $m^*\neq 0$.
}\label{fig:NJLRHSF0}
}
\end{figure}

\begin{figure}[ht]
{
\centerline{\scalebox{0.9}{\includegraphics{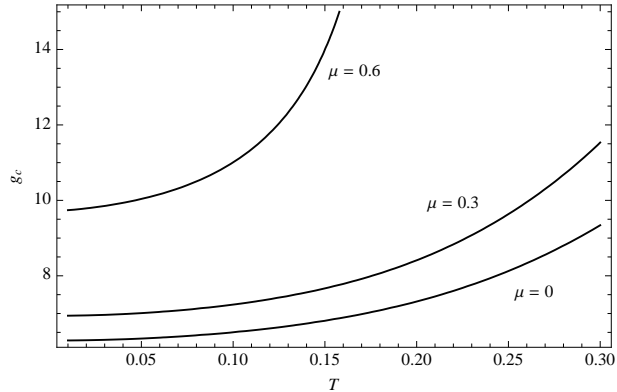}}}
\caption{\footnotesize
  The critical coupling $g_c$ is defined to be the value, below which
  the solution to the tadpole condition does not exist.
  Note that for small $T$ and large $\mu$, $g=g_c$ does not imply $m^*=0$,
  as shown in the right panel of Figure~\ref{fig:NJLRHSF0}.
  The plots show that the larger $T$ and $\mu$ require the larger
  coupling.
}\label{fig:NJLgcF0}
}
\end{figure}

We first plot the right-hand side of the tadpole condition (\ref{eq:tadCondF0})
in Figure~\ref{fig:NJLRHSF0}.
These plots show that there is a critical value of the coupling $g$,
below which the solution to the equation does not exist
(and the only solution to the tadpole condition is $m^*=0$),
and the system is in the chirally symmetric phase.
The critical coupling is plotted against the temperature in 
Figure~\ref{fig:NJLgcF0}.
In particular, the critical coupling for $T=0=\mu$ is the famous
$g_c=2\pi$ ($\pi^{3/2}/g_c^2\approx 0.141$) 
of the original work \cite{Nambu:1961tp}.
We observe that both the temperature and chemical potential tend to
destroy the quark boundstate, as these parameters necessitate stronger
coupling.

\begin{figure}[ht]
{
\centerline{\scalebox{1.0}{\includegraphics{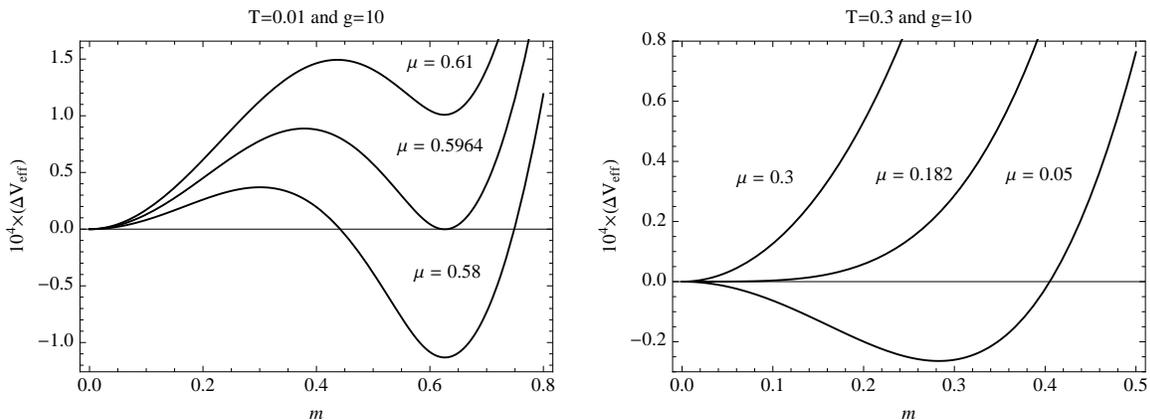}}}
\caption{\footnotesize
  The plot of $\Delta\Veff := \Veff(m)-\Veff(0)$.
  The left panel shows the first order phase transition with respect to
  the chemical potential in the low $T$, high $\mu$ region and the right
  panel shows the second order transition
  in the high $T$, low $\mu$ region.
}\label{fig:NJLVeffF0}
}
\end{figure}

In the right panel of Figure~\ref{fig:NJLRHSF0}, we notice 
the difference between the low and high values of the chemical potential.
For the parameters given in the diagram, we have
$(\pi T)^2-\mu^2>0$ for $(T,\mu)=(0.1,0.3)$ and negative for $(0.1,0.6)$.
Thus, when the chemical potential is large enough (and the temperature is
low enough) to require
the analytic continuation, the system respond in a more complicated way
and in particular, there can be multiple solutions to the tadpole condition
for a given value of the coupling.
One manifestation of such complication is the order of the chiral phase
transition.
In Figure~\ref{fig:NJLVeffF0}, the effective potentials are plotted
against the parameter $m$ for the low $T$, high $\mu$ region and for
the high $T$, low $\mu$.
The reference point of the potential is taken to be the value at $m=0$, which
is always a minimum, and this base value defines the difference potential
$\Delta\Veff$.
Notice that the low $T$, high $\mu$ diagram on the left has a large
range of $m$ where $(\pi T)^2-\mu^2+m^2<0$, and this leads to
the extra structure, the potential barrier,
that causes the first order chiral phase transition.
Meanwhile, the high $T$, low $\mu$ diagram of the right panel has
$(\pi T)^2-\mu^2>0$ for all the values of $\mu$ plotted.
For this case, the analytic continuation does not occur and $\Veff$ is
relatively simple, leading to the second order phase transition.
The complication due to the analytic continuation discussed briefly here
is the generic phenomenon and we will see the further consequences in the
next subsection.

\begin{figure}[ht]
{
\centerline{\scalebox{1.0}{\includegraphics{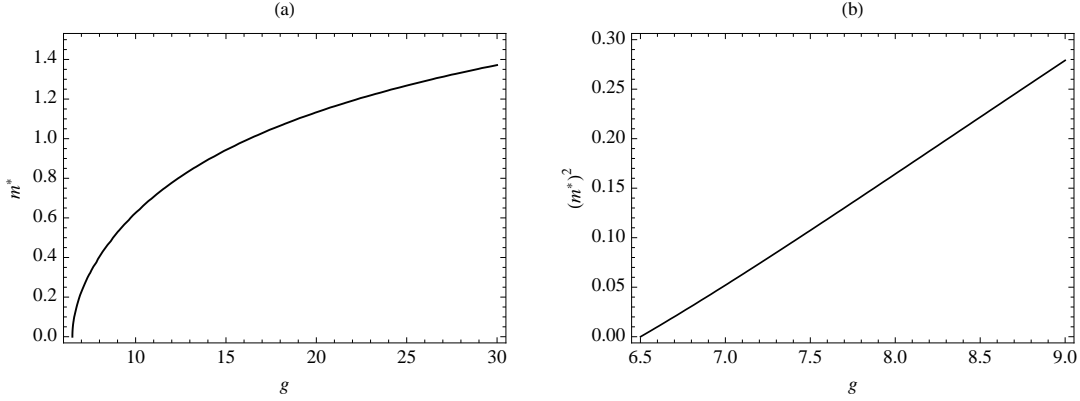}}}
\caption{\footnotesize
  These are plotted for $T=0.1$ and $\mu=0$.
  Diagram (b) is ${m^*}^2$ against the region of $g$ near the critical value
  and shows the relation $m^*\sim\sqrt{g}$.
  The linear behavior of (b) fails in the larger values of $g$.
  The graph for $T=0$ is nearly identical, except that the critical value
  is $g_c=2\pi$ and the graph is slightly shifted to the left.
}\label{fig:NJLm_gF0}
}
\end{figure}

Finally in the absence of the electromagnetic background,
we show the response of $m^*$ with respect to the coupling $g$ in
Figure~\ref{fig:NJLm_gF0}.
As expected, the dynamical mass generally increases with the coupling.
Diagram (b) zooms into the near critical coupling $g_c$, and
shows the relation $m^*\sim\sqrt{g}$ in this region.
This response of the dynamical mass with respect to $g$ will be
contrasted to the dynamical mass of the SS model with respect to
the probe brane separation $L$.

%%%%%%%%%%%%%%%%%%%%%%%%%%%%%%%%%%%%%%%%%%%%%%%%%%%%%%%%%%%%%%%%%%%%%
\subsubsection{Pure $B$ Background}\label{sec:NJLBNum}
We now turn on the $B$ field (and $E=0$) and
the effective potential for this case is given in
Equation~(\ref{eq:fVeffB}).

The salient feature of this effective potential is the last term in
the equation, the oscillation term.
When the condition (\ref{eq:muCond}) is violated, this term comes in
effect, leading to the oscillatory behavior of the effective potential
with respect to the various parameters.
\begin{figure}[p]
{
\centerline{\scalebox{0.9}{\includegraphics{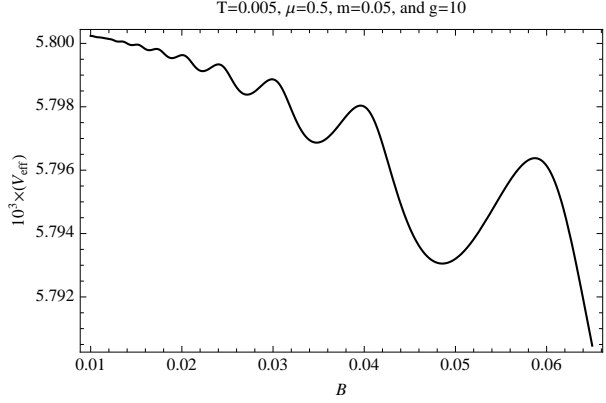}}}
\caption{\footnotesize
  The plot of the effective potential with respect to $B$,
  illustrating the de Haas-van Alphen effect.
}\label{fig:NJLOscl}
}
\end{figure}
\begin{figure}[p]
{
\centerline{\scalebox{0.9}{\includegraphics{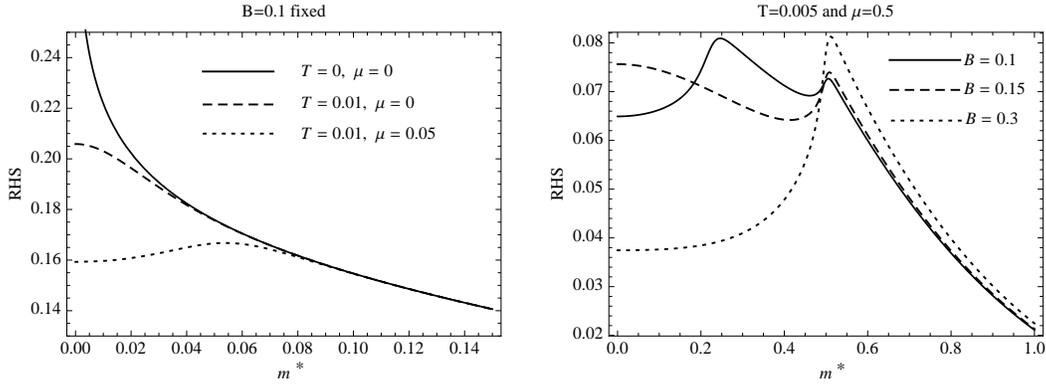}}}
\caption{\footnotesize
  The right-hand side (RHS) of the tadpole condition similar to
  Equation~(\ref{eq:tadCondF0}) is plotted as a function of $m^*$.
  On the left panel with $T=0$, the graph blows up to infinity
  as $m^*\to 0$, indicating $g_c=0$.
  Finite $T$ brings $g_c$ finite.
  The right panel is plotted for the parameters that satisfy
  $(\pi T)^2-\mu^2+{m^*}^2<0$ for $m^*<0.5$.
  Notice the effect of the oscillation that leads to multiple solutions
  to the tadpole condition for a certain range of $g$.
}\label{fig:NJLBRHS}
}
\end{figure}
\begin{figure}[p]
{
\centerline{\scalebox{0.9}{\includegraphics{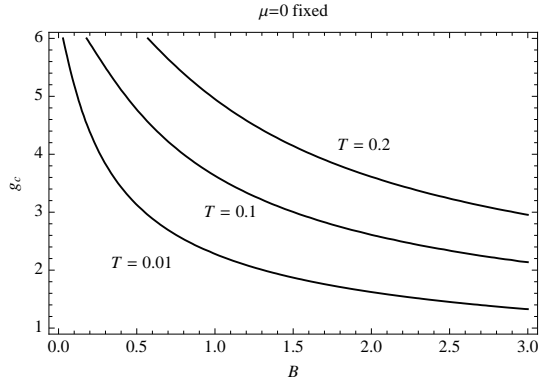}}}
\caption{\footnotesize
  The plots show that the $B$ field is the stabilizer of the boundstate, as
  it requires smaller value of the coupling at high $B$, and as discussed
  previously, the temperature is the destabilizer.
}\label{fig:NJLgcB}
}
\end{figure}
Figure~\ref{fig:NJLOscl} is the plot of the effective potential with
respect to the parameter $B$ and the other parameters fixed.
We clearly see the oscillation, known as the de Haas-van Alphen
effect, and this is caused by the crossings of the fermi sphere through
the Landau levels.
This effect leads to the multiple solutions to the tadpole
condition (\ref{eq:tadCancel}) and complicates the model
in the low $T$, high $\mu$ region of the parameter space.

From Equation~(\ref{eq:fVeffB}), it is easy to work out the tadpole
condition (\ref{eq:tadCancel}) and it can be written in the form
$\pi^{3/2}/g^2 = \text{RHS}$, similar to Equation~(\ref{eq:tadCondF0}).
The right-hand side (RHS) is plotted for various parameters in
Figure~\ref{fig:NJLBRHS}.
The plots of the left panel show the behavior of RHS similar
to Figure~\ref{fig:NJLRHSF0} without the $B$ field.
One important difference, though, is the case with $T=0$ where the presence
of the $B$ field makes the critical coupling $0$, {\it i.e.}, 
the magnetic field stabilizes the quark boundstate and any
finite coupling breaks the chiral symmetry.
For this reason, the magnetic field is sometimes called the ``catalysis''
of the chiral symmetry breaking \cite{Gusynin:1995nb}.
However, even with the $B$ field, any finite temperature brings
the critical coupling nonzero.
Figure~\ref{fig:NJLgcB} shows the plots of critical coupling against
the magnetic field at $\mu=0$.
In the figure, we see the competing effects of $B$ as the stabilizer
and $T$ as the destabilizer of the quark boundstate.

\begin{figure}[t]
{
\centerline{\scalebox{1.0}{\includegraphics{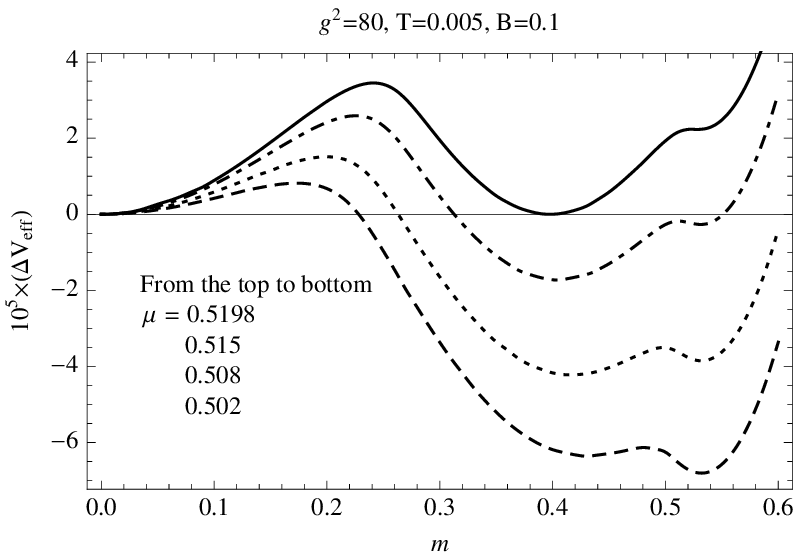}}}
\caption{\footnotesize
  The plot of the effective potential for nonzero $B$-field
  and in the low $T$, high $\mu$ region of parameters.
  In addition to the potential barrier that causes the first
  order chiral phase transition, we have extra structure
  due to the de Haas-van Alphen oscillation
  that causes another first order transition,
  different from the chiral phase transition (the bottom two plots).
}\label{fig:NJLMMuVeff}
}
\end{figure}

Now, the right panel of Figure~\ref{fig:NJLBRHS} is plotted for the parameters
that satisfy $(\pi T)^2-\mu^2+{m^*}^2<0$ for $m^*<0.5$.
We can observe the effect of the oscillation, and for a certain range of
the coupling and $B$ ({\it e.g.}, $B=0.1$ and $\pi^{3/2}/g^2\approx 0.07$
at $g^2=80$),
there are four $m^*\neq 0$ solutions to the tadpole condition.
We must consult with the effective potential to decide which solution is the
stable and energetically preferred one.
Figure~\ref{fig:NJLMMuVeff} is the plot of the potential.
\begin{figure}[!]
{
\centerline{\scalebox{0.8}{\includegraphics{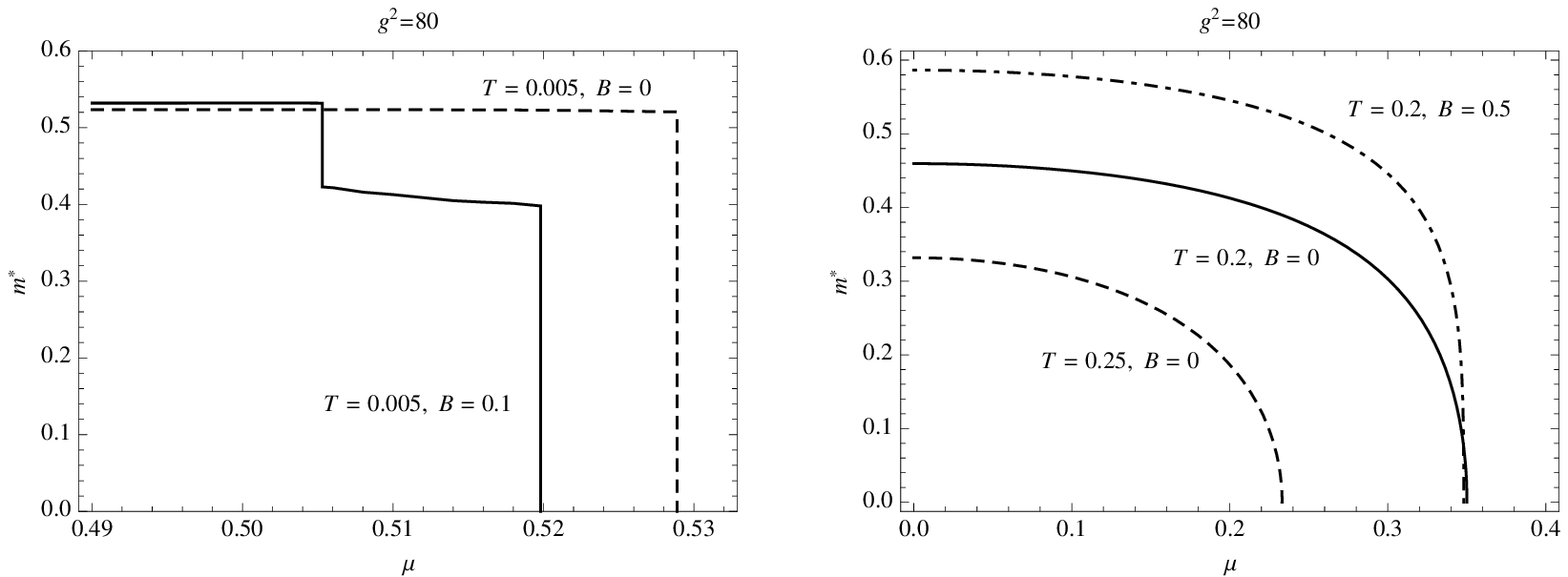}}}
\caption{\footnotesize
  The plot of the dynamical mass against the chemical potential $\mu$.
  The left panel shows the region of low $T$, high $\mu$, where
  the mass goes to zero abruptly, corresponding to the first order
  phase transition.
  For the nonzero $B$-field, we have an additional structure
  due to the de Haas-van Alphen oscillation
  and the system goes through a sequence of first order phase transitions.
  The right panel shows the region of high $T$, low $\mu$, where the
  mass goes to zero smoothly as the chemical potential is increased.
  This corresponds to the second order chiral phase transition.
}\label{fig:NJLMMu}
}
\end{figure}
We indeed observe four peaks and troughs.
The top plot is at the critical value of the chemical potential and
above this value, the mass sharply drops to zero and the
system goes through the chiral phase transition to the symmetric phase.
Curiously, the bottom two curves show that there is another first
order phase transition that is different from the chiral phase
transition.
During this transition, the mass sharply drops to a lower value but
not to zero.
This is the extra structure introduced by the oscillation.

Figure~\ref{fig:NJLMMu} of the $\mu$-$m^*$ graphs clearly summarize
the discussions made so far.
On the left, the graph in the solid line directly corresponds to
the behavior of the effective potential shown in Figure~\ref{fig:NJLMMuVeff}
and reveals the existence of the two first order phase transitions.
In the same diagram, the dashed graph is plotted for $B=0$, indicating
that there is only one first order chiral phase transition
(see the left diagram in Figure~\ref{fig:NJLVeffF0}).
Therefore, the extra first order phase transition is visibly due to
the de Haas-van Alphen effect at $B\neq 0$.%
\footnote{
  We remark that the sharp first order phase transitions shown here
  is slightly different from rather smooth transitions in
  Reference~\cite{Inagaki:2003yi}.
  The smoothness is most likely due to the higher temperature
  in their setup.
}
On the right panel, we have plotted the $\mu$-$m^*$ graph for the higher
temperature so that the condition (\ref{eq:muCond}) is always satisfied.
For this case, the mass smoothly goes down to zero and this is the
second order chiral phase transition, similar to the one shown in
the right panel of Figure~\ref{fig:NJLVeffF0}.
Notice also the role of $B$ and $T$
as the stabilizer and destabilizer, respectively.

\begin{figure}[ht]
{
\centerline{\scalebox{0.8}{\includegraphics{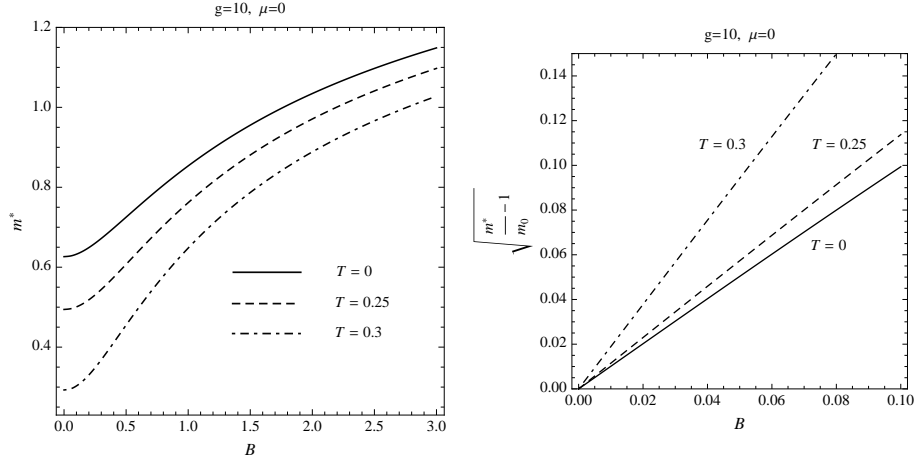}}}
\caption{\footnotesize
  The plot of the dynamical mass with respect to $B$ at $\mu=0$ fixed.
  The temperature here is low enough so that the chiral symmetry is broken
  at $B=0$.
  The right panel shows the relation $m^*\sim B^2$ for the region
  of small $B$.
  The parameter $m_0$ is defined as $m_0 := m^*(B=0)$.
}\label{fig:NJLBM3}
}
\end{figure}

We now discuss the response of the dynamical mass with respect to the $B$
field.
First, Figure~\ref{fig:NJLBM3} shows the simpler case with $\mu=0$.
For this value of the chemical potential,
the mass responds intuitively;
it increases monotonically as the external magnetic field (the stabilizer)
increases.
Notice that for this plot, we have chosen the low enough temperature
so that the chiral symmetry is broken even at $B=0$.
The right panel of the figure shows that the mass increases quadratically
with respect to $B$ in the region of small $B$.
Also the diagram shows that 
the tangent of the graphs increases as the temperature is raised.

\begin{figure}[ht]
{
\centerline{\scalebox{0.9}{\includegraphics{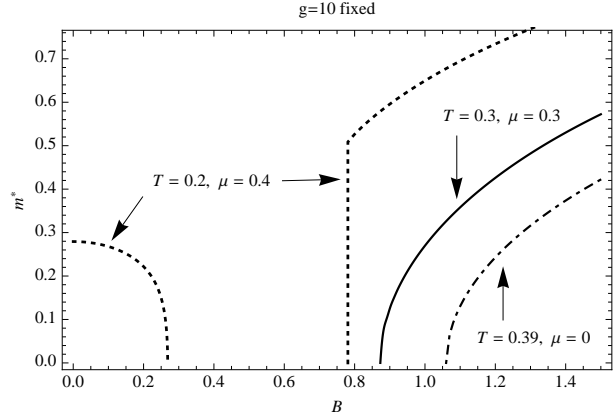}}}
\caption{\footnotesize
  The plot of the dynamical masses with respect to $B$.
  For $(T,\mu)=(0.3,0.3)$ and $(0.39,0)$,
  the masses continuously rise from zero as
  $B$ increases, corresponding to the second order phase transitions.
  For $(T,\mu)=(0.2,0.4)$, the mass goes smoothly goes to zero
  in the region of small $B$ and rises abruptly in the region
  of large $B$.
  They correspond to the second and first order phase transitions
  (see the left panel of Figure~\ref{fig:NJLBM2}).
  We note that for a certain range of parameters, the left part of the
  dotted graph can exhibit a sequence of first order phase transitions
  similar to the one shown in $m^*$-$\mu$ diagram of
  Figure~\ref{fig:NJLMMu}.
}\label{fig:NJLBM1}
}
\end{figure}
\begin{figure}[!]
{
\centerline{\scalebox{0.8}{\includegraphics{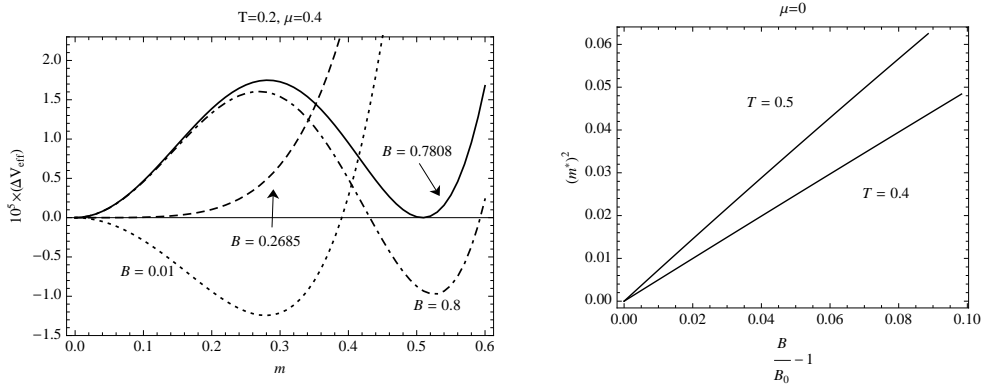}}}
\caption{\footnotesize
  The left panel shows the behavior of the effective potential
  for $(T,\mu)=(0.2,0.4)$, exhibiting the first and second order
  phase transitions (see Figure~\ref{fig:NJLBM1}).
  The right panel is plotted for the high enough temperatures, concentrating
  on the values near $B=B_0$ of the second order phase transition points.
  The plots reveal the relation $m^*\sim \sqrt B$ near $B=B_0$
  and the coefficient increases as temperature goes up.
}\label{fig:NJLBM2}
}
\end{figure}
The $B$-$m^*$ graph with nonzero chemical potential is shown in
Figure~\ref{fig:NJLBM1}.
Let us first observe the dotted curves for $T=0.2$ and $\mu=0.4$.
The response of the mass against $B$ is highly counter-intuitive
for lower $B$: the mass {\it decreases} as the magnetic field
is dialed up.
Moreover, this curve in the low $B$ region can exhibit a sequence of
first order phase transitions that are different from the chiral transition,
due to the de Haas-van Alphen oscillation.
(We have omitted this very complicated graph.)
The further increase in the $B$ field leads to the second order chiral
phase transition around $B\approx 0.27$.
But approximately at 0.78, the system comes back to the broken phase
through the first order transition and the mass increases as $B$ is raised.
The left panel of Figure~\ref{fig:NJLBM2} shows the series of action
in terms of the effective potential.

The complicated behavior here requires a digression.
As argued in Reference~\cite{Gusynin:1995nb}, the stabilizing effect of
the $B$ field directly comes from the dominance of 
the lowest Landau level (LLL).
In Figure~\ref{fig:NJLBM3}, we have plotted for $\mu=0$ and the system
is always in the LLL, leading to the stabilization effect.
When $\mu$ is large and $T$ is relatively small, we have a reasonably
well-defined fermi sphere (smeared due to the temperature) and
the fermi energy can be much larger than the Landau level spacing.
If this is the situation of the system, the effect from the LLL is not
significant and we do not expect the stabilizing effect of the magnetic field.
Even though the finite temperature and the interaction effects prevent
the precise estimate here, this in fact is roughly
the case for the low $B$ region of the dotted curve
in Figure~\ref{fig:NJLBM1}.
Here we observe that the $B$ field acts as a {\it destabilizer} of the
quark boundstate.
As the magnetic field becomes large enough, the Landau level spacing
becomes also large and the stabilizing effect from the LLL takes place.
To summarize the digression, we can conclude that the oscillation effect
and the destabilization effect of $B$ are in general caused by the existence
of the fermi sphere and its relation to the Landau levels.

Let us now turn to the other plots in Figure~\ref{fig:NJLBM1}.
These are the cases where
the temperatures are high and the chemical potentials are low enough
so that the system is in the symmetric phase in the region of low $B$.
The masses rise continuously from zero at critical values $B_0$s,
indicating the second order
chiral phase transition, and increases monotonically with $B$.
The right panel of Figure~\ref{fig:NJLBM2} shows that the mass
rises from zero as $\sqrt{B}$ and the coefficient increases as
the temperature.

\begin{figure}[ht]
{
\centerline{\scalebox{0.8}{\includegraphics{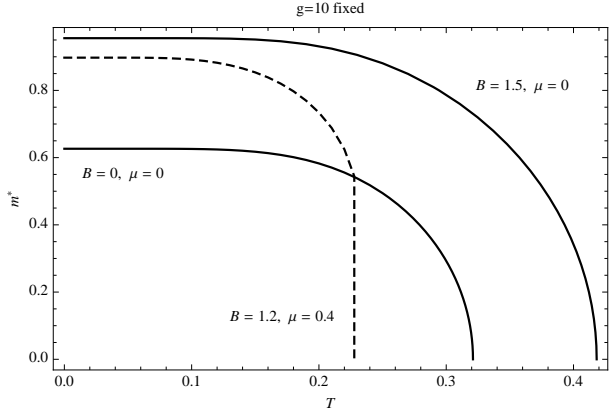}}}
\caption{\footnotesize
  The plot of the dynamical mass as a function of the temperature.
  We see that the larger values of $B$ lead to the larger mass,
  {\it i.e.}, $B$ is the stabilizer.
  For $\mu=0$, the mass smoothly becomes zero as the temperature is
  increased, indicating the second order chiral phase transition.
  Meanwhile, for $\mu=0.4$, the mass drops abruptly to zero
  at a critical temperature, indicating the first order phase
  transition.
}\label{fig:NJLMT}
}
\end{figure}

The mass depends also on the temperature and the $T$-$m^*$ diagram
is shown in Figure~\ref{fig:NJLMT}.
Here we see that as long as $\mu=0$, or when it is small enough,
the mass continuously
goes to zero as the temperature is increased, indicating the second
order chiral phase transition.
We can also observe that the larger the $B$ field is, the larger
the mass is, {\it i.e.}, $B$ acts as the stabilizer of the quark boundstate.
When the chemical potential is high, such as $\mu=0.4$,
the first order phase transition
occurs, as shown as the dashed curve in the diagram.

We now proceed to discuss the phase diagrams.
Though we have seen that there are more than one kind of phase transition,
here, we focus on the chiral phase diagrams.

\begin{figure}[ht]
{
\centerline{\scalebox{0.8}{\includegraphics{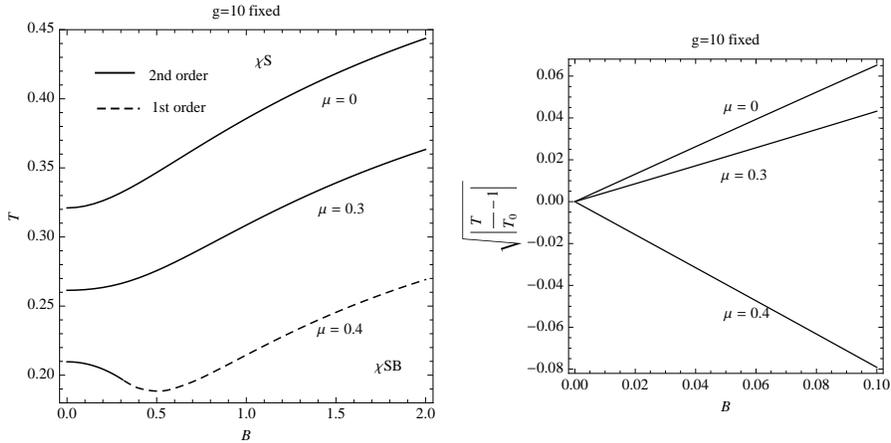}}}
\caption{\footnotesize
  The $B$-$T$ chiral phase diagram.
  The upper temperature regions are the chiral symmetric phase.
  The dotted curve shown on the left panel indicates the first
  order phase transition and the solid ones are second order.
  The right panel plots the small $B$ region against
  $(|T/T_0-1|)^{1/2}$ where $T_0$ is the phase transition temperature
  at $B=0$, showing the relation $T\sim\pm B^2$.
}\label{fig:NJLBT}
}
\end{figure}

The $B$-$T$ phase diagram is shown in Figure~\ref{fig:NJLBT}.
The top two curves for $\mu=0$ and $0.3$ shape in expected way
as $T$ being the destabilizer and $B$ as the stabilizer.
However, the small $B$ region of the bottom curve for the large chemical
potential $\mu=0.4$ is rather surprising.
The critical temperature {\it decreases} as
$B$ is increased, indicating that the magnetic field
is acting as the {\it destabilizer}
in that region of $B$.
When $B$ is further dialed up, the first order transition is
triggered and then $B$ starts acting as the stabilizer.
We have seen the similar phenomenon when we discussed the $B$-$m^*$ diagram
of Figure~\ref{fig:NJLBM1} and noted that this is caused by the existence
of the fermi sphere and the small Landau level spacing.
This effect, in the $B$-$T$ diagram, appears as the ``dent'' in the
chiral phase transition curve.
The right diagram of Figure~\ref{fig:NJLBT} zooms into the region
of small $B$ and shows the relation $T\sim\pm B^2$.
Note that the slops of the plots decrease as the chemical potential increases.

\begin{figure}[t]
{
\centerline{\scalebox{0.8}{\includegraphics{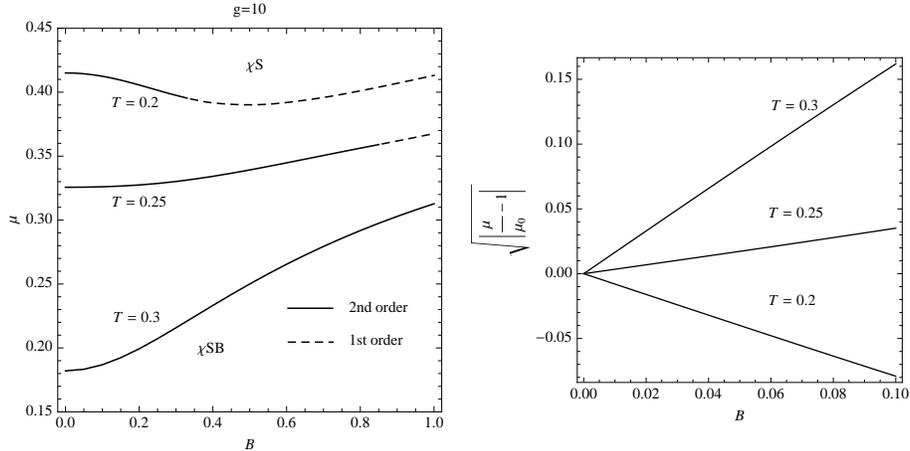}}}
\caption{\footnotesize
  The $B$-$\mu$ phase diagram.
  For sufficiently large $B$, the transition becomes first order.
  The right panel is plotted for the small-$B$ region.
  The parameter $\mu_0$ is the critical value of $\mu$ when $B=0$.
}\label{fig:NJLBMu}
}
\end{figure}

The ``dent'' observed in the $B$-$T$ phase diagram can also be observed
in the $B$-$\mu$ phase diagram of Figure~\ref{fig:NJLBMu}.
\begin{figure}[!]
{
\centerline{\scalebox{1.0}{\includegraphics{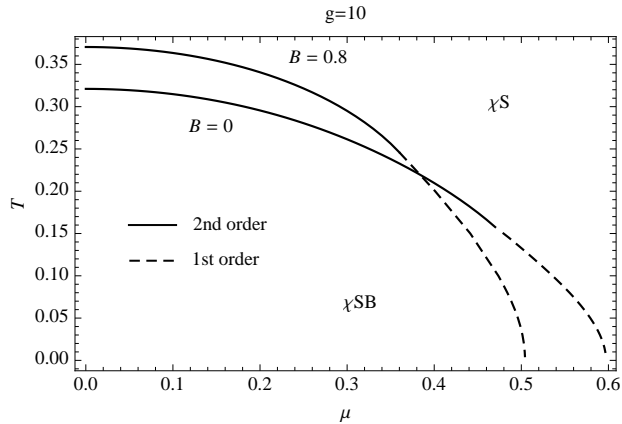}}}
\caption{\footnotesize
  The $\mu$-$T$ phase diagram.
  In the low temperature region, the order of the transition becomes
  first order.
  It is counter-intuitive that the critical value of the chemical
  potential in the low temperature region is smaller for large value
  of $B$, and this is due to the suppression of the LLL effect.
}\label{fig:NJLMuT}
}
\end{figure}
This phase diagram appears to be similar to the $B$-$T$ graph, because
as we recall, $\mu$ acts as the destabilizer of the boundstate just as
$T$ does.
The right panel also shows the relation $\mu\sim\pm B^2$ in the region
of small $B$ and the slopes of the plots increase as the temperature
increases.

Let us look into the $B$-$\mu$ phase diagram a little further.
When the temperature is high, say, $T=0.3$, we have the relation
of the critical values of the chemical potential,
$\mu_c(B=0)<\mu_c(B=0.8)$.
In the low temperature regime, say, $T=0.2$, we have
$\mu_c(B=0)>\mu_c(B=0.8)$.
This implies that the $\mu$-$T$ phase diagram, plotted for $B=0$ and $0.8$
must cross somewhere between $0.2<T<0.3$, that is, the peculiar
phenomenon
due to the fermi sphere must appear as a ``cross'' in the
$\mu$-$T$ phase diagram.
Indeed, the ``cross'' is observed in the $\mu$-$T$ phase diagram shown
in Figure~\ref{fig:NJLMuT}.
In the low temperature region, we observe that the critical chemical
potential for the lower $B$ is {\it higher} and the higher $B$ in this
region of temperature is giving smaller critical $\mu$.
This counters the intuition that $B$ and $\mu$ are the stabilizer and
the destabilizer of the quark boundstate, respectively.
So we learn that when the LLL is not dominant due to the fermi sphere,
this intuition is false.

%%%%%%%%%%%%%%%%%%%%%%%%%%%%%%%%%%%%%%%%%%%%%%%%%%%%%%%%%%%%%%%%%%%%%
\subsubsection{Pure $E$ Background}\label{sec:pureENum}
In this subsection, we examine the constant electric field background.
As remarked in Section~\ref{subsec:pureE}, the essential singularities
in the effective potential (\ref{eq:VeffE}) prevent us from evaluating
this as it appears.
Therefore, we resorted to two approaches, one is the zero temperature limit
and the other is the weak $E$ expansion.
We discuss those in turn.
\\

\noindent {\it Zero Temperature}

\noindent The effective potential of this case (without the chemical
potential) is given in Equation~(\ref{eq:VeffET0}) and we recall that
this is always complex for any $E\neq 0$.
The general idea is that we extract the dynamical mass from the real part
of the effective potential (\ref{eq:ReVeffET0}) and check if the imaginary
part (\ref{eq:ImVeffET0}) is not too large compared to the real part.

\begin{figure}[ht]
{
\centerline{\scalebox{1.0}{\includegraphics{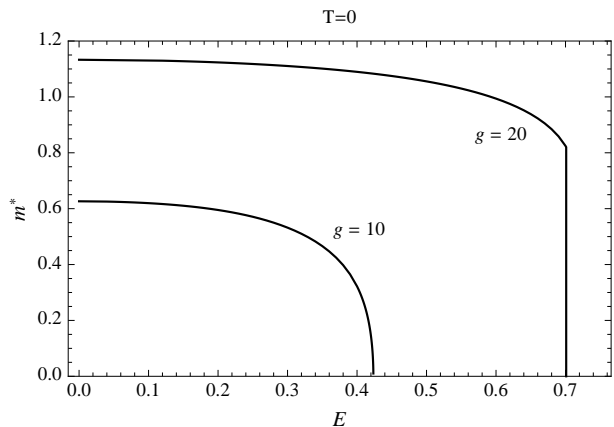}}}
\caption{\footnotesize
  The plot of the dynamical mass with respect to the $E$ field at $T=0$.
  For the coupling $g=10$ and $20$, we have the second and first order
  phase transitions, respectively.
}\label{fig:NJLEMT0}
}
\end{figure}

It is straightforward to obtain the tadpole condition
(\ref{eq:tadCancel}) for the real part of the effective potential
(\ref{eq:ReVeffET0}) and numerically solve for the dynamical
mass $m^*$ for a given $E$.
The result is presented in Figure~\ref{fig:NJLEMT0}
for $g=10$ and $20$.
The diagram indicates that the $E$ field, in general, acts as the destabilizer
of the quark boundstate, just as the temperature and the chemical potential
({\it c.f.}, the pure $B$ background $T$-$m^*$ diagram of 
Figure~\ref{fig:NJLMT} and 
$\mu$-$m^*$ diagram of Figure~\ref{fig:NJLMMu}).
When the coupling is not too large (but above the critical coupling $g_c$
which is finite), the mass continuously goes down to zero, indicating
the second order phase transition.
This second order phase transition was discovered in
Reference~\cite{Klevansky:1989vi}.

What has not been noticed is the existence of the first order phase
transition at a large coupling constant, as shown in the figure for $g=20$.
The behavior of the effective potential is very similar to that of
Figure~\ref{fig:NJLVeffF0}, where the role of the chemical potential in
that figure is replaced with the $E$ field (and of course $T=0$ for the
current discussion).
For this reason, we can expect the very similar result when we include
the chemical potential to this zero temperature system.

\begin{figure}[ht]
{
\centerline{\scalebox{0.9}{\includegraphics{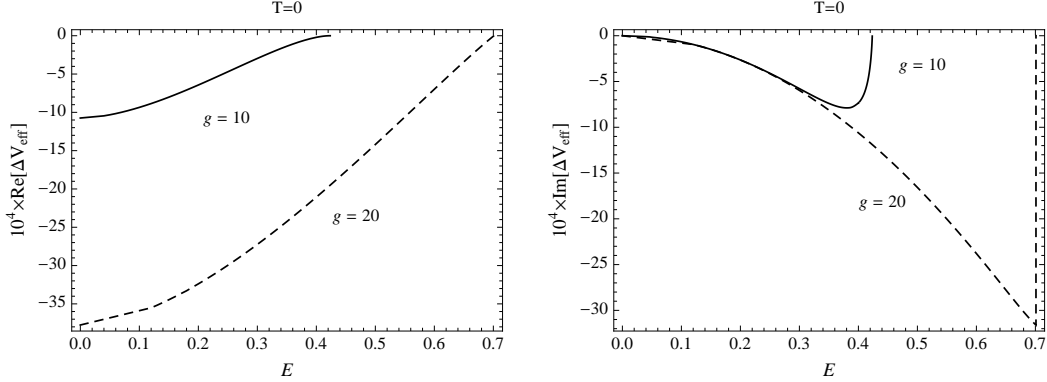}}}
\caption{\footnotesize
  The plots of the real and imaginary parts of the difference action at
  $T=0$ and $m=m^*$.
  The curves terminate at the phase transition point.
  Near the phase transition points, the imaginary part strongly dominates
  over the real part.
}\label{fig:NJLT0VeffE}
}
\end{figure}

We now must check if the results drawn from the real part of the effective
potential above have any physical significance, by examining the imaginary
part of the potential.
Figure~\ref{fig:NJLT0VeffE} shows the on-shell values of the effective
potential, that is, it is evaluated at $m=m^*$, with respect to the $E$ field.
The figure shows that the real part dominates when the $E$ field is small
and away from the critical value.
However, when the electric field is near the critical value, the imaginary
part strongly dominates over the real part.
This could indicate that the information about the phase transition points
is not reliable, due to the breakdown of the stable background by the 
pair creation.

To obtain further insight into this issue, we recall that Schwinger
has worked out the rate of the pair creation in 
Reference~~\cite{Schwinger:1951nm} and it is roughly given by
$\exp[-\pi m^2/E]$.
This means that the results from the real part is reliable
when $m^2\gtrsim E$.
Again, this condition is not so well satisfied around the point of the 
phase transitions.
Therefore, though there are good indications of the first and second
order phase transitions, it is not safe to conclude so.
Most certainly, the values of the critical electric field cannot
be trusted.

As in Reference~\cite{Klevansky:1989vi}, the imaginary part of the
effective potential has not been paid much attention, probably due
to the NJL model's lack of the confinement.
The confinement may fix the problem, but then we are dealing with
different model and the conclusions from the real part still
does not seem to correspond to the real world.
Nonetheless, we have carried out the analysis for both the real and
imaginary parts, because we can compare the results to the deconfined
phase of the SS model.
\\

\noindent {\it Weak $E$ Expansion}

\begin{figure}[ht]
{
\centerline{\scalebox{0.9}{\includegraphics{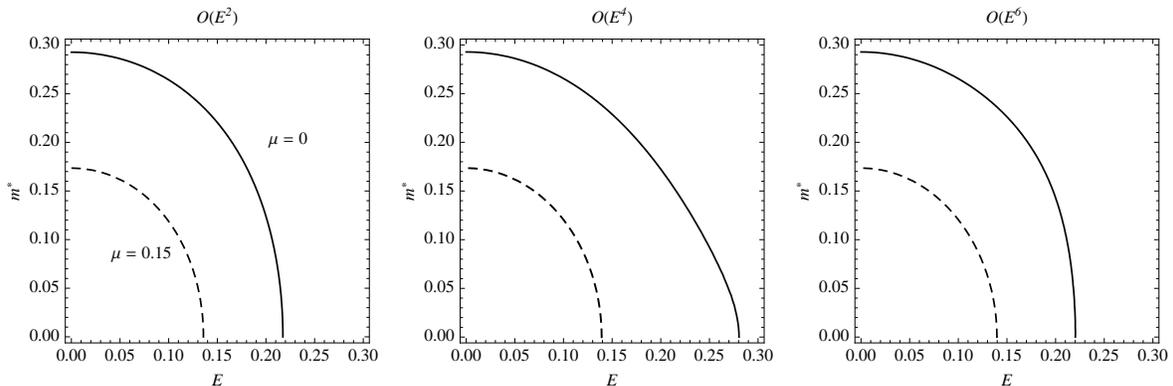}}}
\caption{\footnotesize
  The plot of the dynamical mass against $E$ for $T=0.3$ and $g=10$.
  From the left $\mathcal{O}(E^2)$, $\mathcal{O}(E^4)$ and
  $\mathcal{O}(E^6)$.
  They show that the higher order in the weak $E$ expansion is converging.
  We also observe the general tendency of $\mu$ as the destabilizer.
}\label{fig:NJLEMT}
}
\end{figure}

\noindent We now discuss the weak $E$ expansion of the $T>0$
effective potential (\ref{eq:Eexpansion}).
The expansion completely washes out the singularities and the potential
is real.
One can easily obtain the tadpole condition (\ref{eq:tadCancel}) from
the potential (\ref{eq:Eexpansion}) and solve numerically for the dynamical
mass.
Figure~\ref{fig:NJLEMT} shows the result for $g=10$, $T=0.3$ and up to
$\mathcal{O}(E^6)$.
We observe fairly good convergence of the higher terms and the result
may be trusted.
In particular, the second order phase transition observed at $T=0$
appears to persist at finite temperature.

\begin{figure}[ht]
{
\centerline{\scalebox{0.8}{\includegraphics{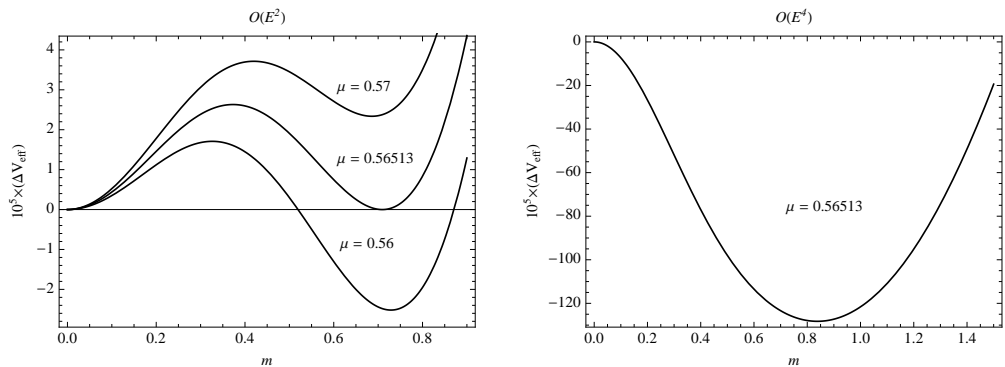}}}
\caption{\footnotesize
  The effective potential at $T=0.3$ and $g=20$.
  The $\mathcal{O}(E^2)$ on the left shows the first order phase transition 
  consistent with the $T=0$ result, 
  however, the inclusion of the higher order terms [shown on the right
  is up to $\mathcal{O}(E^4)$] destroys this behavior.
}\label{fig:NJLVeffET}
}
\end{figure}

It is of interest to see if the first order transition at the zero
temperature can also be observed at finite temperature.
Figure~\ref{fig:NJLVeffET} shows the plot of the effective potential
at $g=20$ and $T=0.3$.
The $\mathcal{O}(E^2)$ effective potential shows the first order phase
transition, but the examination of the higher order terms imply that
this may not be trusted.
On the right panel, the effective potential up to $\mathcal{O}(E^4)$
is shown.
We see that the higher order terms are not converging and invalidates 
the result of $\mathcal{O}(E^2)$.
Therefore, we cannot conclude if the first order phase transition at
$g=20$ and $T=0$ persist in finite temperature.

When the temperature is lower, the convergence of the higher order
terms gets worse.
In all cases, the potential up to $\mathcal{O}(E^2)$ shows all
the expected behavior (as in the previous paragraph), but
one finds that the higher order terms are behaving completely
unphysical way.
We have expanded the $T=0$ effective potential with respect
to small $E$ and found similar defects in the higher order terms,
so the expansion at low temperature is not valid.
For this reason, we do not attempt to draw the $E$-$T$ phase diagram.

%%%%%%%%%%%%%%%%%%%%%%%%%%%%%%%%%%%%%%%%%%%%%%%%%%%%%%%%%%%%%%%%%%%%%
\section{Sakai-Sugimoto Model}\label{sec:SS}
%%%%%%%%%%%%%%%%%%%%%%%%%%%%%%%%%%%%%%%%%%%%%%%%%%%%%%%%%%%%%%%%%%%%%
We analyze the SS model in this section.
We specifically deal with the one-flavor case, but
the generalization to more than one flavor is straightforward.
We first discuss the general features of the model, including the chiral
phase transition and the dynamically generated mass of the model.
Then, after setting up the model with external parameters,
it is further examined separately for the confined
and deconfined phases.
Finally we present the results of the numerical evaluation.

%%%%%%%%%%%%%%%%%%%%%%%%%%%%%%%%%%%%%%%%%%%%%%%%%%%%%%%%%%%%%%%%%%%%%%%%
\subsection{Chiral Phase Transition and Dynamical Mass}
\label{sec:chiPhTrcMass}

The SS model is defined on the IIA background geometry of D$4$ branes
in the near horizon form,
\begin{align}
  ds^2 &= (U/R)^{3/2} \{f_tdX_0^2+(dX_i)^2+f_4(dX_4)^2 \}
        + (R/U)^{3/2}(f^{-1}dU^2 + U^2 d\Omega_4^2)
  \nonumber\\
  \Phi &= \Phi_0 + \frac{3}{4}\ln (U/R)
  \nonumber\\
  C_{(3)} &= -3 g_s^{-1} R^3 \sin^3\theta_1 \sin^2\theta_2 \cos\theta_3
                d\theta_1 \wedge d\theta_2 \wedge d\theta_4
  \;,
\end{align}
where $d\Omega_4$ is the metric for the four-sphere with the coordinates
$\theta_i$, $\Phi$ is the dilaton and $C_{(3)}$ is the RR potential to which
the D$4$-branes couple magnetically.
We also have
\begin{equation}
  g_s := \exp(\Phi_0) \;,\qquad R^3 := \pi g_s N_c l_s^3
  \;,
\end{equation}
and the ``blackening factors'' are
\begin{align}\label{eq:capish}
  f_t = f = 1 - (U_T/U)^3 &\;\text{and}\;
  f_4 = 1
  \;\text{for the high temperature deconfined phase}
  \nonumber\\
  f_4 = f = 1 - (\UKK/U)^3 &\;\text{and}\;
  f_t = 1
  \;\text{for the low temperature confined phase}
  \;.
\end{align}
The $X_4$ direction is compactified and at finite temperature, the Euclidean
$X_0$ direction is also compactified as well.
To avoid the conical singularity from the compactifications, we must
have the circumference of $X_4$-direction, $\beta_4$ as
\begin{align}\label{eq:beta4}
  \beta_4 = \frac{4\pi}{3} \bigg( \frac{R^3}{\UKK} \bigg)^{1/2}
  \;,
\end{align}
in the low temperature confined phase and the circumference
of $X_0$-direction, $\beta$ as
\begin{align}\label{eq:beta}
  \beta = \frac{4\pi}{3} \bigg( \frac{R^3}{U_T} \bigg)^{1/2}
  \;,
\end{align}
in the high temperature deconfined phase.

The system with the {\it non-compact} $X_4$ direction is of interest as well.
In this case, the system is always in the deconfined phase and this
allows us to discuss the low temperature behavior of the model in
the deconfined phase.
In fact, when we discuss the deconfined phase below, we will be cavalier
about the confinement/deconfinement critical temperature.
\\

\noindent {\it Chiral Phase Transition}

\begin{figure}[ht]
{
\centerline{\scalebox{0.6}{\includegraphics{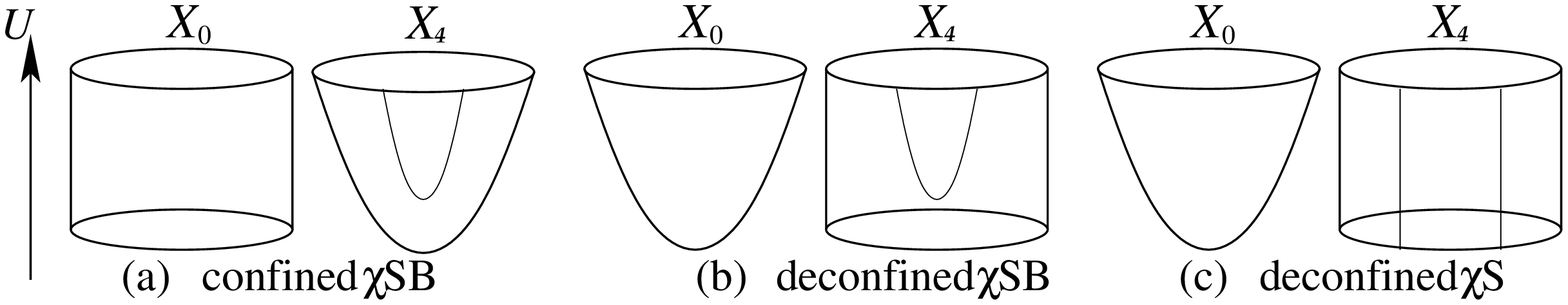}}}
\caption{\footnotesize
  Three phases of the SS model.
  The coordinate $U$ is the radial coordinate transverse to the D$4$ branes.
  The thin lines shown in the $X_4$ circles represent the probe D$8$ branes.
  The diagram (a) represents the low temperature confined phase with broken
  the chiral symmetry ($\chi$SB), (c) is the high temperature deconfined phase
  with the restored chiral symmetry ($\chi$S) and (b) is the intermediate
  deconfined phase with $\chi$SB.
}\label{fig:threePhases}
}
\end{figure}

\noindent The model has the well-known three phases \cite{Aharony:2006da}
shown in Figure~\ref{fig:threePhases}.
The thin lines in the $X_4$ circles are the probe D$8$-branes.
The confinement [Diagram (a)]
and deconfinement [Diagrams (b,c)]
phases are determined only
through the D$4$ background, independent of the probe D$8$,
and the phase transition occurs at $\beta=\beta_4$ \cite{Aharony:2006da}.

The U-shape configurations of D$8$ shown in Diagrams (a,b) represent
the chiral symmetry breaking in this model.
There is a range of temperature and asymptotic separation of the D$8$
branes, denoted by $L$, so that the background is in the deconfined phase
and the chiral symmetry is broken, as shown in the diagram (b) of 
Figure~\ref{fig:threePhases}.
Aharony {\it et al.} \cite{Aharony:2006da} have shown that this intermediate
phase exists when $L<L_c\approx 0.97(\beta_4/2\pi)$ in the deconfined phase.
When $L>L_c$, the intermediate phase does not exist and the chiral phase
transition and the confinement/deconfinement phase transition occurs
simultaneously.
The chiral phase transition in the deconfined phase is determined by
comparing the values of the D$8$ probe action in U- and 
$\parallel$-configurations for a given set of parameters.
Since the probe configuration jumps from one to the other at the phase
transition point, this is always a first order phase transition.
\\

\noindent {\it String Endpoint Mass (SEP)}

\begin{figure}[ht]
{
\centerline{\scalebox{0.7}{\includegraphics{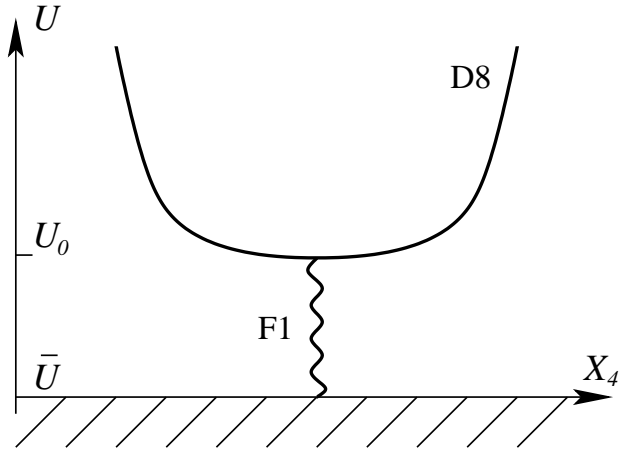}}}
\caption{\footnotesize
  A slice of the string configuration describing the high spin mesons.
  The location $U=U_0$ is the tip of the U-shaped D$8$ and $\bar U = \UKK$
  for the confined phase and $\bar U = U_T$ in the deconfined phase.
  The string is also stretching out of the figure into the field theory
  directions and coming back up again to the D$8$ tip.
}\label{fig:cMass}
}
\end{figure}

\noindent The model in the U-shape configuration exhibits various spectra
which are similar to low energy QCD, such as mesons.
For instance, the fluctuations of the embedding map from D$8$ to the spacetime
represent the scalar mesons and the D$8$ worldvolume gauge fields
(of non-zero modes in a certain mode expansion)
correspond to the vector mesons.
Moreover, the higher spin mesons are described by the spinning 
open string of the following configuration~\cite{Casero:2005se}.
The endpoints of the string are both attached to the the tip of the 
U-shaped D$8$-brane, in the different locations of the field theory
directions, and they are rotating around each other in those directions.
The body of the string is draped all the way down to the wall and
lying there.
Figure~\ref{fig:cMass} represents a slice of the string configuration
and the field theory directions are coming out of the diagram.
The ``wall'' is the tip of the ``$X_4$ cigar'' in the confined phase
and the horizon in the deconfined phase (see Figure~\ref{fig:threePhases}).

The energy of the string segment in the $U$ direction
(the vertical segment in Figure~\ref{fig:cMass}) is given by
\begin{align}\label{eq:constituentM}
  m^* = \frac{1}{2\pi\alpha'} \int_{\bar U}^{U_0}
          \sqrt{g_{00}g_{UU}}dU
        = \frac{1}{2\pi\alpha'} \int_{\bar U}^{U_0}
          \sqrt{f_tf^{-1}}dU
  \;,
\end{align}
where $\bar U = \UKK$ and $U_T$ for the confined and deconfined phases,
respectively.
It is argued quite generally in Reference~\cite{Casero:2005se}
that $m^*$ represents the dynamically generated mass of the quarks.%
\footnote{
  The arguments made in Reference~\cite{Casero:2005se} is especially
  reliable for a very high spin mass, because the semi-classical treatment
  of the string is good in this regime.
  However, the analysis of the Regge trajectory carried out in the reference
  indicates that the description of the mass is fairly
  good even for the low spin mesons.
}
One evidence shown in the reference is that the masses of the 
massive pseudo-scalar and vector mesons increase monotonically with the
energy of the vertical segment and when the vertical part is long enough,
the masses increase linearly.
Also it is pointed out that the model (without a tachyon potential) is
believed to describe a QCD-like system with zero current quark mass,
because the pions are massless.
This leaves the only candidate, the dynamically generated quark mass.
This mass, however, is different from the constituent mass and is
called ``string endpoint mass (SEP)''.%
\footnote{
  Private communication with J. Sonnenschein.
}
The difference mainly comes from the energy of the string segment
that is horizontal to the wall.
For this reason, we refer this mass as the dynamical mass or SEP.

One might argue that the meson spectrum and the dynamical mass would
not make sense in the deconfined phase.
However, we have the fact that the SS model exhibits the chiral
phase transition in the
deconfined phase, and moreover, there are evidences from the lattice simulations
that indicate the existence of the quark boundstate in QCD, significantly
above the deconfinement temperature \cite{de Forcrand:2000jx}.
See Reference~\cite{Peeters:2006iu}
for the holographic study of mesons in the deconfined phase.

In what follows, we will be interested in the dynamical mass
described here and we examine it for both in the confined 
and deconfined phases.

%%%%%%%%%%%%%%%%%%%%%%%%%%%%%%%%%%%%%%%%%%%%%%%%%%%%%%%%%%%%%%%%%%%%%%%%
\subsection{External Parameters}\label{sec:BNSandMu}

\noindent {\it Constant NSNS Field}

\noindent The aim of this section is to investigate the chiral phase
transitions and the dynamical masses under the influence
of external parameters, such as temperature.
One of the external parameters in which we are interested is the background
constant electromagnetic field and
we would like to introduce this into the SS model.
The most natural thing to do is to introduce the background gauge field
of diagonal $U(1)_V$ that comes from $U(1)\times U(1)$ of D$8$ and
$\barDEight$ probes.
This, in fact, behaves exactly like the electromagnetic field for our
one-flavor model.%
\footnote{
  As remarked in Reference~\cite{Bergman:2008sg}, the quarks have different
  couplings to the electromagnetic field for $N_f>1$, but even in this case,
  we should expect that the $U(1)_V$ gauge field acts similar to the real
  background electromagnetic field.
}
Now, recall that the gauge invariant quantity in the D-brane action is
not the field strength tensor of the gauge field, but the combination
of that with the NSNS $B$-field ($\BNS$), pulled back on the brane
with respect to the embedding map of the D-brane.
Therefore, we can freely trade the background gauge field to $\BNS$,
up to the embedding map and the factor of $2\pi\alpha'$.
As we will describe momentarily, we will choose the standard simple
embedding map of the D$8$ brane so that the introduction of the constant
background electromagnetic field is equivalent to the constant $\BNS$
in the IIA supergravity background.
Notice that the constant $\BNS$ does not affect the supergravity background.

Let us then consider which components of $\BNS$ we should turn on.
We are interested in the fields that have direct interpretation in the
dual field theory.
This implies that we should have at least one component in the field theory
direction.
We also avoid turning on $(\BNS)_{i u}$, because according to the general
holographic dictionary, they source the boundary currents in $X_i$-direction
and we do not have such counterpart in the NJL model.
[We will discuss the exceptional case $(\BNS)_{0U}$ in a moment.]
There are interesting possibilities to turn on
$(\BNS)_{\mu 4}$ and $(\BNS)_{4U}$.
However, these terms complicates the probe action and more importantly,
the field theory interpretation of those quantities are not clear,
so in the present work, we will not discuss these possibilities.
We are then left with
\begin{align}\label{eq:constEB}
  E_{i}:=(\BNS)_{0i}
  \quad\text{and}\quad
  B_{i}:=\epsilon_{ijk}(\BNS)_{jk}
  \;,
\end{align}
where $\epsilon_{123}=+1$ for the convention and
we deal with these background fields
in the following.

Before setting out for the further analysis,
we must mention the topological constraints imposed on the constant $\BNS$.
As can be seen in Figure~\ref{fig:threePhases}, the SS model
possesses the degenerate point in the background geometry,
and any constant tensor (of rank$>0$)
cannot make sense as a field at the degenerate point.
This severely restricts the possible constant $\BNS$ and
the situation is summarized in Table~\ref{tbl:bolts}, together with
the excluded cases for the completeness.
\begin{table}[ht]
{
\centerline{
\begin{tabular}{c||ccccccc}
${\BNS}_{(A,B)}$ & $(0,i)$ & $(0,4)$ & $(0,U)$ & $(i,j)$ & $(i,4)$ & $(i,U)$ & $(4,U)$ \\
\hline\hline
conf. & \cm & $\times$ & \cm & \cm & $\times$ & \cm & $\times$ \\
\hline
deconf. & $\times$ & $\times$ & $\times$ & \cm  & \cm  & \cm  & \cm
\end{tabular}
}
\caption {\footnotesize
  The topological constraints on the constant $\BNS$-field.
  The top row is the components of $\BNS$-field.
}
\label{tbl:bolts}
}
\end{table}
In particular, notice that the electric fields $E_i$ are not at all
allowed in the deconfined phase.
As we will see, the free energy of the system
in this case is always imaginary
for any nonzero constant electric field and
this is precisely due to the violation of the topological constraint.
The reader who looked at the previous section for the NJL model would
recall the same imaginary behavior in the effective potential.
There, we carried out the analysis observing the real and imaginary
parts separately.
Thus, we do here the same for the sake of the comparison and we assume
that the imaginary part has the same physical interpretation as in NJL, namely,
the instability due to the pair creation of the quark pair.%
\footnote{
  We do not induce the current {\it a la} Karch
  and O'Bannon \cite{Karch:2007pd}, because we do not have such counterpart
  in the NJL model.
  See Reference~\cite{Bergman:2008sg} for the treatment with the current.
}
\\

\noindent {\it Chemical Potential}

\noindent In addition to the constant electromagnetic field,
we include the quark chemical potential and see how the system respond
to this.
For this, let us consider the $(0,U)$-component of the $\BNS$ field.
The AdS/CFT dictionary tells us that this sources the charge density
of the boundary theory.
Recall that a chemical potential can be thought to be the time-component of
the gauge field that minimally couples to the conjugate charge density.
Thus, this component of $\BNS$-field should be
directly related to the quark chemical potential of our interest.
In fact, on the probe D$8$-brane, this is equivalent to the same
component of the field strength tensor, indicating that $(\BNS)_{0U}$
is the $U$-derivative of the chemical potential.
The treatment of the chemical potential in Reference~\cite{Horigome:2006xu}
actually starts by including the time component of the worldvolume
gauge field and we have just argued that this is equivalent to
the introduction of $(\BNS)_{0U}$.
Thus, in addition to the components (\ref{eq:constEB}),
we include
\begin{align}
  E_U(U) := {\BNS}_{0U}(U)
  \;,
\end{align}
where following Reference~\cite{Horigome:2006xu}, we assume the dependence
of this component on $U$ and this function must be determined by solving
the equations of motion that follow from the probe action.
Finally,
notice that since
\begin{align}
  d [E_U(U) dX_0 \wedge dU] = 0
  \;,
\end{align}
this does not affect the supergravity background.

%%%%%%%%%%%%%%%%%%%%%%%%%%%%%%%%%%%%%%%%%%%%%%%%%%%%%%%%%%%%%%%%%%%%%%%%
\subsection{Probe Action and Equations of Motion}\label{sec:probeEOM}

\noindent {\it Embedding Map}

\noindent Writing down a D-brane action requires the information on
how to embed the D-brane in the spacetime.
We adopt the standard embedding map of the original paper \cite{Sakai:2004cn}.
To straighten out the notations and conventions, we explicitly write
out the map.

Let the spacetime coordinates be
\begin{equation}
  \text{spacetime: }  X_\mu,\quad
  \mu = 0\;,1\;,2\;,3\;,4\;,u\;,\theta_1\;,\theta_2\;,\theta_3\;,\theta_4
  \;,
\end{equation}
and let the D8 worldvolume coordinates be
\begin{equation}
  \text{D8: }  x_\alpha,\quad
  \alpha = 0\;,1\;,2\;,3\;,u\;,\theta_1\;,\theta_2\;,\theta_3\;,\theta_4
  \;.
\end{equation}
Note that $\alpha=4$ is missing.
With the slight abuse of notation, let $X$ denote the embedding map
\begin{equation}
  X : \text{D8}\to\text{spacetime}
  \;,
\end{equation}
and following the standard practice, this is the identity map, except
\begin{align}
  X_4 (x_\alpha) = x_4(x_u)
  \;,
\end{align}
where $x_4(x_u)$ is the profile function.

We define the metric on the D$8$-brane as the pullback of the spacetime
metric two-tensor, with respect to the embedding map,
\begin{align}
  g_{\text{D8}} := X^*g_{\text{st}}
  \;,
\end{align}
and this yields
\begin{align}
  ds_\text{D8}^2 = \bigg( \frac{u}{R} \bigg)^{3/2}
		 \bigg[ f_t dx_0^2+(dx_i)^2
		 +\bigg\{ x_4'^2f_4 + 
		  \bigg( \frac{R}{u} \bigg)^3 f^{-1} \bigg\}du^2
		 \bigg]
		 + \bigg( \frac{R}{u} \bigg)^{3/2} u^2 d\Omega_4^2
  \;,
\end{align}
where we have renamed as $x_u\to u$ and $x_4':=dx_4/du$.
Parenthetically, we also have
\begin{align}
  f_t = f = 1 - (u_T/u)^3 &\;\text{and}\;
  f_4 = 1
  \;\text{for the high temperature deconfined phase}
  \nonumber\\
  f_4 = f = 1 - (\uKK/u)^3 &\;\text{and}\;
  f_t = 1
  \;\text{for the low temperature confined phase}
  \;,
\end{align}
where the spacetime coordinate $U$ in Equations~(\ref{eq:capish})
is changed to the worldvolume
coordinate $u$.
In addition,
we pullback $\BNS$ onto the D$8$ worldvolume, $\tBNS:=X^*\BNS$
and we have
\begin{align}
  \tBNS =
  \begin{pmatrix}
    0 & iE_1 & 0 & 0 & iE_u(u) \\
    -iE_1 & 0 & B_3 & 0 & 0 \\
     0    & -B_3 & 0 & B_1 & 0 \\
     0    &  0  & -B_1 & 0 & 0 \\
    -iE_u(u) & 0 & 0 & 0 & 0
  \end{pmatrix}
  \;,
\end{align}
where we have dropped $E_{2,3}$ and $B_2$, because we will be interested
in the system with $\vec B\; (\perp \text{or} \parallel)\; \vec E$
and for this purpose, the components above are sufficient.
We have also extracted the factor of $i$ and redefined
the real $E_{1,u}$.
\\

\noindent {\it Probe Action}

\noindent For our background
$C_{(3)}$ and $\BNS$, we do not have the Chern-Simons
coupling of the D8-brane, so the probe action consists only of the DBI part
\begin{equation}
  S_\text{DBI} = 2T_8 \int d^9 x e^{-\Phi}
		 \sqrt{\det(g_\text{D8}+\tBNS)}
  \;,
\end{equation}
where $T_8$ is the brane tension and the factor of $2$ comes from
the D$8$ and $\barDEight$ contributions.
We define the ``Lagrangian'' $\mathcal{L}$ through the relation
\begin{align}
  e^{-\Phi}\sqrt{\det(g_\text{D8}+\tBNS)} =
  g_s^{-1} \mathcal{L} \, |\Omega_4|
  \;,
\end{align}
where
\begin{align}
  |\Omega_4|:=\sin^3\theta_1\sin^2\theta_2\sin\theta_3
  \;.
\end{align}

According to the discussion made in Section~\ref{subsec:anomaly},
the cases $\vec B \parallel \vec E$ ($B_3=0$) and 
$\vec B \perp \vec E$ ($B_1=0$) correspond to the $U(1)_A$
anomalous and non-anomalous boundary theory, respectively.
Though we are interested in the non-anomalous case at the end,
we carry out the analysis for both cases up until the numerical evaluation,
because it is easy to generalize to $N_f>1$ 
and we can observe some general similarities and dissimilarities
to the NJL model.
Examining separately for those cases, we obtain
for the anomalous case
\begin{align}\label{eq:LAnom}
  \mathcal{L}_{\text{Anom}} =
  u^4 \bigg[ \big\{ (R/u)^3f_4^{-1} + f {x_4'}^2 \big\}\mathcal{B}_1\mathcal{E}_1
                - (R/u)^3\mathcal{B}_1E_u^2 \bigg]^{1/2}
  \;,
\end{align}
and for the non-anomalous case
\begin{align}\label{eq:LNAnom}
  \mathcal{L}_{\text{NAnom}} =
  u^4 \bigg[ \big\{ (R/u)^3f_4^{-1} + f {x_4'}^2 \big\}\mathcal{A}
                - (R/u)^3\mathcal{B}_3E_u^2 \bigg]^{1/2}
  \;,
\end{align}
where we have defined
\begin{align}
  &\mathcal{B}_i := 1 + (R/u)^3 B_i^2
  \;,\quad
  \mathcal{E}_1 := 1 - (R/u)^3 f_t^{-1}E_1^2
  \nonumber\\
  &\text{and}\quad
  \mathcal{A} := 1 + (R/u)^3 (B_3^2 - f_t^{-1}E_1^2)
  \;.
\end{align}

We see that $\mathcal{L}$s depend only on the coordinate $u$, so
the action can be written as
\begin{align}\label{eq:SDBI}
  S_\text{DBI} = 2\omega_4 V_3 T_8 g_s^{-1}\beta \int du \mathcal{L}
  \;,
\end{align}
where $\beta$ is the integration of Euclidean time,
\begin{align}
  \omega_4 := \int d\theta_1d\theta_2d\theta_3d\theta_4 \,|\Omega_4|
  \;,\quad\text{and}\quad
  V_3 = \int d^3x_i
  \;.
\end{align}
\\

\noindent {\it Equations of Motion}

\noindent The degrees of freedom in the action appears to be $x_4(u)$
and $E_u(u)$.
In the D$8$ worldvolume, however, it is more natural to trade
$\tBNS$ to $2\pi\alpha' F$, that is, we identify
$(\tBNS)_{0u}=E_u(u)$ as the $u$ derivative of
the worldvolume gauge field $A_0$,
and treat the gauge field as the degrees of freedom in the action
(see Reference~\cite{Horigome:2006xu}).
Thus, we replace as
\begin{align}
  E_u(u) \to \mu(u)' := 2\pi\alpha'\partial_uA_0(u)
  = 2\pi\bigg(\frac{l_s}{R}\bigg)^2 \frac{\partial(A_0R)}{\partial(u/R)}
  \;,
\end{align}
where we have emphasized that everything is measured in the units of
the scale $R$ and $\mu$ is dimensionless that involves
the factor of $(l_s/R)^2$ in its definition.

Identifying the degrees of freedom, the equations of motion that follow
from the action are
\begin{align}
  \frac{\partial\mathcal{L}}{\partial{x_4'}} = c_1
  \quad\text{and}\quad
  \frac{\partial\mathcal{L}}{\partial{\mu'}} = -c_2
  \;,
\end{align}
where $c_{1,2}$ are the constants of motion and the minus sign
on $c_2$ is inserted for the later convenience.
For $\mathcal{L}_{\text{NAnom}}$ of Equation~(\ref{eq:LNAnom}),
they lead to
\begin{align}\label{eq:EOMs1}
  c_1 =& u^4f\mathcal{A}x_4'
        \bigg[ \big\{ (R/u)^3f_4^{-1} + f {x_4'}^2 \big\}\mathcal{A}
                - (R/u)^3\mathcal{B}_3{\mu'}^2 \bigg]^{-1/2}
  \nonumber\\
  c_2 =& u^4\mathcal{B}_3(R/u)^3 \mu'
        \bigg[ \big\{ (R/u)^3f_4^{-1} + f {x_4'}^2 \big\}\mathcal{A}
                - (R/u)^3\mathcal{B}_3{\mu'}^2 \bigg]^{-1/2}
  \;.
\end{align}
The equations of motion for $\mathcal{L}_{\text{Anom}}$ of 
Equation~(\ref{eq:LAnom}) can be obtained by the replacements
$\mathcal{A}\to\mathcal{B}_1\mathcal{E}_1$
and $\mathcal{B}_3\to\mathcal{B}_1$.

%%%%%%%%%%%%%%%%%%%%%%%%%%%%%%%%%%%%%%%%%%%%%%%%%%%%%%%%%%%%%%%%%%%%%%%%
\subsection{Confined Phase}\label{subsec:conf}
We now specialize the discussion to the confined phase.
We will concentrate on the $\vec B \perp \vec E$
non-anomalous case,
Equation~(\ref{eq:LNAnom}), because the anomalous case,
Equation~(\ref{eq:LAnom}), can be obtained by the replacement just
mentioned.
For the clarity of expressions, we set $B:=B_3$ and $E:=E_1$,
accordingly, we have $\mathcal{A}=1+(R/u)^3(B^2-E^2)$ and
$\mathcal{B}:=1+(R/u)^3B^2$.
Also we remember that in this phase, $f_t=1$ and
$f_4=f=1-(\uKK/u)^3$.
\\

\noindent {\it The U-Shape Solution and the Free Energy}

\noindent The $x_4$ direction is degenerate at $u=\uKK$ in this phase
(see Figure~\ref{fig:threePhases}) and the D$8$ and $\barDEight$
probes must join somewhere.
Following the common practice, we assume that the probes join smoothly
at $u=u_0$, forming a U shape, and this results in the condition
\begin{align}\label{eq:smoothCond}
  x_4'(u_0) = \infty
  \;.
\end{align}
The equations of motion (\ref{eq:EOMs1}) can be solved
for $x_4'$ and $\mu'$, then we can use the condition (\ref{eq:smoothCond})
to eliminate the constant $c_1$ in favor of $u_0$.
By resetting $c:=c_2$, we obtain
\begin{align}
    {x_4'}^2 =&
        \frac{(R/u)^6f_0(\mathcal{A}_0/\mathcal{B}_0)
              \big\{ u_0^8\mathcal{B}_0+c^2(u_0/R)^3 \big\} }
        { f^2 \big[ (R/u)^3(u^8f\mathcal{A} - u_0^8f_0\mathcal{A}_0)
                + (c^2/u^3) \{ u^3f(\mathcal{A}/\mathcal{B})
                               -u_0^3f_0(\mathcal{A}_0/\mathcal{B}_0) \} \big] }
  \nonumber\\
  {\mu'}^2 =&
        \frac{ (\mathcal{A}/\mathcal{B})^2c^2}
        { (R/u)^3(u^8f\mathcal{A} - u_0^8f_0\mathcal{A}_0)
                + (c^2/u^3) \{ u^3f(\mathcal{A}/\mathcal{B})
                               -u_0^3f_0(\mathcal{A}_0/\mathcal{B}_0) \} }
  \;,
\end{align}
where the quantities $X_0$ are meant to be $X(u_0)$.

It is clear in the equation of $\mu'$ that this blows up at $u=u_0$
leading to a singular $\BNS$.
To avoid the singularity, we must set
\begin{align}\label{eq:cZero}
  c=0
  \;,
\end{align}
for this U-shape solution and
the solution boils down to
\begin{align}\label{eq:confSoln}
  {x_4'}^2 = f^{-2}\bigg(\frac{R}{u}\bigg)^{3}
           \bigg( \frac{u^8f\mathcal{A}}
                       {u_0^8f_0\mathcal{A}_0}
           -1 \bigg)^{-1}
  \quad\text{and}\quad
  \mu' \equiv 0
  \;.
\end{align}
The condition (\ref{eq:cZero}) has the drastic consequence that the system
is completely independent of the chemical potential $\mu$ and in particular,
the dynamical mass will be insensitive to the chemical potential.

Consulting with Equation~(\ref{eq:SDBI}), we define the (dimensionless)
free energy of the system by
\begin{align}
  \hat F := R^{-5} \int_{u_0}^\infty du \mathcal{L}
  \;,
\end{align}
and we are going to evaluate the free energy for the solution
derived above.
Plugging the solution (\ref{eq:confSoln}) in the Lagrangian~(\ref{eq:LNAnom}),
we obtain the on-shell expression
\begin{align}\label{eq:cnfProbeA}
  \mathcal{L} = u^4\bigg(\frac{R}{u}\bigg)^{3/2}f^{-1/2}\mathcal{A}^{1/2}
                \bigg( 1- \frac{u_0^8f_0\mathcal{A}_0}{u^8f\mathcal{A}}
                \bigg)^{-1/2}
  \;.
\end{align}
Following Reference~\cite{Aharony:2006da}, we measure the quantities
in the units of $u_0$ by introducing%
\footnote{
  This scaling by $u_0$ is a temporary prescription and as mentioned
  earlier, we will measure all physical quantities in the units of $R$
  at the end.
}
\begin{align}
  y:= u/u_0
  \;,\quad
  y_R:=R/u_0
  \;,\quad\text{and}\quad
  \yKK:=u_{\text{KK}}/u_0
  \;.
\end{align}
Again, following the same reference, we further set
$z=y^{-3}$, yielding
\begin{align}
  \hat F = \frac{1}{3}
    y_R^{-7/2} \int_0^1 dz z^{-13/6} f^{-1/2} \mathcal{A}^{1/2}
      \bigg(1- z^{8/3} \frac{f_0\mathcal{A}_0}{f\mathcal{A}} \bigg)^{-1/2}
  \;.
\end{align}
From the relation (\ref{eq:beta4}), we have
\begin{align}\label{eq:yRb4yKK}
  y_R = a^{-1}\hat\beta_4^2 \yKK
  \quad\text{with}\quad
  a:=\bigg(\frac{4\pi}{3}\bigg)^2
  \quad\text{and}\quad
  \hat\beta_4:= \beta_4/R
  \;,
\end{align}
and we can eliminate $y_R$ in favor of $\yKK$.
We then have
\begin{align}
  \hat F =& \frac{1}{3} (a^{-1}\hat\beta_4^2)^{-7/2}\yKK^{-7/2}
    \int_0^1 dz z^{-13/6} f^{-1/2} \mathcal{A}^{1/2}
      \bigg(1- z^{8/3} \frac{f_0\mathcal{A}_0}{f\mathcal{A}} \bigg)^{-1/2}
  \nonumber\\
  f=&1-\yKK^3z
  \;,\;
  \mathcal{A} = 1+z\hat\beta_4^6\yKK^3(\bar B^2-\bar E^2)
  \;,
  \nonumber\\
  f_0=&1-\yKK^3
  \;,\;
  \mathcal{A}_0 = 1+\hat\beta_4^6\yKK^3(\bar B^2-\bar E^2)
  \;,
\end{align}
where we set $\bar B:=a^{-3/2}B$ and $\bar E:=a^{-3/2}E$.
We again note that the anomalous case with the
Lagrangian~(\ref{eq:LAnom}) can be obtained by the replacement
$\mathcal{A}\to\mathcal{B}_1\mathcal{E}_1$.

The free energy obtained above actually is ill-defined, for it diverges
at $z=0$, and currently it is not known how to regulate this.
Still, we can draw general conclusions from the free energy.
Consider the case with $B=0$.
Then we see that the electric field may not exceed the critical value
\begin{align}
  E^2 < \bigg( \frac{u_0}{R} \bigg)^3
  \;.
\end{align}
This is the curved space version of the upper limit on the electric
field in string theory.
Beyond this value, the free energy becomes imaginary, indicating
the instability of the system.
Thus, the electric field acts as a sort of ``destabilizer'' in the system.
On the contrary, we note that the $B$ field behaves exactly the opposite;
it is the ``stabilizer'' for the $\vec B\perp \vec E$ non-anomalous case.
In particular,
those effects exactly cancel when $B=E$ and the system is completely
the same as the one without the external electromagnetic fields.
For the anomalous $\vec B \parallel \vec E$ case, those external fields
do not interfere each other and they behave more or less independently.
Later, we will compare these behaviors to those of the NJL model.
\\

\noindent {\it Inter-Brane Distance and SEP}

\noindent The SS model has another parameter $L$ which is the asymptotic
separation of the D$8$ and $\barDEight$.
This is related to the parameter $\yKK:=\uKK/u_0$ and
we prefer $L$ to $\yKK$ as the fundamental parameter of the theory.

For the relation between those parameters,
we start with the equation
\begin{align}
  L = \int dx_4 = 2 \int_{u_0}^\infty {x_4}' du
  \;,
\end{align}
where the factor of $2$ accounts for D$8$ and $\barDEight$.
Using the solution (\ref{eq:confSoln}) and carrying out the
changes of variables $u\to y\to z$ as before, we obtain
\begin{align}\label{eq:ell}
  \ell :=& \hat L / \hat\beta_4
  \nonumber\\
   =& \frac{2}{3}a^{-1/2}\yKK^{1/2}\int_0^1dzf^{-1}
      \bigg( z^{-1}\frac{f\mathcal{A}}{f_0\mathcal{A}_0} -z^{5/3} \bigg)^{-1/2} 
  \;,
\end{align}
where we have defined $\hat L := L/R$ and $\ell$ has the range
$0<\ell\leq 1/2$.

Now the dynamical mass (SEP) discussed in Section~\ref{sec:chiPhTrcMass}
can be written as
\begin{align}
  m^*R = \frac{R^2}{2\pi\alpha'} (u_0/R) \int_{\yKK}^1 f(y)^{-1/2} dy
  \;.
\end{align}
We notice that the mass is totally independent of the temperature and
the chemical potential.
It turns out that the $y$ integral can be carried out to yield a closed
form as a function of $\yKK$.
Trading the parameter $u_0/R$ to $\hat\beta_4$ and $\yKK$
using the relation (\ref{eq:yRb4yKK}),
we get
\begin{align}\label{eq:cnfM}
  \hat m^* :=&  2\pi\bigg( \frac{l_s}{R} \bigg)^2 (m^*R)
  \nonumber\\
    =& a \hat\beta_4^{-2} \big\{ -\sqrt{\pi}\Gamma(2/3)/\Gamma(1/6)
          +\yKK^{-1}\,_2F_1(-1/3,1/2,2/3,\yKK^3) \big\}
  \;.
\end{align}
This dimensionless mass $\hat m^*$
together with Equation~(\ref{eq:ell}) is ready for
the numerical evaluation.

%%%%%%%%%%%%%%%%%%%%%%%%%%%%%%%%%%%%%%%%%%%%%%%%%%%%%%%%%%%%%%%%%%%%%%%%
\subsection{Deconfined Phase}\label{subsec:deconf}
We now proceed to examine the deconfined phase.
Strictly speaking, the discussion here would make sense only above
the confinement/deconfinement critical temperature.
However as mentioned earlier, we can always consider a very large $\beta_4$
or even the non-compact $x_4$ model, so we will freely discuss
the low temperature behavior of this phase without worrying about
the critical temperature.
Like in the confined case,
we will concentrate on the $\vec B \perp \vec E$
non-anomalous case with the Lagrangian
(\ref{eq:LNAnom}).
For the clarity of expressions, we set $B:=B_3$ and $E:=E_1$,
accordingly we have $\mathcal{A}=1+(R/u)^3(B^2-f^{-1}E^2)$ and
$\mathcal{B}:=1+(R/u)^3B^2$.
Also we remember that in this phase, $f_4=1$ and
$f_t=f=1-(u_T/u)^3$.
\\

\noindent {\it The U- and $\parallel$-Shape Solutions}

\noindent In the deconfined phase,
the $x_0$ direction is degenerate at $u=u_T$ but the $x_4$ direction
is not (see Figure~\ref{fig:threePhases}). 
Therefore, the D$8$ and $\barDEight$
probes can either join smoothly in the U shape or go straight down
to the horizon, forming the $\parallel$ shape.
Those configurations correspond to the broken and unbroken chiral
symmetry phases, respectively.

The derivation of the U-shape solution from the equations of motion 
(\ref{eq:EOMs1}) is nearly identical to the confined case and we have
\begin{align}\label{eq:deconfU}
  {x_4'}^2 = f^{-1}\bigg(\frac{R}{u}\bigg)^{3}
           \bigg( \frac{u^8f\mathcal{A}}
                       {u_0^8f_0\mathcal{A}_0}
           -1 \bigg)^{-1}
  \quad\text{and}\quad
  \mu' \equiv 0
  \;.
\end{align}
Again, the system in the U shape
is completely independent of the chemical potential $\mu$ and in particular,
the dynamical mass will be insensitive to the chemical potential.

For the $||$-shape configuration, we have
$x_4'(u) \equiv 0$.
Thus, the equations of motion (\ref{eq:EOMs1}) yield
the $\parallel$ solution
\begin{align}\label{eq:deconfPara}
  x_4' \equiv 0
  \quad\text{and}\quad
  \mu'^2 = \frac{\mathcal{A}c^2}
                {\mathcal{B}\{u^8(R/u)^3\mathcal{B}+c^2\}}
  \;,
\end{align}
where we set $c:=c_2$.
The anomalous $\vec B \parallel \vec E$ solution can be obtained by
the replacements $\mathcal{A}\to\mathcal{B}_1\mathcal{E}_1$
and $\mathcal{B}\to\mathcal{B}_1$.

We notice that for any finite $E$, there is a factor
$f^{-1}E^2$ inside $\mathcal{A}$ (or $\mathcal{E}$)
and this diverges at $u=u_T$.
To avoid this, we must conclude that we need $c=0$, {\it i.e.},
$\mu'=0$ for any $E \neq 0$.
Moreover, since $\mu$ must vanish at the horizon, we get $\mu\equiv 0$.
This strong conclusion has a clear physical interpretation.
In the deconfined and chirally symmetric phase, quarks and anti-quarks
are not bound together and the
chemical potential causes the unbalance in the charge density.
Thus, application of the external electric field immediately
triggers the instability.
Note that the key factor in this instability is $f^{-1}E^2$, which
is the direct consequence from the violation of the topological
constraint discussed in Section~\ref{sec:BNSandMu}.
\\

\noindent {\it The Difference Action}

\noindent Given the two solutions, we must determine the energetically
preferred configuration for a set of external parameters.
For this purpose, we compute the difference action with respect to
the solutions.
As before, we define the dimensionless free energy
\begin{align}
  \hat F := R^{-5} \int du \mathcal{L}
  \;,
\end{align}
where $\mathcal{L}$ is given in Equation~(\ref{eq:LNAnom}).
Evaluating the Lagrangian for the solutions (\ref{eq:deconfU})
and (\ref{eq:deconfPara}), we get
\begin{align}\label{eq:dcnfProbeA}
  \mathcal{L}_U =& u^4 \bigg(\frac{R}{u}\bigg)^{3/2}\mathcal{A}^{1/2}
                \bigg( 1- \frac{u_0^8f_0\mathcal{A}_0}{u^8f\mathcal{A}}
                \bigg)^{-1/2}
  \nonumber\\
  \mathcal{L}_\parallel =& u^4 \bigg(\frac{R}{u}\bigg)^{3/2}\mathcal{A}^{1/2}
                \bigg( 1+ \frac{c^2}{u^8(R/u)^3\mathcal{B}}
                \bigg)^{-1/2}
  \;.
\end{align}
Now the difference of the free energies is
\begin{align}
  \Delta\hat F = R^{-5}\int_{u_0}^{\infty}du\mathcal{L}_U
    - R^{-5}\int_{u_T}^{\infty}du\mathcal{L}_\parallel
  \;,
\end{align}
so that when this is negative (resp. positive), the U shape 
(resp. $\parallel$ shape) is preferred.
For the numerical handiness, we rewrite this as
\begin{align}
  \Delta\hat F =& 
      \int_{u_0}^{\infty}\frac{du}{R} \bigg(\frac{u}{R}\bigg)^{5/2}
      \mathcal{A}^{1/2}
      \bigg\{\bigg(1-\frac{u_0^8f_0\mathcal{A}_0}{u^8f\mathcal{A}} \bigg)^{-1/2}
      - \bigg( 1+ \frac{c^2}{u^8(R/u)^3\mathcal{B}} \bigg)^{-1/2}
      \bigg\}
  \nonumber\\
    &- \int_{u_T}^{u_0}\frac{du}{R} \bigg(\frac{u}{R}\bigg)^{5/2}
       \mathcal{A}^{1/2}
       \bigg( 1+ \frac{c^2}{u^8(R/u)^3\mathcal{B}} \bigg)^{-1/2}
  \;.
\end{align}
We carry out the sequence of the variable changes,
$u\to y\to z$, as was done in Section~\ref{subsec:conf}.
Assuming the relation
\begin{align}
  y_R = a^{-1}\hat\beta^2 y_T
  \quad\text{with}\quad
  a:=\bigg(\frac{4\pi}{3}\bigg)^2
  \quad\text{and}\quad
  \hat\beta:= \beta/R
  \;,
\end{align}
that follows from Equation~(\ref{eq:beta}), we have the expression
of $\Delta\hat F$ in terms of $y_T$ as
\begin{align}\label{eq:SSDeltaF}
  (3y_R^{7/2})\Delta \hat F =& \int_0^1 dz z^{-13/6}\mathcal{A}^{1/2}
      \bigg\{\bigg(1-z^{8/3}\frac{f_0\mathcal{A}_0}{f\mathcal{A}} \bigg)^{-1/2}
      - \bigg( 1+ z^{5/3}\frac{C}{y_R^3\mathcal{B}} \bigg)^{-1/2}
      \bigg\}
  \nonumber\\
    &- \int_1^{y_T^{-3}}dz z^{-13/6}\mathcal{A}^{1/2}
      \bigg( 1+ z^{5/3}\frac{C}{y_R^3\mathcal{B}} \bigg)^{-1/2}
  \;,
\end{align}
where $C:=c^2/u_0^8$ and we have
\begin{align}
  f = 1 - y_T^3z
  \;,\quad
  f_0 = 1 - y_T^3
  \;,
\end{align}
and setting $\bar B:=a^{-3/2}B$ and $\bar E:=a^{-3/2}E$,
\begin{align}
  \mathcal{A}=&1+z\hat\beta^{6}y_T^3(\bar B^2-f^{-1}\bar E^2)
  \;,\;
  \mathcal{B}=1+z\hat\beta^{6}y_T^3\bar B^2
  \;,\;
  \mathcal{E}=1-z\hat\beta^{6}y_T^3f^{-1}\bar E^2
  \;,
  \nonumber\\
  \mathcal{A}_0=&1+\hat\beta^{6}y_T^3(\bar B^2-f_0^{-1}\bar E^2)
  \;,\;
  \mathcal{B}_0=1+\hat\beta^{6}y_T^3\bar B^2
  \;,\;
  \mathcal{E}_0=1-\hat\beta^{6}y_T^3f_0^{-1}\bar E^2
  \;.
\end{align}
As noted before, because of the factor $f^{-1}E^2$ in $\mathcal{A}$
and $\mathcal{E}$, the integrand of the second line in
Equation~(\ref{eq:SSDeltaF}) is divergent at $z=y_T^{-3}$, however,
it turns out that the integral is finite.
When we consider the pure $E$ background, we find that the difference
free energy is complex and the imaginary part comes only from the $\parallel$
configuration.
But since we cannot define the action without taking the difference,
we treat the complex value as representing the nature of the whole system.
\\

\noindent {\it Chemical Potential, Inter-Brane Distance and SEP}

\noindent The $\parallel$ configuration without the $E$ field has
the non-trivial $\mu'$ as shown in Equation~(\ref{eq:deconfPara}).
We would like to set the asymptotic value of $\mu$ to $\bar \mu$,
which should correspond to the chemical potential of the dual field
theory.
We must also require that $\mu(u_T) =0$.
Therefore,
\begin{align}\label{eq:dcnfMu}
  \bar\mu =& \int_{u_T}^\infty du \bigg[
            \frac{\mathcal{A}c^2}
                 {\mathcal{B}\big\{u^8(R/u)^3\mathcal{B}+c^2\big\}} \bigg]^{1/2}
  \nonumber\\
          =& \frac{1}{3}a\hat\beta^{-2}y_T^{-1}\int_0^{y_T^{-3}} dzz^{-1/2} \bigg[
            \frac{\mathcal{A}C}
             {\mathcal{B}\big\{a^{-3}\hat\beta^{6}y_T^3\mathcal{B}
              +Cz^{5/3}\big\}} \bigg]^{1/2}
  \;.
\end{align}

The asymptotic brane distance for the $\parallel$-shape
configuration is a parameter on which no other parameters depend.
That of the U-shape configuration can be worked out in the similar
manner as in the confined case and we get
\begin{align}\label{eq:dcnfL}
  \hat L :=& L/R
  \nonumber\\
  =&\frac{2}{3}\hat\beta a^{-1/2}y_T^{1/2}\int_0^1dzf^{-1/2}
  \bigg( z^{-1}\frac{f\mathcal{A}}{f_0\mathcal{A}_0}
  -z^{5/3} \bigg)^{-1/2}
  \;.
\end{align}

The dynamical mass is simply proportional to the distance
$(u_0-u_T)$, because we have $\sqrt{g_{00}g_{UU}}=1$, and we get
\begin{align}\label{eq:Tneq0mass}
  \hat m^* :=& 2\pi\bigg(\frac{l_s}{R}\bigg)^2 (m^*R)
  \nonumber\\
  =& a \hat\beta^{-2} (y_T^{-1}-1)
  \;.
\end{align}

The expressions for $\Delta\hat F$, $\bar \mu$, $\hat L$ and $\hat m^*$
obtained in this subsection
are ready for the numerical evaluation and they suffice to determine
the dynamical mass and the phase structure of the deconfined phase.
\\

\noindent {\it Zero Temperature}

\noindent For the sake of the comparison with the NJL model, it is
interesting and important to consider the deconfined phase in the
low temperature regime, including at zero temperature.
At zero temperature, the parameters of the theory change slightly.
We must replace the ``blackening factors'' as
$f,f_0 \to 1$
and we have the range
$0 < u < \infty$
where $u_T$ is no longer the parameter of the theory.
We then have
\begin{align}\label{eq:T0Repl}
  \mathcal{A}:= 1+zy_R^{3}a^3(\bar B^2  - \bar E^2)
  \;,\;
  \mathcal{B}:= 1+zy_R^{3}a^3\bar B^2
  \;,\;
  \mathcal{E}:= 1-zy_R^{3}a^3\bar E^2
  \;,
\end{align}
where we have extracted the factor of $a$ by $B^2=a^3(a^{-3}B^2)=a^3\bar B^2$
and similarly for $E$, to be consistent with the finite temperature
parameters.

The difference free energy shown in Equation~(\ref{eq:SSDeltaF}) does not
change, except that the parameter $y_R$ is the parameter of the theory itself
and {\it not} expressed in terms of $y_T$, and the replacement
(\ref{eq:T0Repl}) is assumed.
Examination of the difference free energy shows that the U-shape configuration
is preferred to the $\parallel$ configuration at zero temperature.

The chemical potential and the inter-brane distance for $T=0$ take the
forms
\begin{align}\label{eq:zeroTMuL}
  \bar{\mu} =& \frac{1}{3}y_R^{-1}\int_0^\infty dzz^{-1/2} \bigg[
            \frac{\mathcal{A}C}
             {\mathcal{B}\big\{y_R^3\mathcal{B}+Cz^{5/3}\big\}} \bigg]^{1/2}
  \nonumber\\
  \hat L =& \frac{2}{3}y_R^{1/2}\int_0^1 dz z^{-5/6}
       \bigg( z^{-8/3}\frac{\mathcal{A}}{\mathcal{A}_0}-1 \bigg)^{-1/2}
  \;.
\end{align}
It is also easy to show that the mass becomes
\begin{align}\label{eq:zeroTmass}
  \hat m^* = 2\pi \bigg(\frac{l_s}{R}\bigg)^2 (m^*R) = y_R^{-1}
  \;.
\end{align}

%%%%%%%%%%%%%%%%%%%%%%%%%%%%%%%%%%%%%%%%%%%%%%%%%%%%%%%%%%%%%%%%%%%%%%%%
\subsection{Numerical Evaluation}\label{sec:SSnum}
We now present the numerical results, separately for the confined and
deconfined phases.
In this section, we assume all the quantities are measured in the units
of $R$ and we are going to omit the hats ($\hat X$) on the parameters.

%%%%%%%%%%%%%%%%%%%%%%%%%%%%%%%%%%%%%%%%%%%%%%%%%%%%%%%%%%%%%%%%%%%%%%%%
\subsubsection{Confined Phase}\label{sec:SScnfNum}
We have seen in Section~\ref{subsec:conf} that the system is independent
of the temperature and the chemical potential, except that this phase
is defined below the critical temperature of the confinement/deconfinement
phase transition.
Therefore, the only parameters are the external electromagnetic
field and the inter-brane distance.
As for the numerical evaluation, the relevant equations are
$\ell := L/\beta_4$ of (\ref{eq:ell}) and $m^*$ of (\ref{eq:cnfM}).

\begin{figure}[ht]
{
\centerline{\scalebox{1.0}{\includegraphics{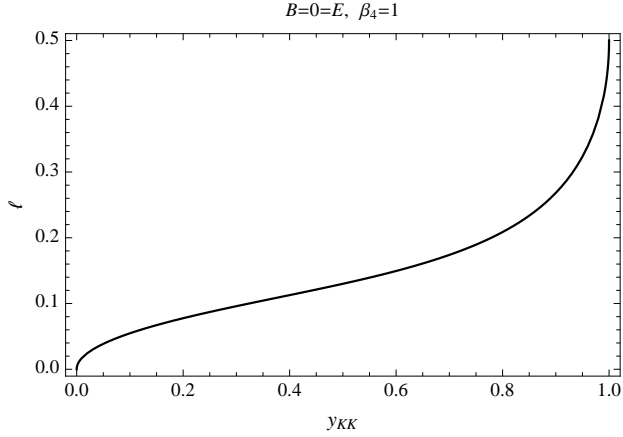}}}
\caption{\footnotesize
  The plot of $\ell$ with respect to the variable $\yKK$ without
  electromagnetic background and $\beta_4=1$.
  This shows that the location of D$8$ and $\barDEight$ can be anywhere
  in the $x_4$ circle.
}\label{fig:SScnfF0yKKEll}
}
\end{figure}

We first examine the properties related to the parameter $\ell$
without the background electromagnetic field.
From Equation~(\ref{eq:ell}), we can plot $\ell$ with respect
to $\yKK$ and it is shown in Figure~\ref{fig:SScnfF0yKKEll}.
Recall that the ranges of $\ell$ and $\yKK$ are $(0,1/2)$ and $(0,1)$,
respectively, and the figure shows that the parameter $\ell$ takes
all the possible values.
We remark that the lower bound $2$ of $\ell^{-1}$ is the geometrical
constraint and not the bound determined by the dynamics (like the NJL
coupling $g_c$).
We will further discuss the comparison between $\ell^{-1}$ and $g$
later in Section~\ref{sec:comparison}.

\begin{figure}[ht]
{
\centerline{\scalebox{1.0}{\includegraphics{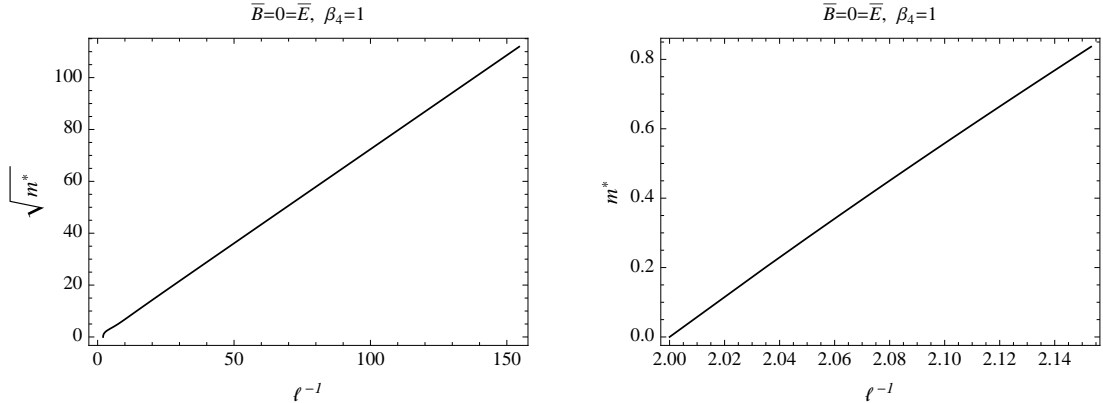}}}
\caption{\footnotesize
  The plots of mass $m^*$ against $\ell^{-1}$ ($2\leq\ell^{-1}<\infty$).
  The left panel implies the relation $m^*\sim(\ell^{-1})^2$ for
  large values of $\ell^{-1}$ and the right panel shows that the relation
  $m^*\sim\ell^{-1}$ is fairly well for small values of $\ell^{-1}$.
}\label{fig:SScnfF0EllM}
}
\end{figure}

Figure~\ref{fig:SScnfF0EllM} shows the response of SEP
with respect to $\ell^{-1}$.
The left panel shows the relation $m^*\sim(\ell^{-1})^2$ for the
range of large $\ell^{-1}$ away from the lower bound and the right
panel shows that near the lower bound $2$, we have the linear
relation $m^*\sim\ell^{-1}$.
\\

\noindent {\it Pure $B$ Background}

\begin{figure}[ht]
{
\centerline{\scalebox{1.0}{\includegraphics{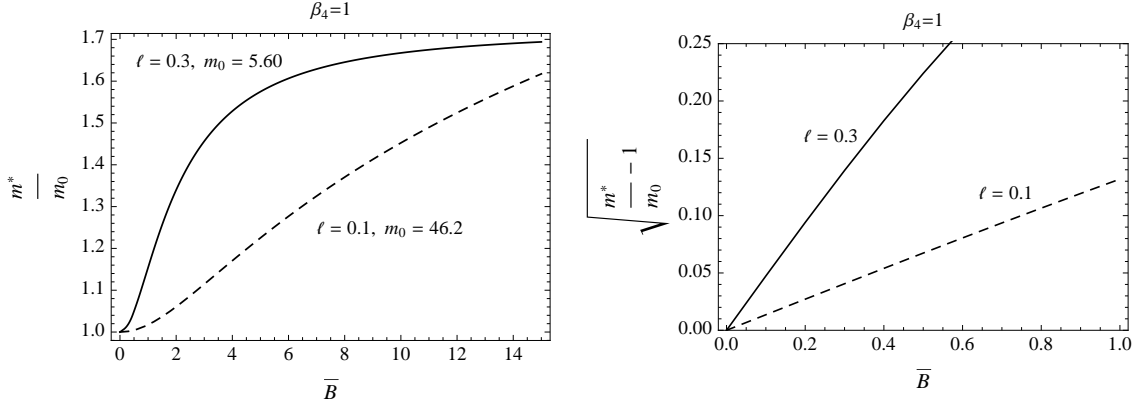}}}
\caption{\footnotesize
  The SEP against the $B$ field.
  We have defined $m_0:=m^*(\bar B=0)$ and notice that
  the absolute value $m^*$ is much larger for $\ell=0.1$.
  The right panel shows the relation $m^*\sim \bar B^2$ for small
  values of $\bar B$
}\label{fig:SScnfBM}
}
\end{figure}

\noindent When the $B$ field is turned on, the $\yKK$-$\ell$ graph
appears essentially the same as Figure~\ref{fig:SScnfF0yKKEll} and
the parameter $\ell$ takes all the possible value $0<\ell\leq1/2$.
The response of the mass with respect to $B$ is shown in
Figure~\ref{fig:SScnfBM}.
The graph shows the general tendency that $B$ acts as the stabilizer
of the boundstate.
The right diagram reveals the relation $m^*\sim\bar B^2$ in the region
of small $\bar B$.
\\

\noindent {\it Pure $E$ Background}

\noindent When we turn on the $E$ field, the $\yKK$-$\ell$ relation
becomes more complicated.
\begin{figure}[ht]
{
\centerline{\scalebox{0.8}{\includegraphics{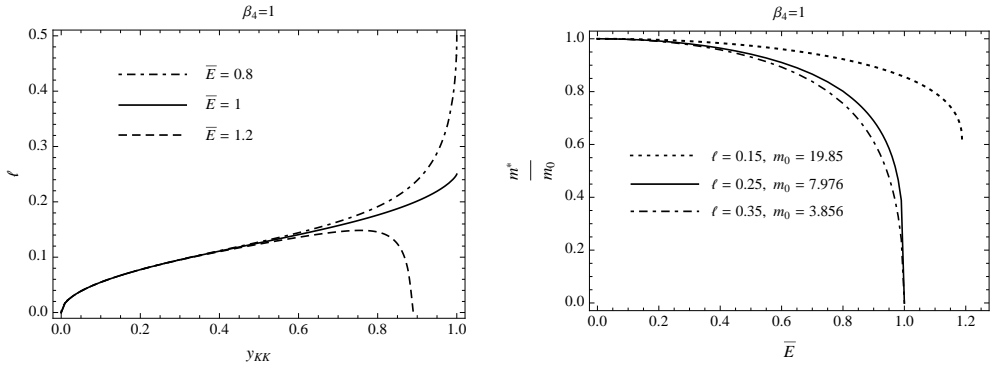}}}
\caption{\footnotesize
  The left panel shows that
  for $\bar E<1$, $\ell$ takes all allowed values $0<\ell<1/2$, and
  for $\bar E>1$, $\ell$ has upper limit, implying that the D$8$
  and $\barDEight$ may not be separated too far.
  Notice in this case that there are two solutions of $\yKK$ for a given $\ell$.
  The critical value $\bar E=1$ shown in this figure depends on the
  value of $\beta_4$, and in general, the critical value is given
  at $(\beta_4^3\bar E)=1$.
  At this value, the curve approaches the point $(\yKK,\ell)=(1,1/4)$.
  The right panel shows the response of the dynamical masses
  with respect to $\bar E$.
  The parameter $m_0$ is the value of $m^*$ at $\bar E=0$.
  For $1/4\leq\ell<1/2$, the dynamical mass goes smoothly to zero
  at $\bar E=1$ and beyond which the system becomes unstable.
  For $\ell<1/4$, the system becomes unstable before the mass
  becomes zero.
}\label{fig:SScnfE}
}
\end{figure}
The left panel of Figure~\ref{fig:SScnfE} shows that there is a critical
value $(\beta_4^3\bar E) = 1$, above which $\ell$ takes all possible values
and below, there is an upper limit which is lower than $1/2$.
Therefore, for sufficiently large value of $(\beta_4^3\bar E)$,
the D$8$ probe system becomes ill-defined.
Since the supergravity background is in the confined phase, the physical
implication of this instability is unclear.

Notice that when the $E$ field is below the critical value, there can
be two solutions of $\yKK$ for a given $\ell$.
One can show along the analysis of Reference~\cite{Sakai:2004cn}
that the lower $\yKK$ solution is the stable configuration and the other
is unstable.
Hence, we always pick the lower $\yKK$ solutions to obtain
the dynamical masses.
The right panel of Figure~\ref{fig:SScnfE} shows how the dynamical mass
respond to the $E$ field.
As expected, the graph shows the general tendency of $E$ as the destabilizer.
As discussed in the previous paragraph, there is a critical value of $\bar E$
and the curves in the diagram terminate there.

%%%%%%%%%%%%%%%%%%%%%%%%%%%%%%%%%%%%%%%%%%%%%%%%%%%%%%%%%%%%%%%%%%%%%%%%
\subsubsection{Deconfined Phase}\label{sec:deconfNum}
As mentioned in Section~\ref{subsec:deconf}, we will be cavalier about
the confinement/deconfinement phase transition temperature and discuss
this phase in all temperature range, including zero temperature.

\begin{figure}[ht]
{
\centerline{\scalebox{0.9}{\includegraphics{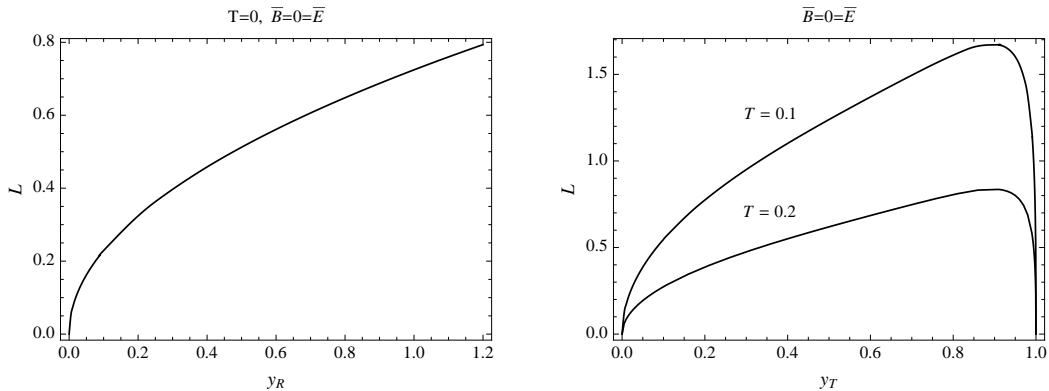}}}
\caption{\footnotesize
  The plots of the inter-brane distance $L$ in the U-shape configuration.
  The left panel is for $T=0$ and this case is assumed to be the non-compact
  model.
  There is no upper limit on $L$ and it increases as $\sqrt{y_R}$.
  The right panel is plotted for finite temperatures.
  Any finite temperature induces the upper limit on $L$, {\it i.e.},
  D$8$-$\barDEight$ inter-distance may not be too large to form the U shape.
}\label{fig:SSdcnfF0yTL}
}
\end{figure}

Let us first examine the range of parameter $L$ without
the electromagnetic background.
Unlike the confined phase, $L$ in the deconfined phase depends on
the temperature, as shown in Equation~(\ref{eq:dcnfL})
and the zero temperature case is given in Equation~(\ref{eq:zeroTMuL}).
Figure~\ref{fig:SSdcnfF0yTL} shows the plots of $L$ for different
temperatures.
\begin{figure}[ht]
{
\centerline{\scalebox{1.0}{\includegraphics{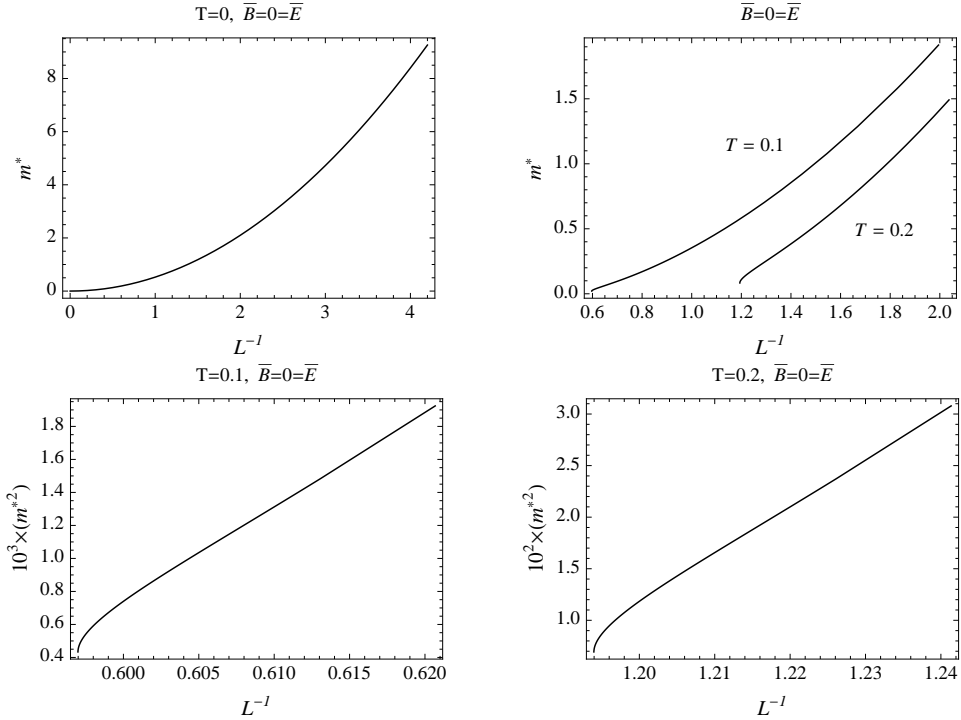}}}
\caption{\footnotesize
  The plots of $m^*$ against $L^{-1}$.
  For $T=0$ (upper left corner),
  the relation $m^*\sim(L^{-1})^2$ holds for all values of $L^{-1}$.
  At finite temperature, the relation is more complicated as shown in
  the upper right corner.
  The bottom two indicate the relation $m^*\sim(L^{-1})^{1/2}$
  in the region of small $L^{-1}$, but slightly away from the
  critical values.
}\label{fig:SSdcnfF0LM}
}
\end{figure}
The left panel of the zero temperature plot shows that there is no upper
limit on $L$ and it increases as $\sqrt{y_R}$.
This is a manifestation of the fact that at zero temperature, the U-shape
configuration is always preferred to the parallel one, as mentioned
in Section~\ref{subsec:deconf}.
The right panel shows that finite temperature introduces the upper limit
on $L$.
For a fixed value of $L$, a sufficiently high temperature yields no
solution for $y_T$ ({\it e.g.}, at $L=1.0$ and $T=0.2$ of the diagram),
implying that the system is actually in the chirally symmetric $\parallel$-shape
configuration.
Here again, the temperature plays the role of the destabilizer of the quark
boundstate.
Meanwhile, a sufficiently low temperature yields two solutions for
$y_T$ ({\it e.g.}, at $L=1.0$ and $T=0.1$ of the diagram).
As discussed in Section~\ref{sec:SScnfNum}, the smaller solution is
the stable configuration and we will always pick this one in what follows.

Before introducing the background electromagnetic field, we would like
to examine the response of SEP with respect to $L^{-1}$.
One can obtain the mass as a function of $L^{-1}$ from 
Equations~(\ref{eq:zeroTMuL},\ref{eq:zeroTmass}) for $T=0$, and for
finite temperature, from Equations~(\ref{eq:dcnfL},\ref{eq:Tneq0mass}).
The results are shown in Figure~\ref{fig:SSdcnfF0LM}.
The upper left panel of the figure shows the $T=0$ case and one
can show analytically for this case that $m^*$ increases quadratically
with respect to $L^{-1}$.
The upper right panel shows the finite temperature cases and these
are more complicated than the zero temperature.
We have observed in Figure~\ref{fig:SSdcnfF0yTL} that there is an upper
bound on $L$ for a given temperature.
This translates to the lower bound on $L^{-1}$ and the plots of the upper
right panel terminates at those critical values of $L^{-1}$.
The diagrams of the bottom row are plotted for the region near the critical
values of $L^{-1}$ and show the fairly good relation $m^*\sim\sqrt{L^{-1}}$
slightly away from the critical values.
\\

\noindent {\it Pure $B$ Background}

\noindent We now consider the effects of the constant background $B$ field.
At $T=0$ without the $B$ field,
we have seen that there is no upper limit on $L$, and
finite $B$-field does not change this conclusion.
Neither does it change the general observation that finite temperature
introduces an upper limit on $L$.
In short, $L$ behaves similar to Figure~\ref{fig:SSdcnfF0yTL},
except that the larger $B$ allows the larger separation
$L$ for a given $y_{R,T}$.
This, of course, implies that the $B$ field acts as the stabilizer
of the quark boundstate.

\begin{figure}[p]
{
\centerline{\scalebox{0.8}{\includegraphics{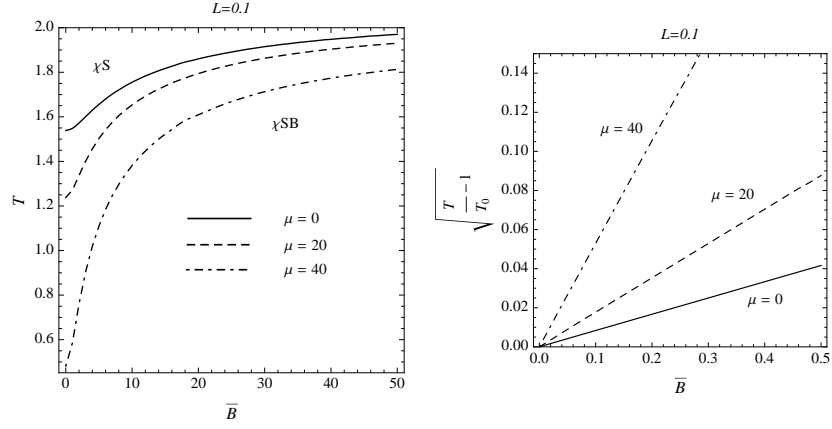}}}
\caption{\footnotesize
  The $B$-$T$ phase diagram  at $L=0.1$.
  The right panel shows the region of small $\bar B$ and the parameter $T_0$
  is the critical temperature at $\bar B=0$.
  It shows the relation $T\sim\bar B^2$ in the region of small $\bar B$.
}\label{fig:SSBT}
}
\end{figure}
\begin{figure}[p]
{
\centerline{\scalebox{0.85}{\includegraphics{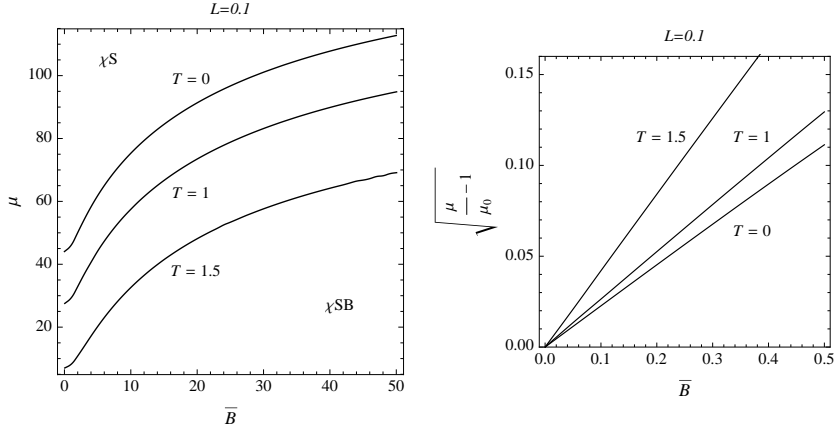}}}
\caption{\footnotesize
  The $B$-$\mu$ phase diagram  at $L=0.1$.
  The right panel shows the region of small $\bar B$ and 
  the parameter $\mu_0$
  is the critical value of the chemical potential at $\bar B=0$.
  It shows the relation $\mu\sim\bar B^2$ in the region of small $\bar B$.
}\label{fig:SSBMu}
}
\end{figure}
\begin{figure}[p]
{
\centerline{\scalebox{0.9}{\includegraphics{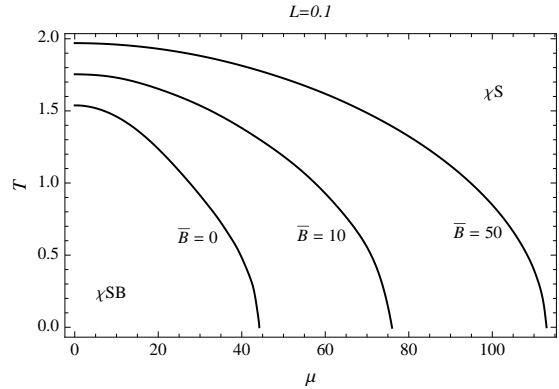}}}
\caption{\footnotesize
  The $\mu$-$T$ phase diagram at $L=0.1$.
  The case with $\bar B=0$ reproduces the result of 
  Reference~\cite{Horigome:2006xu} (the reference plots
  $TL$ against $\mu L^2$).
}\label{fig:SSMuT}
}
\end{figure}
In the deconfined phase, the system exhibits the chiral phase transition
from the U-shape to the $\parallel$-shape configuration, which cannot
take place in the confined phase.
Unlike the way we presented the NJL results, we first show the phase diagram
before the response of the dynamical mass with respect to the external
parameters.
This is because we must consult with the chiral phase diagrams to determine
exactly where the mass drops to zero at the phase transition point.

The phase diagrams can be all worked out from 
Equations~(\ref{eq:SSDeltaF},\ref{eq:dcnfMu},\ref{eq:dcnfL}) and
Figures~\ref{fig:SSBT}, \ref{fig:SSBMu} and \ref{fig:SSMuT}
show the $B$-$T$, $B$-$\mu$ and $\mu$-$T$ phase diagrams, respectively.
The transitions are all first order as the configuration ``jumps'' from
U- to $\parallel$-shape.
We observe the general trend that $B$ acts as the stabilizer and
$T$, $\mu$ as the destabilizer.
This fact appears as the general similarity in the $B$-$T$ and
$B$-$\mu$ graphs.

\begin{figure}[h]
{
\centerline{\scalebox{0.8}{\includegraphics{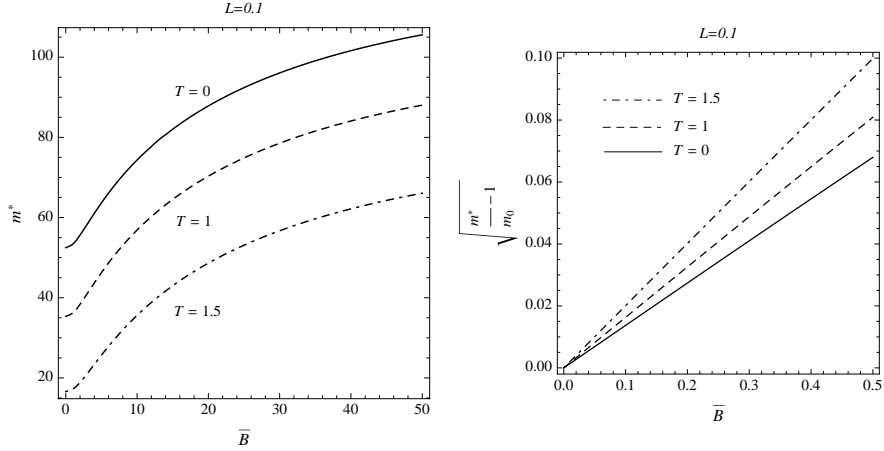}}}
\caption{\footnotesize
  The SEP plotted against $B$-field.
  The right panel exhibits the small $\bar B$ region and 
  the parameter $m_0$ is defined as the mass at $\bar B=0$.
  The diagram shows the relation $m^*\sim\bar B^2$ in the region of
  small $\bar B$.
  Typically at high temperature,
  the lower $B$ region of these curves may sharply drop to zero as
  the chiral phase transition sets in and
  the point of the drop depends on the temperature and the chemical
  potential.
}\label{fig:SSBM}
}
\end{figure}

Let us now investigate the response of the dynamical mass with respect
to the external parameters.
Figure~\ref{fig:SSBM} shows the $B$-$m^*$ graph.
\begin{figure}[h]
{
\centerline{\scalebox{1.0}{\includegraphics{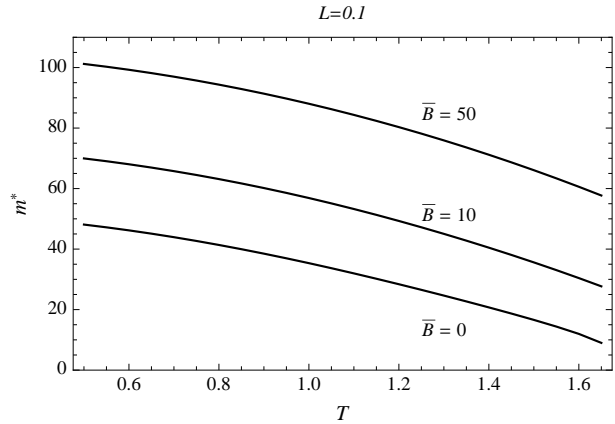}}}
\caption{\footnotesize
  The plot of mass against temperature.
  One must be aware that the low temperature region is subject to
  the confinement/deconfinement phase transition, unless $\beta_4$
  is large enough, and the high temperature region typically drops
  to the zero mass sharply due to the chiral phase transition.
  The point of the chiral phase transition depends on the chemical
  potential as well.
}\label{fig:SSTM}
}
\end{figure}
As expected, the mass increases as $\bar B$ is raised and higher
temperature results in the smaller mass.
It appears that the dependence on the chemical potential is missing,
but one must be aware that the chiral phase transition can occur
when $\bar B$ is small.
For instance at $T=1$ and $\mu=20$, the phase diagrams tell us that
there is no chiral phase transition for all values of $\bar B$,
so the dashed curve of the diagram is not modified.
However at $T=1$ and $\mu=40$, the chiral phase transition sets in
at $\bar B\approx 4$ and the dashed curve for $T=1$ in the diagram
must be truncated below this value and $m^*$ sharply drops down to
zero at that point.
Therefore, the $B$-$m^*$ diagram depends on $\mu$ implicitly
and $\mu$ determines where the mass drops to zero.

Figure~\ref{fig:SSTM} shows the $T$-$m^*$ diagram.
The lower temperature region of the diagram is subject to the truncation
by the confinement/deconfinement phase transition at $\beta=\beta_4$.
Thus, for example, if we have $\beta_4=1$, then the diagram must end
at $T=1$.
The higher temperature region can end with the chiral phase transition,
by dropping sharply to the zero mass.
The chiral phase diagrams tell us, for example, that the $\bar B=10$
curve must be truncated at $T=1$ for $\mu\approx 57.6$ and at
$T=1.5$ for $\mu\approx 32.5$.
So again, the truncation point depends on the value of the chemical potential.
\\

\noindent {\it Pure $E$ Background}

\noindent We now discuss the constant $E$ background.
As mentioned in Section~\ref{subsec:deconf}, the introduction of the
chemical potential is not allowed for this case.
Moreover, the difference free energy $\Delta F$ in Equation~(\ref{eq:SSDeltaF})
is complex and we follow what we have done for the NJL model:
we treat the real and imaginary parts of the free energy separately.
We extract the information about the phase diagram
from the real part, and we make crude estimate on whether such information
can be reliable, by checking if the imaginary part is subdominant.
We are mainly interested in the zero temperature limit, since in the NJL
counterpart, we had most information at zero temperature.

\begin{figure}[ht]
{
\centerline{\scalebox{1.0}{\includegraphics{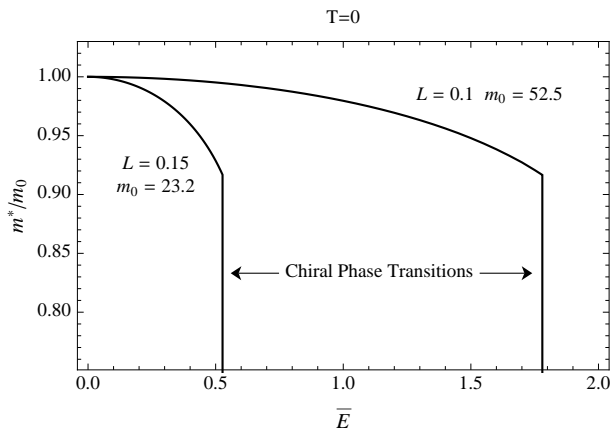}}}
\caption{\footnotesize
  The SEP as a function of the $E$ field at $T=0$.
  The parameter $m_0$ is the $m^*$ at $\bar E=0$.
  The curves at higher value of $\bar E$ is terminated at the chiral phase
  transition points.
}\label{fig:SSEMT0}
}
\end{figure}

\begin{figure}[ht]
{
\centerline{\scalebox{0.8}{\includegraphics{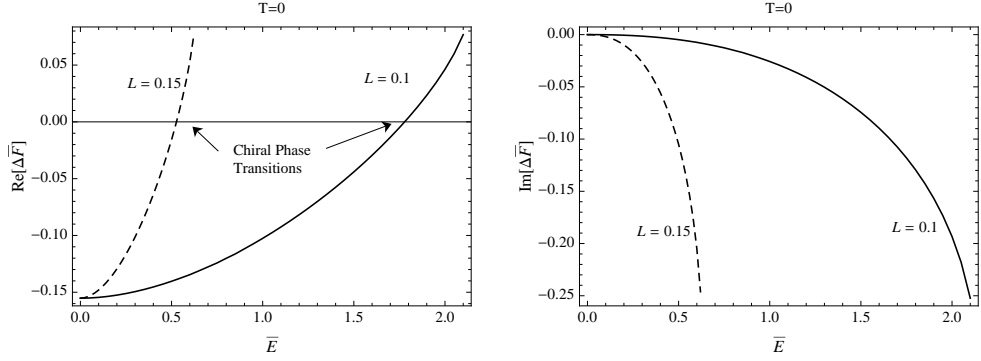}}}
\caption{\footnotesize
  The free energy plotted separately for the real and the imaginary parts.
  The quantity $\bar F$ is the right-hand side of Equation~(\ref{eq:SSDeltaF})
  where the factor of $(3y_R^{7/2})$ is omitted from the free energy.
}\label{fig:SST0FE}
}
\end{figure}

From Equations~(\ref{eq:zeroTMuL},\ref{eq:zeroTmass}) for $T=0$, we can obtain
the dynamical mass as a function of $E$, and the result is shown
in Figure~\ref{fig:SSEMT0}.
We see that $E$ generally is a destabilizer of the boundstate, as expected,
and also the curves always terminate at the first order chiral phase
transitions.
Figure~\ref{fig:SST0FE} shows the corresponding plots of the free energy
in real and imaginary parts.
This figure shows that the imaginary part is always comparable to the real
part, especially near the phase transition points.
Hence there is a doubt on the validity of the masses and the phase transitions
shown in Figure~\ref{fig:SSEMT0}.
However, notice that ${m^*}^2$ is considerably larger than $E$.
Recall that in the NJL counterpart, we also estimated the stability of the
system by considering Schwinger's pair creation rate formula.
The rate is exponentially suppressed by the factor ${m^*}^2/E$
and this indicates that the system of the SS model may have very slow
pair creation rate.
Therefore, it is reasonable to assume that the instability takes place
slowly and the information about the mass and the phase transition can
be meaningful.

\begin{figure}[ht]
{
\centerline{\scalebox{0.9}{\includegraphics{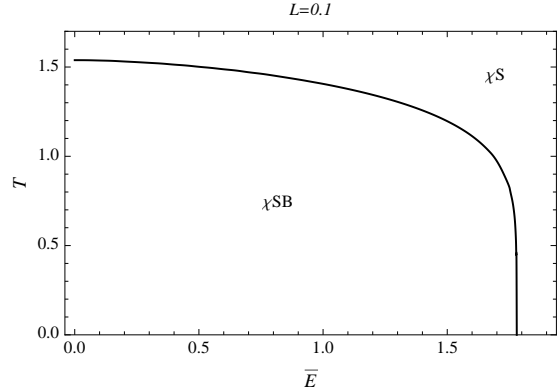}}}
\caption{\footnotesize
  The $E$-$T$ chiral phase diagram obtained from the real part
  of the free energy.
}\label{fig:SSET}
}
\end{figure}

The system at a finite temperature behaves as expected and it 
acts as a destabilizer.
The masses at a certain value of $\bar E$ is smaller at finite temperature
than the zero temperature case and the first order chiral phase transition
sets in at a smaller value of $\bar E$.
The $E$-$T$ phase diagram extracted from the real part of the difference
free energy is plotted in Figure~\ref{fig:SSET}.
%
%

%%%%%%%%%%%%%%%%%%%%%%%%%%%%%%%%%%%%%%%%%%%%%%%%%%%%%%%%%%%%%%%%%%%%%
\section{Comparison}\label{sec:comparison}
%%%%%%%%%%%%%%%%%%%%%%%%%%%%%%%%%%%%%%%%%%%%%%%%%%%%%%%%%%%%%%%%%%%%%
We now have the data for the comparison between the NJL and SS models.
In this section, we summarize the results by putting the NJL and SS
data side by side.

%%%%%%%%%%%%%%%%%%%%%%%%%%%%%%%%%%%%%%%%%%%%%%%%%%%%%%%%%%%%%%%%%%%%%
\subsection{General Observation}\label{sec:compGen}
We first make general observations which are not specialized to
purely $B$ or $E$ backgrounds.

Let us compare the NJL effective potential (\ref{eq:zeroTF})
and the probe action of the SS model in the confined phase 
with the Lagrangian (\ref{eq:cnfProbeA}).
Both expressions are specific for the
non-anomalous $\vec B\perp\vec E$ background configuration.
The former is the expression for $T=0$ and the latter is independent
of the temperature, as long as the system is in the confined phase.
We have already remarked under Equation~(\ref{eq:zeroTF}) that
for the cases with $|\vec B|>|\vec E|$ and
$|\vec B|<|\vec E|$, the NJL system behaves as if there is only
$B$- and $E$-field, respectively.
The case with $|\vec B|=|\vec E|$ is special and the system is equivalent
to the one without the electromagnetic background.
Also the pure $E$-field case can be obtained from
the pure $B$-field case just by sending $B\to iE$, and {\it vice versa}.
As we can see clearly in Equation~(\ref{eq:cnfProbeA}) that the same
is true for the SS model probe action.
A short inspection reveals that the zero temperature limit of the 
SS deconfined
phase discussed near the end of Section~\ref{subsec:deconf} also shows
the same behavior. (One must recall that $\mu=0$ for $E\neq 0$.)

The finite temperature NJL effective potential (\ref{eq:postResum})
shows that the lack of Lorenz invariance destroys the features mentioned
above.
Also the probe Lagrangians
of the SS model in the deconfined phase (\ref{eq:dcnfProbeA})
lack the properties
of the previous paragraph, because of the general appearance of the factor
$f^{-1}E^2$ in $\mathcal{A}$.

Therefore, we see that even though the actions of the theories are very
different, the effects of Lorentz invariance (at $T=0$) and non-invariance
(at $T\neq 0$) appear the same for both theories in a very general sense.
However, we remark that for the anomalous case $\vec B\parallel\vec E$,
apparent similarity is not observed.

We now look into the relation between the NJL coupling $g$ and the
inter-brane distance $L$ of the SS model.
The possible ranges of those parameters are shown in
Figure~\ref{fig:NJLRHSF0} for the NJL coupling (plotted for $\pi^{3/2}/g^2$),
Figure~\ref{fig:SScnfF0yKKEll} for $\ell:=L/\beta_4$ of 
the SS model in the confined phase
and Figure~\ref{fig:SSdcnfF0yTL} for $\hat L:=L/R$ of the SS model
in the deconfined phase.
In the absence of $B$-field, the NJL model always has the lower limit
on the coupling and below this critical coupling, the system is in
the chirally symmetric phase.
For example, the critical coupling for the zero temperature 
and zero chemical potential is $2\pi$.
The SS model in the confined phase also has a lower limit on $\ell^{-1}$,
namely, $\ell^{-1}=2$.
However, this lower limit is different in nature: it is the geometrical
constraint rather than dynamical, and furthermore, at the critical
point the chiral symmetry is still broken (the configuration is
in the U shape) and the region $\ell^{-1}<2$ is not defined, of course.
The SS model in the deconfined phase at zero temperature does not have
a lower limit on $\hat L^{-1}$ and the chiral symmetry is always broken.
However at finite temperature, there is a lower limit on $\hat L^{-1}$
and below this critical value, the system is in the symmetric phase.
So it seems that the NJL coupling $g$ and
$\ell^{-1}$ of the SS model in the confined phase behave rather differently,
and $g$ and $\hat L^{-1}$ of the SS deconfined phase have general
similarity at finite temperature.

To investigate further into the relation among those parameters,
we have shown the response of the dynamical mass with respect
to the parameters.
Figure~\ref{fig:NJLm_gF0} is the plot for the NJL model and
Figure~\ref{fig:SScnfF0EllM} is the SS model in the confined phase
and Figure~\ref{fig:SSdcnfF0LM} is the SS model in the deconfined phase.
In general, the mass respond very differently, except that the higher
$g$ or $L^{-1}$ gives the higher mass.
However, examination into the region near the critical points of these
parameters shows that there is a good similarity between $g$ and
$\hat L^{-1}$ of the SS model in the deconfined phase at finite temperature.
In those cases, the mass grows as square-root of the parameters.

The conclusion out of this comparison is that the parameter $g$ of the
NJL model and $\hat L^{-1}$ of the SS model in deconfined phase at finite
temperature correspond to each other fairly well, especially
near the critical values, and in all cases, the increase in the parameters
result in the higher masses.
However we observed that they are not in one-to-one correspondence
and especially at zero temperature, they show the qualitative differences.

In what follows, we focus on the comparison between the NJL model and
the SS model {\it in the deconfined phase}.
This is because we have already seen that the correspondence is rather
poor for the SS confined phase.
Moreover, the confined phase does not have the dependence on the temperature
nor on the chemical potential, and it does not even have
the chiral phase transition.
These facts make the confined phase rather meaningless in the comparison
with the NJL model.

%%%%%%%%%%%%%%%%%%%%%%%%%%%%%%%%%%%%%%%%%%%%%%%%%%%%%%%%%%%%%%%%%%%%%
\subsection{Pure $B$ Background}\label{sec:compB}
When we turn on the $B$ field at $T=0$, 
the lower limit of the NJL coupling becomes
zero and any finite temperature brings this to a finite value.
(See Figure~\ref{fig:NJLBRHS}).
This makes the correspondence of the parameter $g$ of NJL and $\hat L^{-1}$
of SS even better.

Let us proceed to consider the dynamical mass
in the presence of the background $B$ field.
Compare the $B$-$m^*$ diagram of the NJL model at $\mu=0$
in Figure~\ref{fig:NJLBM3}
and that of the deconfined SS model in Figure~\ref{fig:SSBM}.
We see the striking similarity in how the masses respond to the $B$ field
between the models.
In the small $B$ region, they both show the relation $m^*\sim B^2$
and this fact is more than just the general tendency of $B$ as the 
stabilizer of the boundstate.
The figures also show that the ways the masses in the models respond to 
the temperature are also similar.

We have, however, only looked at the $\mu=0$ case so far.
The finite chemical potential reveals the glaring differences between
the models.
The NJL model exhibits a very complicated behavior at finite
chemical potential due to the de Haas-van Alphen oscillation
and the destabilizing effect of $B$ at large $\mu$, as shown
in Figure~\ref{fig:NJLBM1} (see also Figure~\ref{fig:NJLBM2}).
In contrast, the Figure~\ref{fig:SSBM} of the SS model does not change
the shape at finite chemical potential, except that the lower $B$ part of the
curves can be truncated to zero mass as the chiral phase transition
sets in.
The dynamical mass of the SS model, SEP, is independent of the chemical
potential (except at the chiral phase transition point), and this
is why the $\mu$-$m^*$ diagram of Figure~\ref{fig:NJLMMu} 
for the NJL model does not
have the counterpart in the SS model.

We can also see the similarities and dissimilarities due to the chemical
potential in the phase diagrams.
For example, we can look at the $B$-$T$ phase diagrams of
Figure~\ref{fig:NJLBT} for the NJL model and
Figure~\ref{fig:SSBT} for the SS model.
When the value of the chemical potential is small,
the NJL diagram appears very similar to the SS diagram.
However, the ``dent'' in Figure~\ref{fig:NJLBT} of NJL is due to the
destabilization effect of $B$ at large $\mu$
and this is not observed in the NJL diagram in
Figure~\ref{fig:SSBT}.
Accordingly, the ``cross'' of the $T$-$\mu$ phase diagram in
Figure~\ref{fig:NJLMuT} does not have the counterpart in
the SS diagram in Figure~\ref{fig:SSMuT}.

Therefore, the models in the background $B$ field behave very similar
to each other when the chemical potential is zero or small,
but as soon as the chemical potential becomes large enough to
trigger the oscillation and the $B$-field destabilization,
the models behave differently.

We also point out that the order of the chiral phase transition in
the SS model is, by definition, all first order, because it is defined
to be the point where the configuration jumps from U to $\parallel$
or the other way around.
Meanwhile in the NJL model, the transition is in general second order,
as long as the chemical potential is low enough.
When the chemical potential becomes large,
there is a mixture of the first and second order phase transitions
in the phase diagrams.

%%%%%%%%%%%%%%%%%%%%%%%%%%%%%%%%%%%%%%%%%%%%%%%%%%%%%%%%%%%%%%%%%%%%%
\subsection{Pure $E$ Background}\label{sec:compE}
Turning on the background $E$ field introduces the lower bound for
both $g$ and $\hat L^{-1}$ even at $T=0$,
and the effect is similar to the finite
temperature discussed in Section~\ref{sec:compGen}.
Thus in this case, the qualitative agreement between $g$ and $\hat L^{-1}$
is good at all temperature.

Just like the temperature, the background electric field acts as the
destabilizer in the both theories and this can be seen in
$E$-$m^*$ graph of Figure~\ref{fig:NJLEMT0} for the NJL model and
Figure~\ref{fig:SSEMT0} for the SS model.
The NJL model shows both first and second order phase transitions
depending on the coupling, and the SS model, by definition, shows only
the first order transition.
We noted that the ${m^*}^2$ to $E$ ratio is very large for the SS model
and this implies the slow rate of the pair creation.
Therefore, the phase diagram of the SS model is more reliable than that
of the NJL model.

In both models, the free energy is complex for any finite $E$, implying
that the models are unstable in the $E$ field background.
We have seen that the real and imaginary parts are comparable to
each other in both models.
They are shown in Figure~\ref{fig:NJLT0VeffE} for NJL and
Figure~\ref{fig:SST0FE} for SS.

%%%%%%%%%%%%%%%%%%%%%%%%%%%%%%%%%%%%%%%%%%%%%%%%%%%%%%%%%%%%%%%%%%%%%
\section{Conclusion}\label{sec:conclusion}
%%%%%%%%%%%%%%%%%%%%%%%%%%%%%%%%%%%%%%%%%%%%%%%%%%%%%%%%%%%%%%%%%%%%%
We have seen good agreements between the NJL model and 
the {\it deconfined} phase of the SS model.
Because the NJL model is nonrenormalizable, the cutoff must be introduced
by hand and this makes the energy scale of the model unclear.
One popularly advocated interpretation is that the model describes
the very low energy regime of QCD and the cutoff is around
$\Lambda_{\text{QCD}}$.
Another interpretation is that since the model lacks the confinement,
it describes QCD in the deconfined phase, but not too far from
the phase transition point.
As far as the comparison with the SS model is concerned, our study
strongly implies the latter interpretation.

We have also seen the general discrepancy of the phase transition orders,
and the
most prominent differences between the models are
the de Haas-van Alphen effect and the destabilizing effect of $B$
at large $\mu$, observed in NJL but not in SS.
The cause of the differences can be reduced to the general fact that
the SS model in the U configuration is totally independent of the chemical
potential.
Physically, if a model exhibits hadrons as boundstates of the quarks,
the spectrum should certainly depend on the chemical potential.
The SS model of the form we studied in this work does not show this behavior.
Moreover, the lack of the de Haas-van Alphen and the destabilizing
$B$ effects imply that the SS model
fails to incorporate the fermi sphere.
Therefore, it appears that we need extra ingredients to the SS model,
such as different brane configurations (which corresponds to additional
phases) and extra terms in the probe action.
Further investigation in this direction would deepen the understanding
of the model's universality.

In this work, we have concentrated on the case with $N_f=1$ and compared
the dynamical mass of NJL and SEP of SS.
It is of great interest to generalize the comparison to $N_f>1$ and
investigate the hadron spectrum, even quantitatively,
and the effects of other features, such as
the $U(1)_A$ breaking terms in both models.

%\pagebreak

%%%%%%%%%%%%%%%%%%%%%%%%%%%%%%%%%%%%%%%%%%%%%%%%%%%%%%%%%%%%%%%%%%%%%%%%
\section*{Acknowledgments}
%%%%%%%%%%%%%%%%%%%%%%%%%%%%%%%%%%%%%%%%%%%%%%%%%%%%%%%%%%%%%%%%%%%%%%%%
I would like to thank Shmuel Elitzur, Andy O'Bannon and Jacob Sonnenschein
for the useful discussions.
This work was supported by the Center of Excellence of the ISF.

\bigskip
\bigskip

%%%%%%%%%%%%%%%%%%%%%%%%%%%%%%%%%%%%%%%%%%%%%%%%%%%%%%%%%%%%%%%%%%%%%
%%%%%%%%%%%%%%%%%%%%%%%%%%%%%%%%%%%%%%%%%%%%%%%%%%%%%%%%%%%%%%%%%%%%%
\appendix
%%%%%%%%%%%%%%%%%%%%%%%%%%%%%%%%%%%%%%%%%%%%%%%%%%%%%%%%%%%%%%%%%%%%%
%%%%%%%%%%%%%%%%%%%%%%%%%%%%%%%%%%%%%%%%%%%%%%%%%%%%%%%%%%%%%%%%%%%%%

%%%%%%%%%%%%%%%%%%%%%%%%%%%%%%%%%%%%%%%%%%%%%%%%%%%%%%%%%%%%%%%%%%%%%
\section{Path Integrals of the Worldline Fields}\label{sec:IBIF}
%%%%%%%%%%%%%%%%%%%%%%%%%%%%%%%%%%%%%%%%%%%%%%%%%%%%%%%%%%%%%%%%%%%%%
We are going to evaluate the path integrals $I_B$ and $I_F$ defined
in the beginning of Subsection~\ref{subsec:Veff}.
The derivation is not original and can be found elsewhere.
We largely collect and reproduce the relevant parts
of Reference~\cite{Reuter:1996zm} for the completeness,
and specialize to the finite temperature system at the end.

We first consider the worldline fermionic part $I_F$.
We split the field $\psi$ into the part that satisfies the equation of motion
$\Psi$ (classical part) and the fluctuation part $\eta$ as
\begin{align}
  \psi_\mu(\tau) = \Psi_\mu(\tau) + \eta_\mu(\tau)
  \;,
\end{align}
where they both satisfy the anti-periodic condition.
The classical part $\Psi(\tau)$ satisfy
\begin{align}
  (\partial_\tau-2iF)_{\mu\nu}\Psi_\nu(\tau) = 0
  \;.
\end{align}
We can easily solve the equations of motion to get
\begin{align}
  \Psi_\mu (\tau) = \big( e^{2iF\tau} \big)_{\mu\nu}C_\nu
  \;,
\end{align}
where $C_\nu$ are some constants.
Since we have
\begin{align}
  \Psi_\mu(0) = C_\mu
  \quad\text{and}\quad
  \Psi_\mu(s) = \big( e^{2iFs} \big)_{\mu\nu}C_\nu
  \;,
\end{align}
the anti-periodic condition $\Psi(0)=-\Psi(s)$ implies that
\begin{align}
  \big(\mathbf{1} + e^{2iFs} \big)_{\mu\nu}C_\nu = 0
  \;.
\end{align}
Assuming that the first factor is invertible, we get $C_\mu = 0$.
Hence, we conclude that%
\footnote{
  As in Reference~\cite{Reuter:1996zm}, the action $\mathcal{L}_\psi$
  must be augmented by a surface term in general.
  However, in our case, the classical part vanishes and the surface
  term does not affect our discussion, hence omitted from the beginning.
}
\begin{align}
  \Psi_\mu(\tau) = 0
  \;.
\end{align}
We are therefore left with
\begin{align}
  I_F =& \int_A \mathcal{D}\psi_\mu(\tau)
        \exp[-\frac{1}{2}\int_0^sd\tau \eta\cdot(\partial_\tau-2iF)\cdot\eta]
      = ({\det}_A[\partial_\tau-2iF])^{1/2}
  \nonumber\\
      =& ({\det}_A[\partial_\tau])^{1/2}
         ({\det}_A[\mathbf{1}-2iF\partial_\tau^{-1}])^{1/2}
  \;,
\end{align}
where the subscript ``$A$'' of the determinant reminds us that 
the field is anti-periodic on the time circle and this accordingly
affects the momentum.
Notice that since the field $\eta(\tau)$ is real, the functional determinants
have the power of $1/2$.
Following Strassler~\cite{Strassler:1992zr}, we adopt the normalization
\begin{align}
  ({\det}_A[\partial_\tau])^{1/2} = \sum_{\alpha=1}^4\langle\alpha|\alpha\rangle
  = 4
  \;.
\end{align}
For the other factor, we quote the result from
Reference~\cite{Reuter:1996zm},
\begin{align}
  ({\det}_A[\mathbf{1}-2iF\partial_\tau^{-1}])^{1/2}
  = ({\det}_L[\cos(Fs)])^{1/2}
  \;,
\end{align}
where ${\det}_L$ is the determinant with respect to the Lorentz
structure.
Putting those results together, we obtain the expression in
Equation~(\ref{eq:IFfinal}).

Let us now deal with the bosonic part $I_B$.
As it always happens in a finite temperature field theory, the zero winding
mode of bosonic sector needs a special care.
As shown in Figure~\ref{fig:paths}, this mode has closed paths.
Among the paths, there are constant functions $x(\tau)\equiv x_0$ and 
this makes the functional determinant ill-defined.
Therefore, the constants $x_0$ should be separated from the path integral
and they all should be summed (or rather integrated) over.
However, this integral is precisely the one extracted in the definition
of the effective potential in Equation~(\ref{eq:Gamma}).
So in practice,
we do not have to worry about these constant paths and we can
proceed by assuming that they are absent.

By virtue of the Schwinger-Fock gauge, the Lagrangian $\mathcal{L}_{xSF}$
is quadratic and we rewrite it as
\begin{align}
  I_{B} =& \mathcal{N}\int_{x(0)=\bar x}^{x(s)=\bar y} \mathcal{D}x(\tau)
           \exp[-\frac{1}{4}\int_0^sd\tau 
                x \cdot (-\partial_\tau^2+2ieF\partial_\tau) \cdot x
  \nonumber\\
        & -\frac{1}{4}\{x(s)\cdot\dot x(s)-x(0)\cdot\dot x(0)\} ]
  \;,
\end{align}
where we kept the surface terms, because they will have the contribution
from the classical paths.
We split the field into the classical part $X(\tau)$ and the fluctuation
$z(\tau)$ as
\begin{align}
  x(\tau) = X(\tau) + z(\tau)
  \;,
\end{align}
where they satisfy
\begin{align}\label{eq:Xeom}
  (-\partial_\tau^2 + 2ieF\partial_\tau)_{\mu\nu} X_\nu(\tau) = 0
  \;,
\end{align}
and
\begin{align}
  X(0)=\bar x
  \;,\quad
  X(s)=\bar y
  \quad\text{and}\quad
  z(0)=0=z(s)
  \;.
\end{align}
The equations of motion (\ref{eq:Xeom}) with the boundary conditions can
be easily solved and yield
\begin{align}
  X(\tau) = \bar x + \frac{1-\exp(2ieF\tau)}{1-\exp(2ieFs)} (\bar y-\bar x)
  \;.
\end{align}
This gives the contribution to the surface terms of $I_{B}$,
\begin{align}
  -\frac{1}{4}\{x(s)\cdot\dot x(s)-x(0)\cdot\dot x(0)\}
  = -\frac{1}{4}(\bar y-\bar x)\cdot eF\cot(eFs)\cdot(\bar y-\bar x)
  \;.
\end{align}
The path integral now is over the fluctuation $z(\tau)$ and this gives
\begin{align}\label{eq:xdet}
  &\mathcal{N}({\det}'_P [-\partial_\tau^2+2ieF\partial_\tau])^{-1/2}
  = \mathcal{N}({\det}'_P [-\partial_\tau^2]^{-1/2})
    ({\det}'_P [\mathbf{1}-2ieF\partial_\tau^{-1}]^{-1/2})
  \nonumber\\
  &= (4\pi s)^{-d/2} \bigg({\det}_L\bigg[\frac{\sin(eFs)}{eFs}\bigg]\bigg)^{-1/2}
  \;,
\end{align}
where the prime on the determinant indicates that we omit the constant
paths, we have used Equation~(\ref{eq:Nrelation}) and the evaluation
of the last determinant in the first line is quoted from
Reference~\cite{Reuter:1996zm}.
We thus have
\begin{align}
  I_{B} = (4\pi s)^{-d/2} 
          \exp\bigg[-\frac{1}{4}(\bar y-\bar x)\cdot eF\cot(eFs)
                           \cdot(\bar y-\bar x)\bigg]
          \bigg({\det}_L\bigg[\frac{\sin(eFs)}{eFs}\bigg]\bigg)^{-1/2}
  \;.
\end{align}

Now according to the finite temperature prescription (\ref{eq:FTPriscrip}),
we should set
\begin{align}
  \bar y = \bar x + n\beta \hat x_0
  \;,
\end{align}
where $\hat x_0$ is the unit vector in the time direction.
We then have
\begin{align}
  I_B = (4\pi s)^{-d/2} 
          \exp\big[ -(n\beta/2)^2 \{eF\cot(eFs)\}_{00} \big]
          \bigg({\det}_L\bigg[\frac{\sin(eFs)}{eFs}\bigg]\bigg)^{-1/2}
  \;.
\end{align}

%%%%%%%%%%%%%%%%%%%%%%%%%%%%%%%%%%%%%%%%%%%%%%%%%%%%%%%%%%%%%%%%%%%%%
\section{Computation of the Functions of $F_{\mu\nu}$}\label{sec:F}
%%%%%%%%%%%%%%%%%%%%%%%%%%%%%%%%%%%%%%%%%%%%%%%%%%%%%%%%%%%%%%%%%%%%%
In this appendix, we compute $\big\{{\det}_L\big[(Fs)\cot(Fs)\big]\big\}^{1/2}$
and $\{(Fs)\cot(Fs)\}_{00}$.
The former can be evaluated along the method of
Schwinger~\cite{Schwinger:1951nm}
and the latter is difficult and we adopt the method of
Reference~\cite{Batalin:1971au} (see also \cite{Gusynin:1998bt}).
Both discussions are recast to the Euclidean version in below.

We have the Euclidean version of the field strength
\begin{align}
  F_{\mu\nu} = \begin{pmatrix}
                0 & -i E_1 & -i E_2 & -i E_3 \\
                iE_1 & 0 & B_3 & -B_2 \\
                iE_2 & -B_3 & 0 & B_1 \\
                iE_3 & B_2 & -B_1 & 0
              \end{pmatrix}
  \;,
\end{align}
and in the four spacetime dimension, we have the dual of the same rank
\begin{align}
  \tilde F_{\mu\nu} :=& \frac{1}{2}\epsilon_{\mu\nu\alpha\beta}F_{\alpha\beta}
  \nonumber\\
  =& \begin{pmatrix}
       0 & B_1 & B_2 & B_3 \\
       -B_1 & 0 & -iE_3 & iE_2 \\
       -B_2 & iE_3 & 0 & -iE_1 \\
       -B_3 & -iE_2 & iE_1 & 0
     \end{pmatrix}
  \;,
\end{align}
where we have used the convention $\epsilon_{0123}=+1$.
We then have the invariants
\begin{align}
  \mathcal{F} := -\frac{1}{4}F_{\mu\nu}F_{\mu\nu}
               = \frac{1}{2} (\vec B^2 - \vec E^2)
  \quad\text{and}\quad
  \mathcal{G} := -\frac{i}{4} F_{\mu\nu}\tilde F_{\mu\nu}
               = \vec E \cdot \vec B
  \;.
\end{align}

We first calculate the determinant.
The important identities are
\begin{align}\label{eq:iden1}
  F_{\mu\lambda}F_{\lambda\nu} + \tilde F_{\mu\lambda}\tilde F_{\lambda\nu}
  = -2 \mathcal{F} \delta_{\mu\nu}
  \;,
\end{align}
and
\begin{align}
  F_{\mu\lambda} \tilde F_{\lambda\nu} = i \mathcal{G} \delta_{\mu\nu}
  \;.
\end{align}
Using those, we derive the eigenvalues
of $F_{\mu\nu}$ in terms of $\mathcal{F}$ and $\mathcal{G}$.
Let $\psi_\mu$ and $f$ be the eigenvector and eigenvalue of $F_{\mu\nu}$,
respectively.
We then have
\begin{align}
  \tilde F_{\mu\nu} F_{\nu\lambda} \psi_{\lambda}
  = f \tilde F_{\mu\nu} \psi_{\nu}
  = i \mathcal{G} \psi_\mu
  \;,
\end{align}
and this implies that
\begin{align}
  \tilde F_{\mu\nu} \psi_{\nu} = i f^{-1} \mathcal{G} \psi_\mu
  \;.
\end{align}
By iteration, we have
\begin{align}
  F_{\mu\nu}F_{\nu\lambda} \psi_{\lambda} = f^2 \psi_{\mu}
  \quad\text{and}\quad
  \tilde F_{\mu\nu} \tilde F_{\nu\lambda} \psi_{\lambda} 
  = - f^{-2} \mathcal{G}^2 \psi_{\mu}
  \;.
\end{align}
Then the identity (\ref{eq:iden1}) gives the eigenvalue equation
\begin{align}
  f^4 + 2\mathcal{F}f^2 - \mathcal{G} = 0
  \;,
\end{align}
with the solutions $\pm f_\pm$
(signs are arbitrary, {\it i.e.}, 4 values)
where
\begin{align}
  f_\pm = i \sqrt{\mathcal{F}\pm\sqrt{\mathcal{F}^2+\mathcal{G}^2}}
  \;.
\end{align}

Note that any function of a matrix, $\Phi(F)$, is defined as the formal
power series.
Therefore, the argument of a determinant can be diagonalized and
the determinant reduces to the product of the functions
$\Pi_i\Phi(f_i)$, where $f_i$ are the eigenvalues of $F$.
We thus have
\begin{align}
  {\det}_L\big[(Fs)\cot(Fs)\big] =
  \big\{ s^2 f_+f_- \cot(f_+s)\cot(f_-s) \big\}^2
  \;.
\end{align}
Since we have
\begin{align}
  (f_+f_-)^2 = - \mathcal{G}^2
  \;,
\end{align}
we immediately get the result
\begin{align}
  &\big\{{\det}_L\big[(Fs)\cot(Fs)\big]\big\}^{1/2} =
  \nonumber\\
       &= s^2 \left|  \mathcal{G}
       \coth\bigg(s\sqrt{\mathcal{F}+\sqrt{\mathcal{F}^2+\mathcal{G}^2}}\bigg)
       \cot\bigg(s\sqrt{-\mathcal{F}+\sqrt{\mathcal{F}^2+\mathcal{G}^2}}\bigg)
       \right|
  \;.
\end{align}
\\

We now evaluate $\{(Fs)\cot(Fs)\}_{00}$.
We define
\begin{align}
  \begin{matrix}
  f_1 &:=& f_+ &=& i \sqrt{\mathcal{F}+\sqrt{\mathcal{F}^2+\mathcal{G}^2}}
  \;,\qquad &
  \bar f_1 &:=& if_- &=& - \sqrt{\mathcal{F}-\sqrt{\mathcal{F}^2+\mathcal{G}^2}}
  \;,
  \nonumber\\
  f_2 &:=& -f_+ &=& -i \sqrt{\mathcal{F}+\sqrt{\mathcal{F}^2+\mathcal{G}^2}}
  \;,\qquad &
  \bar f_2 &:=& -if_- &=&  \sqrt{\mathcal{F}-\sqrt{\mathcal{F}^2+\mathcal{G}^2}}
  \;,
  \nonumber\\
  f_3 &:=& -f_- &=& -i \sqrt{\mathcal{F}-\sqrt{\mathcal{F}^2+\mathcal{G}^2}}
  \;,\qquad &
  \bar f_3 &:=& -if_+ &=& \sqrt{\mathcal{F}+\sqrt{\mathcal{F}^2+\mathcal{G}^2}}
  \;,
  \nonumber\\
  f_4 &:=& f_- &=& i \sqrt{\mathcal{F}-\sqrt{\mathcal{F}^2+\mathcal{G}^2}}
  \;,\qquad &
  \bar f_4 &:=& if_+ &=& - \sqrt{\mathcal{F}+\sqrt{\mathcal{F}^2+\mathcal{G}^2}}
  \;,
  \end{matrix}
\end{align}
and a sign
\begin{align}
  s := \mathcal{G}/|\mathcal{G}| = |\mathcal{G}|/\mathcal{G}
  \;.
\end{align}
Then, the matrices
\begin{align}
  A_{(i)\mu\nu} := \frac{1}{2( f_i^2 + \bar f_i^2 )}
           (\bar f_i^2 \delta_{\mu\nu}
           + f_i F_{\mu\nu}
           + F_{\mu\lambda}F_{\lambda\nu}
           -is \bar f_i \tilde F_{\mu\nu} )
\end{align}
satisfy the relations
\begin{align}
  F_{\mu\lambda}A_{(i)\lambda\nu} = f_i A_{(i)\mu\nu}
  = A_{(i)\mu\lambda}F_{\lambda\nu}
  \quad\text{and}\quad
  \sum_i A_{(i)\mu\nu} = \delta_{\mu\nu}
  \;.
\end{align}
Using these two relations, we have for a function $\Phi(F)$,
\begin{align}\label{eq:key}
  \Phi(F)_{\mu\nu} = \sum_i A_{(i)\mu\nu}\Phi(f_i)
  \;.
\end{align}
This is the key point of introducing the matrices $A_{(i)}$.
Now, from this relation, we have
\begin{align}
  \{ (Fs)\cot(eFs) \}_{00}
  = \sum_i  A_{(i)00} (f_is)\cot(f_is) 
  \;.
\end{align}
Since
\begin{align}
  F_{00} = 0 = \tilde F_{00}
  \quad\text{and}\quad
  F_{0\lambda}F_{\lambda 0} = \vec E^2
  \,
\end{align}
we have
\begin{align}
  A_{(i)00} = \frac{\bar f_i^2 + \vec E^2}{2(f_i^2+\bar f_i^2)}
  \;,
\end{align}
and thus
\begin{align}
  \{ (Fs)\cot(Fs) \}_{00}
  = \frac{s}{f_-^2-f_+^2}
    \big\{ (f_-^2-\vec E^2)f_+\cot(f_+s) - (f_+^2-\vec E^2)f_-\cot(f_-s) 
    \big\}
  \;.
\end{align}
We have
\begin{align}
  f_-^2-f_+^2 = 2\sqrt{\mathcal{F}^2+\mathcal{G}^2}
  \;.
\end{align}
and
\begin{align}
  f_\pm^2 - \vec E^2 = -\frac{1}{2}(\vec B^2+\vec E^2)
                    \mp \sqrt{\mathcal{F}^2+\mathcal{G}^2}
  \;.
\end{align}
We also have
\begin{align}
  f_+\cot(f_+s) =&
        \sqrt{\mathcal{F}+\sqrt{\mathcal{F}^2+\mathcal{G}^2}}
        \coth(s\sqrt{\mathcal{F}+\sqrt{\mathcal{F}^2+\mathcal{G}^2}})
  \nonumber\\
  f_-\cot(f_-s) =&
        \sqrt{-\mathcal{F}+\sqrt{\mathcal{F}^2+\mathcal{G}^2}}
        \cot(s\sqrt{-\mathcal{F}+\sqrt{\mathcal{F}^2+\mathcal{G}^2}})
  \;.
\end{align}
So we get
\begin{align}
  &\{ eF\cot(eFs) \}_{00} =\frac{1}{2\sqrt{\mathcal{F}^2+\mathcal{G}^2}}
    \bigg[ 
  \nonumber\\
        &\bigg\{-\frac{1}{2}(\vec B^2+\vec E^2) 
                 + \sqrt{\mathcal{F}^2+\mathcal{G}^2} 
           \bigg\}
        e\sqrt{\mathcal{F}+\sqrt{\mathcal{F}^2+\mathcal{G}^2}}
        \coth\bigg(es\sqrt{\mathcal{F}+\sqrt{\mathcal{F}^2+\mathcal{G}^2}}\bigg)
  \nonumber\\
        &+
        \bigg\{\frac{1}{2}(\vec B^2+\vec E^2) 
                 + \sqrt{\mathcal{F}^2+\mathcal{G}^2} 
           \bigg\}
        e\sqrt{-\mathcal{F}+\sqrt{\mathcal{F}^2+\mathcal{G}^2}}
        \cot\bigg(es\sqrt{-\mathcal{F}+\sqrt{\mathcal{F}^2+\mathcal{G}^2}}\bigg)
   \bigg]
  \;.
\end{align}

%%%%%%%%%%%%%%%%%%%%%%%%%%%%%%%%%%%%%%%%%%%%%%%%%%%%%%%%%%%%%%%%%%%%%
\section{Treatment of Large Chemical Potential}\label{sec:largeMu}
In Section~\ref{subsec:pureB}, we have pointed out that a special treatment
is necessary when the chemical potential of the system is too large.
In this appendix, we give the details of how to handle this case.
The discussion here is somewhat similar to Reference~\cite{Inagaki:2003ac}.
For simplicity, we first consider the system without the constant
background field, hence the goal is to evaluate
\begin{align}\label{eq:sIntegral}
     \frac{T}{4\pi^{3/2}}\int_1^\infty ds s^{-5/2}
     \sum_{l\in\mathbb{Z}_{1/2}}
     \exp\big[ -s \big\{ (2\pi Tl-i\mu)^2 + m^2 \big\} \big]
  \;,
\end{align}
for a general value of $\mu$.
The inclusion of $B$-field (and $E$-field) can be done in relative ease.

\begin{figure}[ht]
{
\centerline{\scalebox{1.0}{\includegraphics{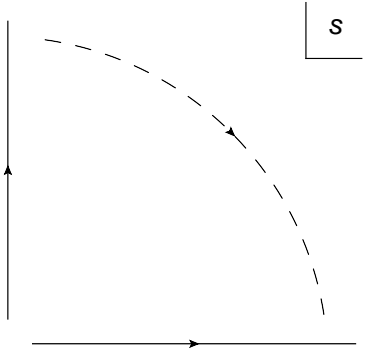}}}
\caption{\footnotesize
  Analytic continuation of real to imaginary propertime.
  Since our notation is that $s$ is the imaginary propertime,
  the vertical axis corresponds to the real propertime.
}\label{fig:ACont1}
}
\end{figure}

We first discuss the analytic continuation of the propertime in general.
Recall that we have dealt with the ``Euclidean worldline Lagrangian''
appearing in Equations~(\ref{eq:pathInt},\ref{eq:EucL}).
This stems from our choice of the propertime variable in
Equation~(\ref{eq:imPTime}).
There, we chose it so that the exponential appears as real, whereas
Schwinger chooses in Ref.~\cite{Schwinger:1951nm} as
\begin{align}
  \ln X = - \int_0^\infty \frac{e^{-iXt}}{t} dt
  \;.
\end{align}
This choice leads to the evolution equation of the {\ real}
propertime.
Therefore, we can think of our choice as the analytically continued
version, {\it i.e.}, the {\it imaginary} propertime $s$.
See Figure~\ref{fig:ACont1}.

\begin{figure}[ht]
{
\centerline{\scalebox{0.8}{\includegraphics{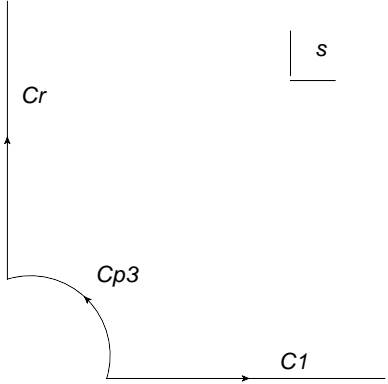}}}
\caption{\footnotesize
  With the introduction of the cutoff, the real and imaginary propertimes
  have difference due to the quarter circle.
  We therefore, {\it define} the real counterpart with the quarter-circle
  contribution: $C_1 \Leftrightarrow C_r + C_{p3}$.
}\label{fig:ACont2}
}
\end{figure}

We note that the situation is a little more complicated when
we must introduce the cutoff at $s=1$.
In this case, it seems as if we have discrepancy with the real propertime
case due to the quarter circle contribution.
See Figure~\ref{fig:ACont2}.
However, the cutoff is designed to make the imaginary
propertime integral well-defined, and it is nonsensical in the real one.
Therefore, we {\it define} the world of real propertime as 
the analytic continuation of the imaginary propertime with the cutoff.
This implies that the contour $C_1$ in Figure~\ref{fig:ACont2}
corresponds to $C_r+C_{p3}$.

\begin{figure}[ht]
{
\centerline{\scalebox{0.8}{\includegraphics{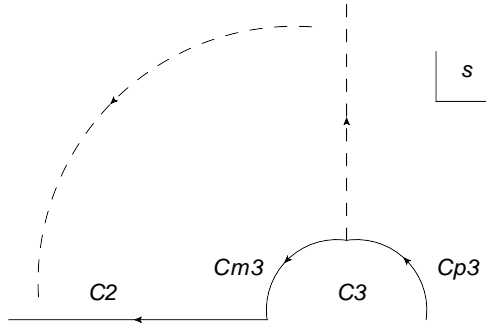}}}
\caption{\footnotesize
  The analytic continuation from positive to negative real axis.
  The contour $C_1$ now corresponds to $C_2+C_3$, where
  $C_3:=C_{p3}+C_{m3}$.
}\label{fig:ACont3}
}
\end{figure}

Now, we notice that we have choices to analytically continue from the
real to imaginary propertime: $t\to s$ or $t\to -s$,
in other words, it is also possible to analytically continue
to the left of Figure~\ref{fig:ACont1}.
The former choice has been made in Equation~(\ref{eq:imPTime})
and we found that this choice is not suitable when the chemical
potential is too large.
When this is the case, we must pick the other choice.
This amounts to the analytic continuation of the variable $s$
in Equation~(\ref{eq:sIntegral}) from the positive to
the negative direction as shown in Figure~\ref{fig:ACont3}.

\begin{figure}[ht]
{
\centerline{\scalebox{0.6}{\includegraphics{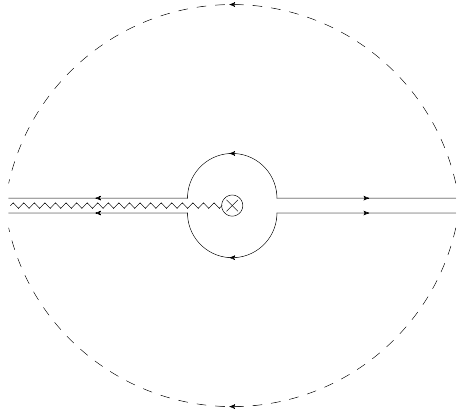}}}
\caption{\footnotesize
  Analytic continuation of the variable $s$ from positive to
  negative real axis.
  The upper (resp. lower) contour should be chosen for positive
  (resp. negative) $l$.
  The branch cut comes from the factor of $s^{-5/2}$ in 
  Equation~(\ref{eq:sIntegral}).
  The branch cut must run along the negative real axis to avoid
  the cross-over with the contours.
  Our convention is that when we take the lower contour, the value
  of $s^{-5/2}$ is taken to be on the second branch for the negative
  real axis.
}\label{fig:ACont4}
}
\end{figure}

We must be careful about which way we should close the contour.
The imaginary part of the exponential in Equation~(\ref{eq:sIntegral})
is $2\pi i\mu Tls$ and this tells us
that we should close the contour above when $l$ is positive
and below when $l$ is negative,
so that the large circles do not contribute to the integral.
See Figure~\ref{fig:ACont4}.
Finally, we must be careful with the branch cut that arise from the factor of
$s^{-5/2}$ in Equation~(\ref{eq:sIntegral}).
The contour should not cross the cut and this is possible only when
the cut runs along the negative horizontal axis.
This causes the ambiguity in choosing the branch of $s^{-5/2}$ on
the negative real axis.
We take the convention that when we close the contour below, we choose
the second branch value.
Later, we will show that this is the consistent choice.

When we have the $B$ field, we have poles along the vertical
axis in the $s$-plane and we must pick them up, as they are within
the contour.
We will deal with this later.
\\

Let us actually evaluate Equation~(\ref{eq:sIntegral}).
As mentioned in Section~\ref{subsec:pureB},
the integral converges only when
\begin{align}
  (2\pi Tl)^2 - \mu^2 + m^2 >0
  \;,
\end{align}
is satisfied.
Without loss of generality, let us assume that $\mu\geq 0$ and $m\geq 0$.
When $\mu<m$, this condition is satisfied for all $l$, but when
$\mu>m$, it is satisfied only for $|l|>\bar l$ with
\begin{align}
  \bar l := \big[ (2\pi T)^{-1}\theta(\mu-m)\sqrt{\mu^2-m^2} \big]_G
  \;,
\end{align}
where $\theta(x)$ is the Heaviside theta and the symbol
$[x]_G$ is the half-integer version of the Gauss symbol, {\it i.e.},
it is the largest half integer less than or equal to $x$.
Notice, in particular, that when $\mu<m$, we have $\bar l=-1/2$
and $l$ runs over all half integers.
According to the previous discussion,
Equation~(\ref{eq:sIntegral}) must be split at $\bar l$
and for positive and negative $l$.
Thus, we write
\begin{align}\label{eq:splitIntSum}
  \text{Eqn.~(\ref{eq:sIntegral})}
   &=\frac{T}{4\pi^{3/2}}\int_{C_1} ds s^{-5/2}
    \sum_{|l|>\bar l}\exp\big[ -s \big\{ (2\pi Tl-i\mu)^2+m^2 \big\} \big]
  \nonumber\\
   &+\frac{T}{4\pi^{3/2}}\int_{(C_2+C_3)\text{up}} ds s^{-5/2}
    \sum_{l=1/2}^{\bar l}\exp\big[ -s \big\{ (2\pi Tl-i\mu)^2+m^2 \big\} \big]
  \nonumber\\
   &+\frac{T}{4\pi^{3/2}}\int_{(C_2+C_3)\text{down}} ds s^{-5/2}
    \sum_{l=-\bar l}^{-1/2}\exp\big[ -s \big\{ (2\pi Tl-i\mu)^2+m^2 \big\} \big]
  \;,
\end{align}
where ``up'' and ``down'' indicated on the integration contour implies the way
the contour is closed as in Figure~\ref{fig:ACont4}.
Consider the integral in the last line.
Since the both integral and sum are well-defined now, we may exchange them.
Also we send the summation variable $l$ to $-l$.
We then have
\begin{align}
  \frac{T}{4\pi^{3/2}}&\sum_{l=1/2}^{\bar l}
  \int_{(C_2+C_3)\text{down}} ds s^{-5/2}
  \exp\big[ -s \big\{ (2\pi Tl+i\mu)^2+m^2 \big\} \big]
  \nonumber\\
  =&\frac{T}{4\pi^{3/2}}\sum_{l=1/2}^{\bar l}
  \int_{(C_2)\text{down}} ds s^{-5/2}
  \exp\big[ -s \big\{ (2\pi Tl+i\mu)^2+m^2 \big\} \big]
  \nonumber\\
  &+\frac{T}{4\pi^{3/2}}\sum_{l=1/2}^{\bar l}
  \int_{(C_3)\text{down}} ds s^{-5/2}
  \exp\big[ -s \big\{ (2\pi Tl+i\mu)^2+m^2 \big\} \big]
  \nonumber\\
  =&-i\frac{T}{4\pi^{3/2}}\sum_{l=1/2}^{\bar l}
  \int_1^\infty ds s^{-5/2}
  \exp\big[ s \big\{ (2\pi Tl+i\mu)^2+m^2 \big\} \big]
  \nonumber\\
  &-i\frac{T}{4\pi^{3/2}}\sum_{l=1/2}^{\bar l}
  \int_0^\pi d\phi e^{\frac{3}{2}i\phi}
  \exp\big[ -e^{-i\phi} \big\{ (2\pi Tl+i\mu)^2+m^2 \big\} \big]
  \;,
\end{align}
where we have taken the second branch value for the negative real axis
as our convention and the sign of the integral variable $\phi$ was
reversed so that now the integration range is $(0,\pi)$ rather than
$(0,-\pi)$.
A little inspection shows that this is exactly the complex conjugate of
the second line in Equation~(\ref{eq:splitIntSum}).

We can also split the first line in Equation~(\ref{eq:splitIntSum}) into
positive and negative sums and it is easy to see that they are complex
conjugate of each other.
Therefore, we can write
\begin{align}\label{eq:fWOF}
  \text{Eqn.~(\ref{eq:sIntegral})} &=
  \frac{T}{2\pi^{3/2}}\opname{Re} \bigg[
    \sum_{l>\bar l}^\infty \int_1^\infty ds s^{-5/2}
    \exp\big[ -s \big\{ (2\pi Tl-i\mu)^2+m^2 \big\} \big]
  \nonumber\\
   &+i\sum_{l=1/2}^{\bar l}\int_1^\infty ds s^{-5/2}
   \exp\big[ s \big\{ (2\pi Tl-i\mu)^2+m^2 \big\} \big]
  \nonumber\\
   &+i\sum_{l=1/2}^{\bar l}\int_0^\pi d\phi e^{-\frac{3}{2}i\phi}
   \exp\big[ -e^{i\phi} \big\{ (2\pi Tl-i\mu)^2+m^2 \big\} \big]
  \bigg]
  \;.
\end{align}

After so much ado, we point out the following.
In Equation~(\ref{eq:sIntegral}), assuming that $|l|>\bar l$,
we can, in principle, carry out $s$-integral before the sum.
The result is a function of $l$ and valid only when $|l|>\bar l$.
{\it However, we can analytically continue} the integral beyond
the validity range
of $l$ to cover all half integers, without worrying about the case by
case examination that we have done.
In fact, one can find a closed form of the $s$ integral in
Equation~(\ref{eq:sIntegral}) and verify this claim by comparing
with the result (\ref{eq:fWOF}).
This in turn implies the validity of the choice of the branch cut
for the lower contour discussed before.
In practice, when the $B$ field is present, we are not able to
find the closed form of $s$-integral and must resort to
the case by case integral.
In what follows, we discuss this case more.
\\

As mentioned earlier, the case with the pure $B$-field has simple poles
along the imaginary axis in the $s$-plane, due to the factor of
hyperbolic cosine in Equation~(\ref{eq:VeffB}).
Therefore, when we must analytically continue $s$ to the negative
horizontal axis, we need to pick up the poles.
The condition for $\bar l$, which determines the point for the analytic
continuation, is not affected by the presence of the hyperbolic cosine,
because it approaches $1$ at large $s$.
Hence the expressions for this case is similar to the previous case
except for the extra contributions from the poles.
\begin{align}
  &\hatVeff =
  \nonumber\\
  &\frac{m^2}{2g^2}
  + \frac{T}{2\pi^{3/2}}\opname{Re} \bigg[
    \int_1^\infty ds (sB)\coth(sB) s^{-5/2}\sum_{l>\bar l}
    \exp\big[ -s \big\{ (2\pi Tl-i\mu)^2+m^2 \big\} \big]
  \nonumber\\
   &+i\int_1^\infty ds (sB)\coth(sB) s^{-5/2}\sum_{l=1/2}^{\bar l}
   \exp\big[ s \big\{ (2\pi Tl-i\mu)^2+m^2 \big\} \big]
  \nonumber\\
   &+i\int_0^\pi d\phi (e^{i\phi}B)\coth(e^{i\phi}B)
   e^{-\frac{3}{2}i\phi}\sum_{l=1/2}^{\bar l}
   \exp\big[ -e^{i\phi} \big\{ (2\pi Tl-i\mu)^2+m^2 \big\} \big]
  \nonumber\\
  &+(2\pi i)\sum_{l=1/2}^{\bar l}\sum_{k=k_b}^\infty e^{-\frac{3}{4}\pi i}
    \bigg(\frac{B}{k\pi}\bigg)^{3/2}
    \exp\big[ -ik\frac{\pi}{B} \big\{ (2\pi Tl-i\mu)^2+m^2 \big\} \big]
  \bigg]
  \;.
\end{align}
Since the location of the poles are at
\begin{align}
  s = in\pi/B
  \quad\text{for all}\quad
  n\in \mathbb{Z}
  \;,
\end{align}
for sufficiently large $B$, the pole at $n=1$ (and more) comes inside the unit
circle that we are excluding.
Thus naively, we might exclude those poles that are inside the unit circle,
{\it i.e.}, we set
\begin{align}
  k_b = \bigg[\frac{B}{\pi}+1\bigg]_G
  \;,
\end{align}
where $[y]_G$ here is the usual integer Gauss symbol.
However, numerical evaluation reveals that this cross over results in
a huge jump in the effective potential and it does not appear physical.
This is the indication that the $B$ field may not be much larger than
the cutoff so that the poles of $n>0$ should not cross the circle.
This means that we must have $B<\pi$ and $k_b = 1$.

\pagebreak

%%%%%%%%%%%%%%%%%%%%%%%%%%%%%%%%%%%%%%%%%%%%%%%%%%%%%%%%%%%%%%%%%%%%%

\end{document}